\documentclass[12pt]{article}

\usepackage{verbatim,color,amssymb}
\usepackage{amsmath}					
\usepackage{amsthm}					
\usepackage{natbib}
\usepackage{multirow}
\usepackage{setspace}
\usepackage[mathscr]{euscript}
\usepackage{fancyhdr}
\usepackage{enumitem}
\usepackage{graphicx}
\usepackage{lineno}
\usepackage{hyperref}
\usepackage{array,booktabs}
\usepackage{subcaption}

\usepackage{multibib}
\newcites{latex}{References}

\usepackage{lscape}

\usepackage{setspace}
\usepackage{tikz}
\usetikzlibrary{arrows}
\usepackage{multirow}
\usepackage{wrapfig}

\newtheorem*{Proof*}{Proof}

\def\cC{\mathbb{C}}

\def\eE{\mathbb{E}}

\def\C{{\cal C}}

\def\K{{\cal K}}

\def\diag{\hbox{diag}}

\def\wh{\widehat}
\def\wt{\widetilde}

\def\diag{\hbox{diag}}

\def\var{\hbox{var}}
\def\cov{\hbox{cov}}

\def\Bern{\hbox{Bernoulli}}

\def\Dir{\hbox{Dir}}

\def\IG{\hbox{Inv-Ga}}

\def\Laplace{\hbox{Laplace}}
\def\MVN{\hbox{MVN}}

\def\Normal{\hbox{Normal}}
\def\TN{\hbox{TN}}
\def\Unif{\hbox{Unif}}
\def\Mult{\hbox{Mult}}

\def\P_25_ICML{{\it Proceedings of the 25th international conference on Machine learning}}

\def\bse{\begin{eqnarray*}}
\def\ese{\end{eqnarray*}}
\def\be{\begin{eqnarray}}
\def\ee{\end{eqnarray}}
\def\bq{\begin{equation}}
\def\eq{\end{equation}}

\def\wh{\widehat}

\def\trans{^{\rm T}}

\def\th{^{th}}

\def\b1e{{\mathbf e}}

\def\bB{{\mathbf B}}
\def\bc{{\mathbf c}}

\def\bD{{\mathbf D}}
\def\bG{{\mathbf G}}
\def\bI{{\mathbf I}}
\def\bk{{\mathbf k}}

\def\bM{{\mathbf M}}

\def\bP{{\mathbf P}}
\def\bR{{\mathbf R}}
\def\bs{{\mathbf s}}
\def\bS{{\mathbf S}}
\def\bt{{\mathbf t}}

\def\bu{{\mathbf u}}
\def\bU{{\mathbf U}}
\def\bv{{\mathbf v}}
\def\bV{{\mathbf V}}
\def\bw{{\mathbf w}}

\def\bx{{\mathbf x}}

\def\by{{\mathbf y}}
\def\bY{{\mathbf Y}}
\def\bz{{\mathbf z}}

\def\bS{{\mathbf S}}
\def\bzero{{\mathbf 0}}

\newcommand{\bmu}{\mbox{\boldmath $\mu$}}

\newcommand{\bpi}{\mbox{\boldmath $\pi$}}

\newcommand{\bvartheta}{\mbox{\boldmath $\vartheta$}}
\newcommand{\bepsilon}{\mbox{\boldmath $\epsilon$}}
\newcommand{\btheta}{\mbox{\boldmath $\theta$}}

\newcommand{\bzeta}{\mbox{\boldmath $\zeta$}}
\newcommand{\bsigma}{\mbox{\boldmath $\sigma$}}
\newcommand{\bSigma}{\mbox{\boldmath $\Sigma$}}

\newcommand{\blambda}{\mbox{\boldmath $\lambda$}}
\newcommand{\bLambda}{\mbox{\boldmath $\Lambda$}}

\newcommand{\abs}[1]{\left\vert#1\right\vert}

\renewcommand\footnoterule{\kern-3pt \hrule \textwidth 2in \kern 2.6pt}

\def\boxit#1{\vbox{\hrule\hbox{\vrule\kern6pt \vbox{\kern6pt \textcolor{blue}{#1}\kern6pt}\kern6pt\vrule}\hrule}}

\def\authorfootnote#1{{\let\thefootnote\relax\footnotetext{#1}}}

\newcommand{\mybox}[4]{
    \begin{figure}[h]
     \small 
     \vskip -35pt
     \centering
    \begin{tikzpicture}
        \node[anchor=text,text width=\columnwidth-1.2cm, draw, rounded corners, line width=1pt, fill=#3, inner sep=5mm] (big) {\\#4};
        \node[draw, rounded corners, line width=.5pt, fill=#2, anchor=west, xshift=5mm] (small) at (big.north west) {#1};
    \end{tikzpicture}
    \end{figure}
}

\allowdisplaybreaks
\begin{document}
\thispagestyle{empty}
\baselineskip=28pt

\begin{center}
{\LARGE{\bf 
%Bayesian Multivariate Density Deconvolution in the Presence of Covariates
Bayesian Semiparametric \\
Covariate Informed 
Multivariate Density Deconvolution
%for Nutritional Epidemiology
}}
\end{center}
\baselineskip=12pt

\vskip 2mm
\begin{center}
Abhra Sarkar\\
abhra.sarkar@utexas.edu \\
Department of Statistics and Data Sciences,
The University of Texas at Austin\\
2317 Speedway D9800, Austin, TX 78712-1823, USA\\
\end{center}

\vskip 8mm
\begin{center}
{\Large{\bf Abstract}} 
\end{center}
Estimating the marginal and joint densities of the long-term average intakes of different dietary components is an important problem in nutritional epidemiology. 
Since these variables cannot be directly measured, data are usually collected in the form of 24-hour recalls of the intakes. 
The problem of estimating the density of the latent long-term average intakes from their observed but error contaminated recalls 
then becomes a problem of multivariate deconvolution of densities. 
The underlying densities could potentially vary with the subjects' demographic characteristics such as sex, ethnicity, age, etc. 
The problem of density deconvolution in the presence of associated precisely measured covariates has, however, never been considered before, 
not even in the univariate setting.  
We present a flexible Bayesian semiparametric approach to covariate informed multivariate deconvolution. 
Building on recent advances on copula deconvolution and 
conditional tensor factorization techniques, 
our proposed method 
not only allows the joint and the marginal densities to vary flexibly with the associated predictors 
but also allows automatic selection of the most influential predictors. 
Importantly, the method also allows the density of interest and the density of the measurement errors to vary with potentially different sets of predictors.  
We design Markov chain Monte Carlo algorithms that enable efficient posterior inference, 
appropriately accommodating uncertainty in all aspects of our analysis. 
The empirical efficacy of the proposed method is illustrated through simulation experiments. 
Its practical utility is demonstrated in the afore-described nutritional epidemiology applications in estimating covariate adjusted long term intakes of different dietary components. 
An important by-product of the approach is a solution to covariate informed ordinary multivariate density estimation. 
Supplementary materials include substantive additional details and R codes are also available online. 

\baselineskip=12pt

\vskip 8mm
\baselineskip=12pt
%\par\vfill
\noindent\underline{\bf Some Key Words}: Copula, Covariates, 
Multivariate density regression, 
Multivariate density deconvolution, 
Measurement error, 
Nutritional epidemiology,
Tensor factorization.

\par\medskip\noindent
\underline{\bf Short/Running Title}: Covariate Informed Multivariate Deconvolution

%\par\medskip\noindent
%\underline{\bf Corresponding Author}: Abhra Sarkar (abhra.sarkar@utexas.edu) 

\clearpage\pagebreak\newpage
\pagenumbering{arabic}
\newlength{\gnat}
\setlength{\gnat}{16pt}  % single spacing = 14 its
\baselineskip=\gnat

\section{Introduction}

The distribution of the dietary intakes can provide answers to important questions 
such as what proportion of the population consume certain dietary components above, between or below certain amounts etc. 
The last question is particularly important as it relates to the proportion of the population that are deficient in certain dietary components. 
Estimating the long-term average intakes of different dietary components $\bx$ and their marginal and joint distributions 
is thus a fundamentally important problem in nutritional epidemiology.

By the very nature of the problem, $\bx$ can never be observed directly. 
Data are thus often collected in the form of 24-hour recalls of the intakes. 
Treating the recalls $\bw$, shown in Table \ref{tab: EATS data structure}, 
to be surrogates for the latent $\bx$ contaminated with additive measurement errors $\bu$ generated as $\bw=\bx+\bu$, 
the problem of estimating the joint and marginal distributions of $\bx$ from the recalls $\bw$ then becomes a problem of multivariate deconvolution of densities.

Dietary intakes may potentially vary with additional precisely measured demographic covariates $\bc$ such as sex, ethnicity and age. 
Women, for example, consume practically all dietary components in lesser amounts compared to men, on average. 
To our knowledge, however, the problem of deconvolution in the presence of covariates has never been considered in the literature, 
not even in the univariate setting, not at least in a statistically principled manner. 
This article attempts to address this gap, developing a novel Bayesian semiparametric approach 
that not only allows robust estimation of the density of $\bx$ as it varies with $\bc$
while also letting the density of the measurement errors $\bu$ to depend flexibly on both $\bx$ and $\bc$ 
but also additionally selects the most important predictors influencing the distributions of $\bx$ and $\bu$ from the set of all available predictors $\bc$. 

We adopt the following generic notation for marginal, joint and conditional densities, respectively. 
For random vectors $\bs$ and $\bt$, we denote the marginal density of $\bs$, 
the joint density of $(\bs,\bt)$, 
and the conditional density of $\bs$ given $\bt$, 
by the generic notation $f_{\bs}, f_{\bs,\bt}$ and $f_{\bs\mid \bt}$, respectively.
Likewise, for univariate random variables $s$ and $t$, the corresponding densities are denoted by $f_{s},f_{s,t}$ and $f_{s\mid t}$, respectively. 
To avoid introducing more notation, with some abuse, 
barring few exceptions, for any random variable $s$ or vector $\bs$, 
their specific values would also be denoted by the same notation, i.e., $s$ and $\bs$. 

{\bf The EATS Data Set:} 
The Eating at America's Table Study (EATS) \citep{Subar2001} is a large scale epidemiological study conducted by the National Cancer Institute 
in which $i=1,\dots,n = 965$ participants were interviewed $j=1,\dots,m_{i}=4$ times over the course of a year 
and, for many different dietary components $\ell$, their 24-hour dietary recalls $w_{\ell,i,j}$were recorded. 
Error free demographic covariates $\bc_{i}=(c_{1,i},c_{2,i},c_{3,i})\trans \equiv \text{(sex, ethnicity, age)}\trans$ are additionally available for each individual $i$.

\begin{table}[!ht]\footnotesize
\begin{center}
\begin{tabular}{|c|c|c|c|c c c c|c c c c|}
\hline
\multirow{2}{30pt}{Subject} & \multirow{2}{20pt}{Sex} & \multirow{2}{20pt}{Ethn} & \multirow{2}{20pt}{Age} & \multicolumn{8}{|c|}{24-hour recalls} \\ \cline{5-12}
   					  &					 &					 & 					& \multicolumn{4}{|c|}{Dietary Component 1}  &  \multicolumn{4}{|c|}{Dietary Component 2}	\\ \hline\hline
1 & $c_{1,1}$ & $c_{2,1}$ & $c_{3,1}$	& $w_{1,1,1}$ 	& $w_{1,1,2}$  & $w_{1,1,3}$ 	& $w_{1,1,4}$ 		& $w_{2,1,1}$ 	& $w_{2,1,2}$ 	& $w_{2,1,3}$ 	& $w_{2,1,4}$\\ \hline
2 & $c_{1,2}$ & $c_{2,2}$ & $c_{3,2}$	& $w_{1,2,1}$ 	& $w_{1,2,2}$ 	& $w_{1,2,3}$ 	& $w_{1,2,4}$ 		& $w_{2,1,1}$ 	& $w_{2,2,2}$ 	& $w_{2,2,3}$ 	& $w_{2,2,4}$\\ \hline
$\vdots$ & $\vdots$ & $\vdots$ & $\vdots$ & $\vdots$ 	& $\vdots$ 	& $\vdots$ 	& $\vdots$ 		& $\vdots$ 	& $\vdots$ 	& $\vdots$ 	& $\vdots$ \\ \hline
n & $c_{1,n}$ & $c_{2,n}$ & $c_{3,n}$	& $w_{1,n,1}$ 	& $w_{1,n,2}$ 	& $w_{1,n,3}$ 	& $w_{1,n,4}$ 		& $w_{2,n,1}$ 	& $w_{2,n,2}$ 	& $w_{2,n,3}$ 	& $w_{2,n,4}$ \\
\hline
\end{tabular}
\caption{\baselineskip=10pt The EATS data set showing the recalls for two regularly consumed dietary components and associated subject-specific predictors. 
Here $w_{\ell,i,j}$ is the reported intake for the $j\th$ recall of the $i\th$ individual for the $\ell\th$ dietary component; 
$c_{h,i}$ is the value of the $h\th$ predictor for the $i\th$ individual. 
\vspace{-20pt}
}
\label{tab: EATS data structure}
\end{center}
\end{table}

\begin{figure}[!ht]
\begin{center}
\includegraphics[height=4.8cm, trim=0cm 0cm 0cm 0cm, clip=true]{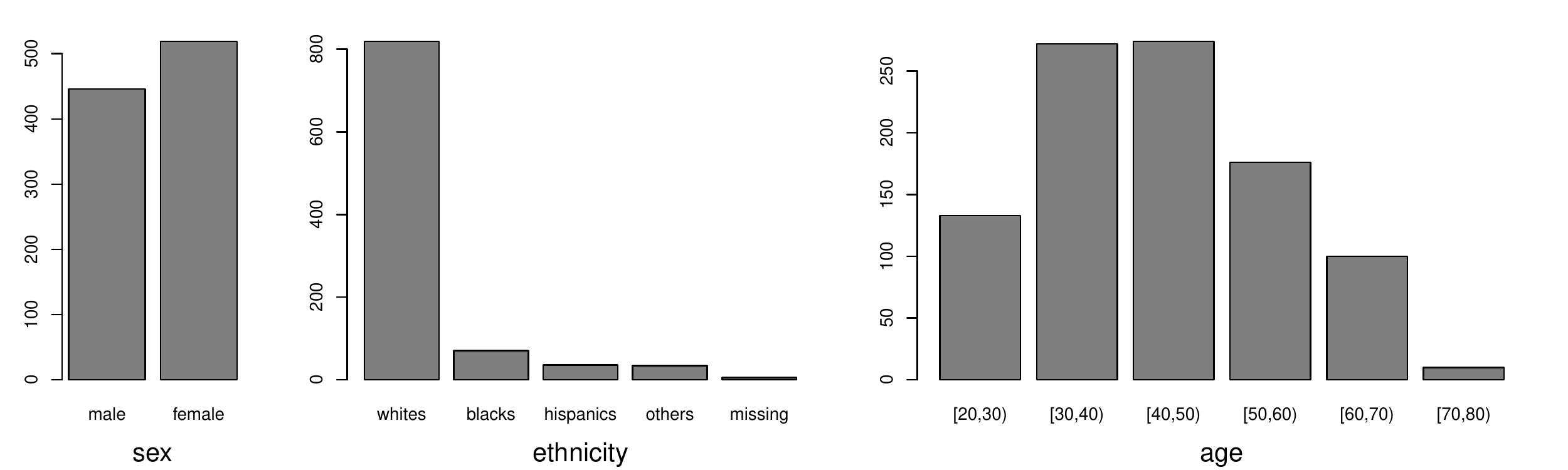}
\end{center}
\vspace{-15pt}
\caption{\baselineskip=10pt Observed distributions of the demographic predictors in the EATS data set. 
}
\label{fig: EATS exploratory cat dists}
\end{figure}

\begin{figure}[!ht]
\begin{center}
\includegraphics[height=16.5cm, trim=0cm 0cm 0cm 0cm, clip=true]{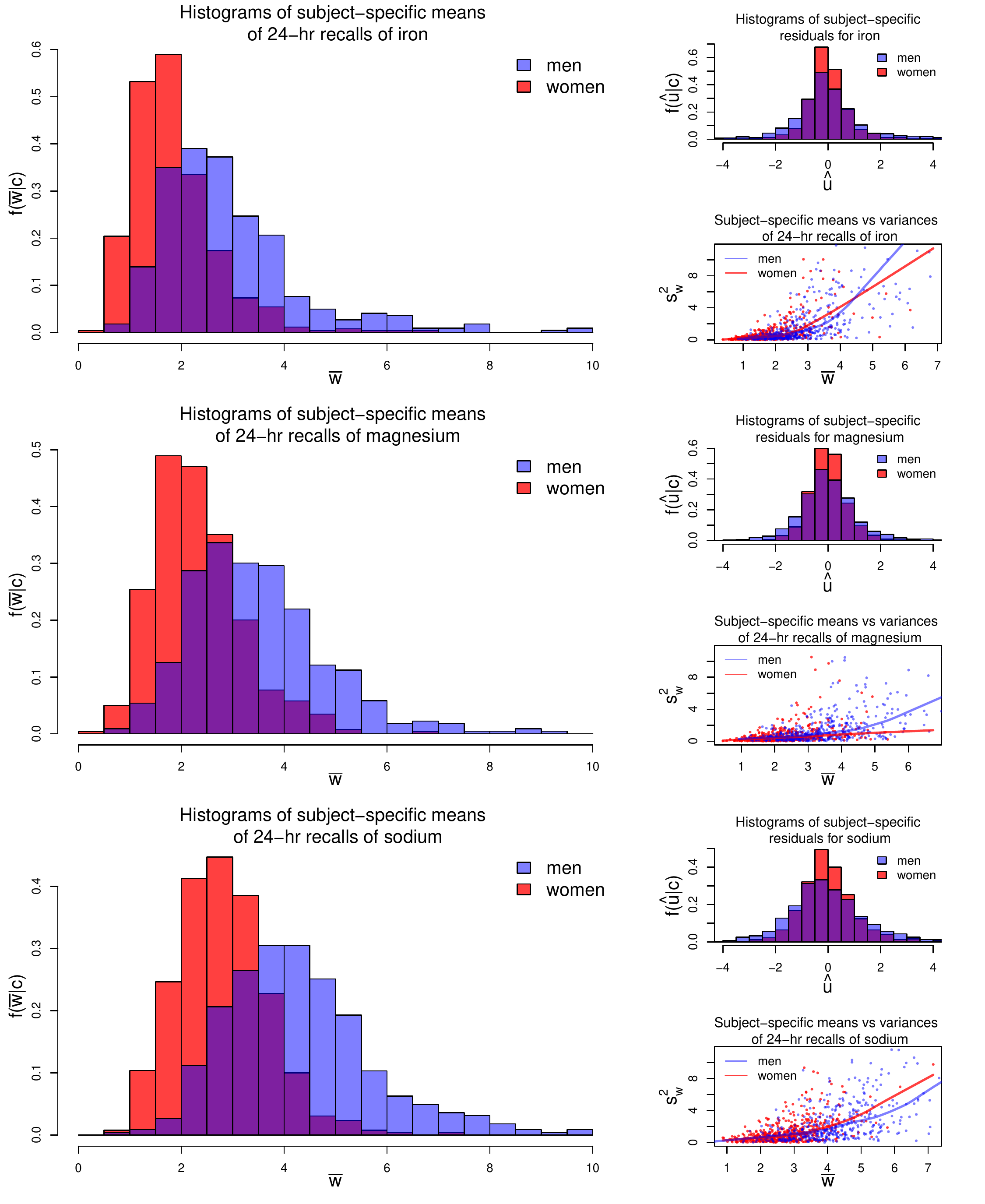}
\end{center}
\caption{\baselineskip=10pt Exploratory plots. 
Left panels: histograms of subject-specific means $\overline{w}_{\ell,i}$, crude estimates of $x_{\ell,i}$; 
right upper panels: histograms of `residuals' $\wh{u}_{\ell,i,j} = (w_{\ell,i,j} - \overline{w}_{\ell,i})$, crude estimates of $u_{\ell,i,j}$; 
right lower panels: subject-specific means $\overline{w}_{\ell,i}$ vs variances $s_{w,\ell,i}^{2}$, crude estimates of $\var(u_{\ell,i,j} \mid x_{\ell,i})$, superimposed with lowess fits. 
}
\label{fig: EATS exploratory sex}
\end{figure}

\begin{figure}[!ht]
\begin{center}
\includegraphics[height=16.5cm, trim=0cm 0cm 0cm 0cm, clip=true]{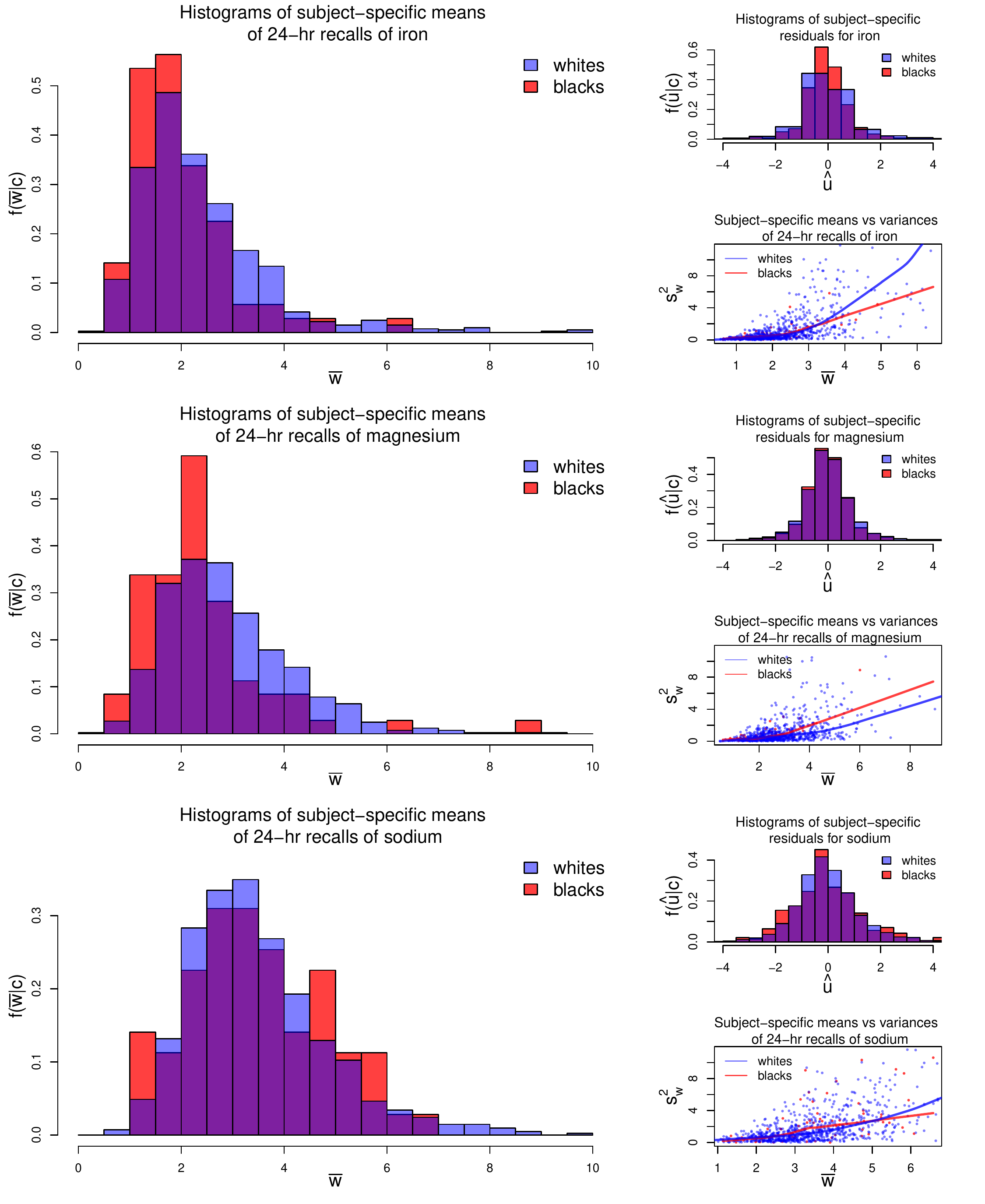}
\end{center}
\caption{\baselineskip=10pt Exploratory plots. 
Left panels: histograms of subject-specific means $\overline{w}_{\ell,i}$, crude estimates of $x_{\ell,i}$; 
right upper panels: histograms of `residuals' $\wh{u}_{\ell,i,j} = (w_{\ell,i,j} - \overline{w}_{\ell,i})$, crude estimates of $u_{\ell,i,j}$; 
right lower panels: subject-specific means $\overline{w}_{\ell,i}$ vs variances $s_{w,\ell,i}^{2}$, crude estimates of $\var(u_{\ell,i,j} \mid x_{\ell,i})$, superimposed with lowess fits. 
}
\label{fig: EATS exploratory race}
\end{figure}

The long term average intakes may vary between different combinations of the levels of the predictors. 
The left panels of Figure \ref{fig: EATS exploratory sex}, for example, show 
the histograms of the subject-specific means $\overline{w}_{\ell,i}=\sum_{j=1}^{m_{i}}w_{\ell,i,j}/4$ for three different minerals, namely iron, magnesium and sodium, separately for men and women but superimposed on each other.
The consumptions for men tend to be higher on average and also have much heavier right tails compared to women for all three dietary components. 
The right upper panels of Figure \ref{fig: EATS exploratory sex} show the histograms of `measurement error residuals' $w_{\ell,i,j} - \overline{w}_{\ell,i}$ for men and women. 
The histograms are all right skewed and the histograms for women are slightly more concentrated around zero compared to men.  
The right lower panels of Figure \ref{fig: EATS exploratory sex} show $\overline{w}_{\ell,i}$ vs the subject-specific variances $s_{w,\ell,i}^{2}=\sum_{j=1}^{m_{i}}(w_{\ell,i,j}-\overline{w}_{\ell,i})^{2}/3$ 
for the 24-hour recalls, 
providing crude estimates of the conditional variances $\var(u_{\ell,i,j} \mid x_{\ell,i})$, 
suggesting strongly that $\var(u \mid x)$ increases as $x$ increases for both men and women, 
although the patterns may not be significantly different between the two gender categories. 
Not all demographic variables may actually be important. 
Figure \ref{fig: EATS exploratory race} summarizes similar exploratory analysis but for the predictor race, specifically the groups `whites' and `blacks'.
Unlike the two gender categories, in this case however, the consumptions do not seem to vary significantly between the two levels. 
Comparison between race groups `whites' and `missing', presented in Figure \ref{fig: EATS exploratory race 2} in the supplementary material, 
may indicate stark differences in consumption patterns at a quick glance 
but this may just be an artifact of the sparse representation of the `missing' group (5 subjects only) in the EATS data set. 
Treating the subjects with missing race labels to come from a separate specific racial group is certainly a bit ad-hoc %if not plain wrong 
but will be instructive in illustrating the robustness of our proposed approach to the presence of small outlying groups in the data. 
Overall, these exploratory analyses illustrate the need for sophisticated density deconvolution methods 
that can accommodate the available demographic covariates 
and can also formally assess their statistical importance in influencing the long-term average consumptions.

{\bf Existing Methods:}  
The literature on univariate density deconvolution, 
in which context we denote the variable of interest by $x$ and the measurement errors by $u$, and the surrogates by $w$, is massive. 
The classical literature, reviews of which can be found in \cite{Carroll2006} and \cite{Buonaccorsi2010}, 
mostly focused on the additive model 
$w=x+u$ subject to $\eE(u) = 0$ with restrictive assumptions, 
such as known $f_{u}$, homoscedasticity of $u$, independence of $u$ from $x$, etc. 
These assumptions are often highly unrealistic, 
especially in nutritional epidemiology applications.

Recent works by \cite{Staudenmayer_etal:2008}, \cite{su2020nonparametric}, \cite{sarkar2014bayesian, sarkar2018bayesian, sarkar2021bayesian} have shown that 
Bayesian hierarchical frameworks and associated computational machinery can %accommodate measurement errors through natural hierarchies, 
provide powerful tools for solving complex deconvolution problems under more realistic scenarios, 
including when the errors $u$ can be conditionally heteroscedastic. 
In their seminal work, \cite{Staudenmayer_etal:2008} considered the model $w=x+u$ with $(u \mid x) \sim \Normal\{0,s^{2}(x)\}$, %but allowed the variability of $u$ to depend on $x$, 
utilizing mixtures of B-splines to estimate $f_{x}$ as well the conditional variability $\var(u \mid x) = s^{2}(x)$.
 \cite{sarkar2014bayesian} relaxed the assumption of normality of $u$, 
employing flexible mixtures of normals \citep{Escobar_West:1995, fruhwirth2006finite} to model both $f_{x}$ and $f_{u \mid x}$. 
\cite{sarkar2018bayesian} extended the methods to multivariate settings $\bw=\bx+\bu$ subject to $\eE(\bu \mid \bx) = \bzero$, 
modeling $f_{\bx}$ and $f_{\bu\mid \bx}$ using mixtures of multivariate normals. 
\cite{sarkar2021bayesian} adopted a complimentary approach, modeling the marginals $f_{x_{\ell}}$ and $f_{u_{\ell}\mid x_{\ell}}$ first 
and then building the joint distributions $f_{\bx}$ and $f_{\bu\mid\bx}$ by modeling the dependence structures separately using Gaussian copulas.  

To the best of our knowledge, however, the problem of deconvoluting $f_{\bx \mid \bc}$ and $f_{\bu \mid \bx,\bc}$ 
in the presence of precisely measured covariates $\bc$ 
from surrogates generated as $\bw=\bx+\bu$ subject to $\eE(\bu \mid \bx,\bc) = \bzero$
has never been considered in the literature, not even in the univariate setting. 
The only practical solution we can mention in this context is the multi-stage pseudo-Bayesian approach of \cite{Zhang2011b}, 
where component-wise Box-Cox transformed \citep{box1964analysis} recalls were assumed to follow a linear mixed model, 
comprising a subject specific random effect component 
and a covariate dependent linear fixed effects component with no interaction terms and an error component.  
The error and the random effects components were then both modeled using single component multivariate normal distributions. 
Multivariate normal priors were also assumed for the fixed effects coefficients. 
Estimates of the long-term intakes were then obtained via individual transformations back to the original scale. 
The density of interest is then obtained by applying a separate off-the-shelf kernel density estimation method on the estimated intakes. 
As shown in \cite{sarkar2014bayesian}, Box-Cox transformations for surrogate observations have severe limitations, 
including almost never being able to produce transformed surrogates that conform to normality, homoscedasticity, and independence. 
Single component multivariate normal models are thus often highly inadequate for the densities even after transformations \citep{sarkar2021bayesian}.

{\bf Our Proposed Approach:}
In this article, we develop a Bayesian semiparametric approach to covariate dependent multivariate density deconvolution, 
carefully accommodating the prominent features of nutritional epidemiology data sets, including conditional heteroscedasticity, departures from normality, etc. 
Our proposed approach not only allows $f_{\bx \mid \bc}$ and $f_{\bu \mid \bx, \bc}$ to vary flexibly with the predictors $\bc$ 
but also allows us to determine which predictors among $\bc$ are the most influential ones for $f_{\bx \mid \bc}$ and $f_{\bu \mid \bx, \bc}$, 
including accommodating the possibility that 
the sets of influential predictors can be different for $f_{\bx \mid \bc}$ and $f_{\bu \mid \bx, \bc}$.

Following \cite{Staudenmayer_etal:2008} and \cite{sarkar2014bayesian, sarkar2018bayesian, sarkar2021bayesian}, 
we begin with the assumption that the measurement errors $u_{\ell}$ decompose into a variance function $v_{\ell}$ that explains their conditional heteroscedasticity 
and a scaled error component $\epsilon_{\ell}$ that captures their general distributional shapes and other properties. 
Building on \cite{sarkar2021bayesian}, 
we model the joint densities using a copula based approach 
with the marginal densities $f_{x_{\ell}}$ and $f_{\epsilon_{\ell}}$ and the variance functions $v_{\ell}$ characterized as flexible mixtures of dictionary functions 
shared between all univariate components and all predictor level combinations. 
Unlike previous approaches, however, we now allow the mixture probabilities to vary with the predictors.  
A predictor is thus considered important if the mixture probabilities vary significantly between its levels, 
thereby significantly altering the densities.  
Viewing these mixture probabilities as a conditional probability tensor 
and relying on tensor factorization techniques \citep{yang_dunson:2015}, 
we then parameterize the mixture probabilities themselves as mixtures of `core' probability kernels 
with mixture weights depending on the level combinations of the predictors. 
The parameterization allows explicit identification of the set of important predictors 
while also implicitly capturing complex higher order interactions between them in a parsimonious manner.  
The elimination of the redundant predictors and the implicit modeling of the interactions among the important ones 
lead to a significant two fold reduction in the effective number parameters required to flexibly characterize the mixture probabilities.  
The daunting challenge of implementing a tensor factorization model separately for each dietary component is avoided 
via a simple innovation of treating the component labels to comprise the levels of an additional categorical predictor. 
We assign sparsity inducing priors that favor such lower dimensional representations. 
We assign a hierarchical Dirichlet prior on the core probability kernels, 
encouraging the model to shrink further towards lower dimensional structures by borrowing strength across these components as well. 
We develop a Markov chain Monte Carlo (MCMC) algorithm to approximately sample from the posterior. 
Importantly, our proposed method allows the density of interest $f_{\bx \mid \bc}$ 
and the density of the scaled measurement errors $f_{\bepsilon \mid \bc}$ to vary with potentially different sets of covariates.  
Applied to our motivating EATS data set, 
the proposed method estimates the distributions of long-term consumptions of different dietary components 
for different level combinations of the predictors, while also selecting the important predictors, 
providing novel insights into how the distributions of the intakes as well as the distributions of the associated measurement errors vary with the available subject specific demographic covariates.

{\bf Outline of the Article:} 
The rest of the article is organized as follows. 
Section \ref{sec: models} details the proposed Bayesian hierarchical framework.  
Section \ref{sec: applications} presents results of our proposed method applied to the motivating nutritional epidemiology problems. 
Section \ref{sec: simulation studies} presents the results of some synthetic experiments. 
Section \ref{sec: discussion} concludes with a discussion.
A brief review of copula and conditional tensor factorization models, 
the Markov chain Monte Carlo (MCMC) algorithm we used to sample from the posterior, 
results of synthetic numerical experiments, 
and some additional results are included in the supplementary material.

\section{Deconvolution Models} \label{sec: models}

We are interested in estimating the unknown joint density of a $d$-dimensional continuous random vector $\bx$ 
in the presence of associated $p$-dimensional categorical covariate $\bc$, the $h\th$ component $c_{h}$ taking $d_{h}$ different categorical values $\{1,\dots,d_{h}\}$. 
There are $i=1,\dots,n$ subjects.
The covariates $\bc_{i}$ are precisely measured for each subject $i$.
For $\bx_{i}$, however, only replicated proxies $\bw_{i,j}$
contaminated with measurement errors $\bu_{i,j}$ are available for $j=1,\dots,m_{i}$ for each subject $i$. 
The density of $\bx$ as well as the density of $\bu$ may both potentially vary with $\bc$. 
The replicates are assumed to be generated by the model
\vspace{-5ex}\\
\be
\bw_{i,j} &=& \bx_{i} + \bu_{i,j}~~~\text{subject to}~~\eE(\bu_{ij} \mid \bx_{i}, \bc_{i}) = \bzero.  \label{eq: additive error}
\ee
\vspace{-5ex}

To accommodate conditional heteroscedasticity in the measurement errors, 
adapting ideas from \cite{sarkar2018bayesian}, we let 
\vspace{-5ex}\\
\bse
&(\bu_{i,j} \mid \bx_{i})= \bS(\bx_{i})\bepsilon_{i,j},~~~\hbox{with}~~~\eE(\bepsilon_{i,j} \mid \bc_{i}) = \bzero,\\
&\text{and}~~~\bS(\bx_{i}) = \diag\{s_{1}(x_{1,i}),\dots,s_{d}(x_{d,i})\}.  	\label{eq: multiplicative structure}
\ese
\vspace{-5ex}\\
The model implies that $\cov(\bu_{i,j}\mid \bx_{i}, \bc_{i}) = \bS(\bx_{i})~ \cov(\bepsilon_{i,j} \mid \bc_{i}) ~ \bS(\bx_{i})$ 
and marginally $\var(u_{\ell,i,j}\mid x_{\ell,i}, \bc_{i}) = s_{\ell}^{2}(x_{\ell,i})\var(\epsilon_{\ell,i,j} \mid \bc_{i})$. 
Other features of the predictor dependent distribution of $\bu$, including its shape and correlation structure, are derived from $f_{\bepsilon \mid \bc}$.

\begin{figure}[!ht]
  \begin{center}
  	\vskip -5pt
	\begin{subfigure}[t]{0.3\textwidth}
  	\hskip 20pt\includegraphics[scale=0.74, trim=1.2cm 1.2cm 1.2cm 1.2cm, clip=true]{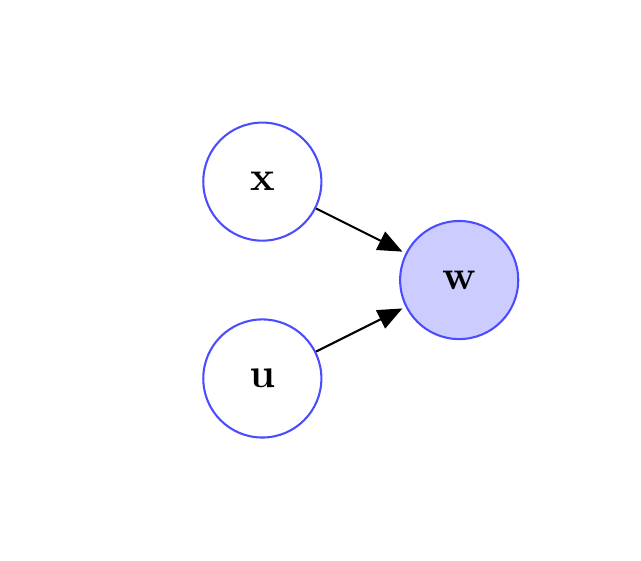}
  	\caption{Classical methods}
	\end{subfigure}
	\begin{subfigure}[t]{0.3\textwidth}
  	\quad\quad\includegraphics[scale=0.74, trim=1.2cm 1.2cm 1.2cm 1.2cm, clip=true]{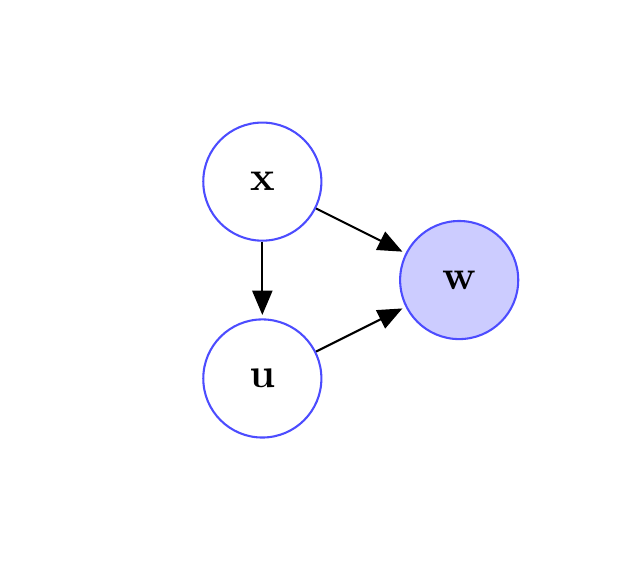} 
  	\caption{Our previous research}
	\end{subfigure}
	\begin{subfigure}[t]{0.33\textwidth}
  	\quad\quad\quad\includegraphics[scale=0.74, trim=1.2cm 1.2cm 1.2cm 1.2cm, clip=true]{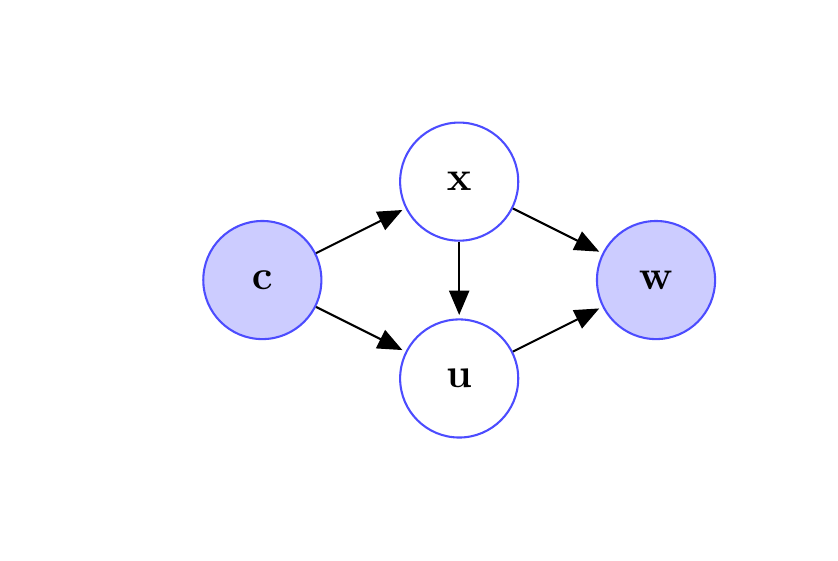}
 	\caption{Our proposed model}
	\end{subfigure}
  \end{center}
  \vskip -10pt
  \caption{\baselineskip=10pt Graphical model depicting the data generating processes considered in classical density deconvolution methods; 
  	in our previous research in \cite{sarkar2014bayesian,sarkar2018bayesian,sarkar2021bayesian};  
	and our proposed model described in Section \ref{sec: models}. 
	The unfilled and the shaded nodes depict latent and observable variables, respectively. 
	Subject and replicate subscripts ($i$ and $j$, respectively) are suppressed for clarity.} 
   \label{fig: graphical model 1}
  %\skip -5pt
\end{figure}

As discussed in detail in \cite{sarkar2018bayesian}, the above model arises naturally for conditionally heteroscedastic multivariate measurement errors. 
Additionally, the model also automatically accommodates multiplicative measurement errors: 
Suppressing the covariates $\bc_{i}$ and setting $s(x_{\ell,i})=x_{\ell,i}$ and $\epsilon_{\ell,i,j} =  (\wt{u}_{\ell,i,j}-1)$, 
we have $w_{\ell,i,j} = x_{\ell,i} + s(x_{\ell,i})\epsilon_{\ell,i,j} = x_{\ell,i} + x_{\ell,i} (\wt{u}_{\ell,i,j}-1) = x_{\ell,i} \wt{u}_{\ell,i,j}$ with $\wt{u}_{\ell,i,j}$ independent of $x_{\ell,i}$ and $\eE(\wt{u}_{\ell,i,j})=1$.

Importantly, in our formulation, the covariates $\bc$ may influence not only the density of $\bx$ 
but also the density of the scaled errors $\bepsilon$ (Figure \ref{fig: graphical model 1}). 
The actual sets of influential predictors may be proper subsets of $\bc$ and are allowed to be different for $\bx$ and $\bepsilon$.

It is possible 
that the variance function components $v_{\ell}(x)$ also vary with $\bc$. 
Exploratory analysis, however, suggest that the functions $\var(u_{\ell} \mid x_{\ell},\bc)$ are very similar for different values of $\bc$.  
Since $\var(u_{\ell} \mid x_{\ell},\bc)$ are already allowed to vary flexibly with $\bc$ through $\var(\epsilon_{\ell} \mid \bc)$, 
it thus becomes difficult to separate the influence of $\bc$ on $v_{\ell}(x)$. 
For nutritional epidemiology applications, our proposed model seems to provide a sufficient compromise.

Cast in a Bayesian hierarchical framework, 
the problem reduces to one of flexibly modeling $f_{\bx \mid \bc}$, $f_{\bepsilon \mid \bc}$ and $v_{\ell}(x_{\ell})$ 
while also selecting the set of most influential covariates for $f_{\bx \mid \bc}$ and $f_{\bepsilon \mid \bc}$.
The methodology developed below addresses these daunting statistical challenges. 
In what follows, the component, subject and replicate subscripts $\ell,i,j$ are often omitted and assumed to be implicitly understood to keep the notation simple.

\subsection{Modeling the Density $f_{\bx \mid \bc}$} \label{sec: predictor dependent density of interest}
The joint density $f_{\bx \mid \bc}$ is specified using a Gaussian copula density model 
\vspace{-5ex}\\
\bse
\textstyle f_{\bx\mid \bc}(\bx \mid  \bc) = |\bR_{\bx}|^{-\frac{1}{2}} \exp\left\{-\frac{1}{2}\by_{\bx}\trans(\bR_{\bx}^{-1}- \bI_{d})\by_{\bx}\right\}  \prod_{\ell=1}^{d} f_{x,\ell \mid \bc}(x_{\ell} \mid \bc),  \label{eq: copula mixture for f_X} 
\ese
\vspace{-5ex}\\
where 
$F_{x,\ell \mid \bc}(x_{\ell} \mid \bc) = \Phi(y_{x,\ell})$, 
$F_{x,\ell \mid \bc}$ being the cdf corresponding to $f_{x,\ell \mid \bc}$;
$\by_{\bx} = (y_{x,1},\dots,y_{x,d})\trans$;
$\Phi(\cdot)$ denotes the cdf of a standard normal distribution; 
and $\bR_{\bx}$ is the correlation matrix between the $d$ components of $\bx$ for all $\bc$.
The Gaussian copula maps $\bx$ to $\by_{\bx}$ such that $\by_{\bx} \sim \MVN_{d}(\bzero,\bR_{\bx})$, 
which allows the dependence relationships between the components of $\bx$ be conveniently modeled separately from their marginals $f_{x,\ell \mid \bc}(x_{\ell} \mid \bc)$.

As in \cite{sarkar2021bayesian}, we model the marginal densities using mixtures of truncated normal distributions with atoms shared across the different dimensions. 
To model the influence of the associated observed covariates, the mixture probabilities are now, however, allowed to vary flexibly with the covariates. 
Specifically, we let 
\vspace{-5ex}\\
\bse
&  f_{x,\ell \mid \bc} (x_{\ell} \mid c_{1},\dots,c_{p}) = \sum_{k=1}^{k_{x}} P_{x,\ell \mid \bc}(k \mid c_{1},\dots,c_{p}) ~ \TN(x_{\ell} \mid \mu_{x,k},\sigma_{x,k}^{2}, [A,B]),
\ese
\vspace{-5ex}\\
where $\TN(x \mid \mu,\sigma^{2}, [A,B])$'s are truncated normal mixture kernels with location $\mu$, scale $\sigma$ and support restricted to the interval $[A,B]$; 
$P_{x,\ell \mid \bc}(k \mid c_{1},\dots,c_{p})$'s are the associated predictor dependent mixture probabilities. 
The corresponding cdfs  $F_{x,\ell \mid \bc}(x_{\ell} \mid \bc)$ are thus given by 
\vspace{-5ex}\\
\bse
& F_{x,\ell \mid \bc}(x_{\ell} \mid \bc) = \sum_{k=1}^{k_{x}} P_{x,\ell \mid \bc}(k \mid c_{1},\dots,c_{p}) ~ \int_{A}^{x_{\ell}} \TN(x \mid \mu_{x,k},\sigma_{x,k}^{2}, [A,B]) dx.
\ese
\vspace{-5ex}

Sharing the mixture components across different predictor combinations here allows efficient estimation of the atoms $\{(\mu_{x,k},\sigma_{x,k}^{2})\}_{k=1}^{k_{x}}$ borrowing information across these combinations.  
This way, since the dependence of the densities on the associated predictors is modeled entirely through the mixture probabilities, 
a predictor will be important if the mixture probabilities vary significantly between its levels. 

Mixtures of truncated normal kernels are just as flexible as mixtures of normals 
but also make the support of the densities $f_{x,\ell \mid \bc}(x_{\ell} \mid \bc)$ consistent with 
the support of $s^{2}(x)$ which we model shortly in Section \ref{sec: predictor dependent var functions} using mixtures of B-splines 
which by construction are finitely supported, here on the interval $[A,B]$. 
The choices of the truncation limits $A$ and $B$ are discussed in Section S.3 in the supplementary material.

Modeling the mixtures probabilities $P_{x,\ell \mid \bc}(k \mid c_{1},\dots,c_{p})$ separately for each dimension $\ell$ would, however, be an extremely challenging task. 
We solve this issue using a simple trick - by including the component label itself as a separate categorical predictor $c_{0} \in \{1,\dots,d_{0}\}$. 
Here $d_{0}$ clearly just equals $d$, the dimension of $\bx$, 
but is additionally introduced to be consistent with the notation $d_{h}$ denoting the number of categories of $c_{h}$. 
Unlike the other $c_{h}$'s which take a single value for each individual $i$ (e.g., sex), 
the predictor $c_{0}$ however takes each value in $ \{1,\dots,d_{0}\}$ for each $i$ depending on which component we are looking at. 
More explicitly, we have $c_{0,\ell,i}=\ell$ for all $(\ell,i)$. 
With $c_{0}$ defined in this manner and $f_{x,\ell \mid \bc} (x \mid c_{1},\dots,c_{p}) =  f_{x \mid \bc} (x \mid c_{0}=\ell,c_{1},\dots,c_{p})$, 
where, with some abuse in notation, we let $\bc$ include $c_{0}$ as $\bc=(c_{0},c_{1},\dots,c_{p})\trans$, 
we model the marginal densities as  
\vspace{-5ex}\\
\be
\textstyle f_{x \mid \bc} (x \mid c_{0},c_{1},\dots,c_{p}) = \sum_{k=1}^{k_{x}} P_{x \mid \bc}(k \mid c_{0},c_{1},\dots,c_{p}) ~ \TN(x \mid \mu_{x,k},\sigma_{x,k}^{2}, [A,B]).  \label{eq: model1a}
\ee
\vspace{-5ex}\\
The mixture kernels are now shared not just between the associated external predictors $c_{1},\dots,c_{p}$ but also across different dimensions $c_{0}=\ell$. 

\begin{figure}[!ht]
\centering
\includegraphics[width=11.5cm, trim=1cm 1cm 1cm 1cm, clip=true]{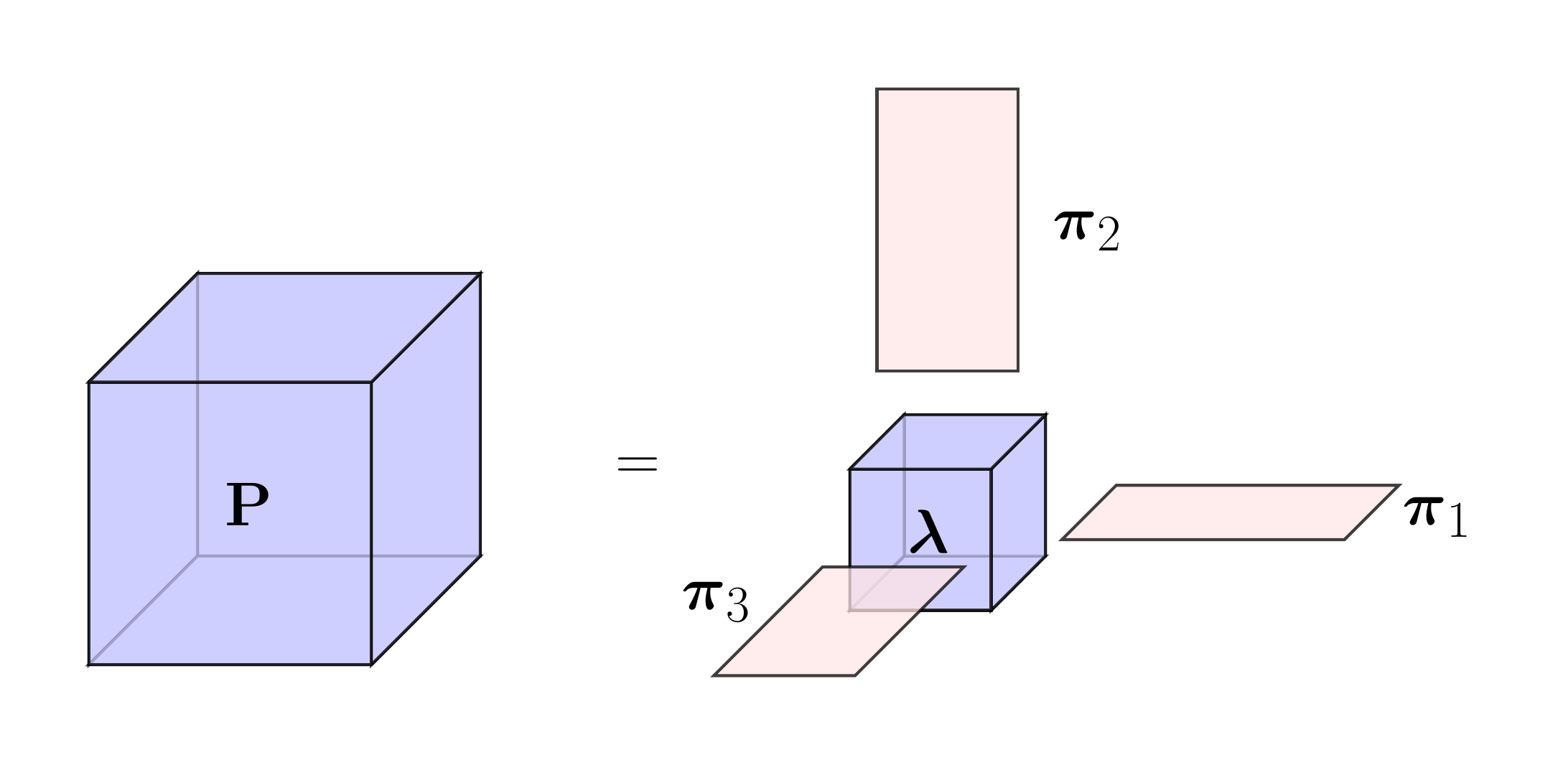}
\caption{Pictorial representation of the factorization of a conditional probability tensor $\bP$ with $3$ categorical covariates with core tensor $\blambda$ and mode matrices $\bpi^{(j)}, j=1,2,3$.}
\label{fig: HOSVD1}
\end{figure}

Modeling the conditional mixture probabilities $P_{x \mid \bc}(k \mid c_{0}, c_{1},\dots,c_{p})$ is still a daunting challenge. 
Being indexed by the $\prod_{h=0}^{p}d_{h}$ different possible values of the predictors $c_{0},c_{1},\dots,c_{p}$, 
a completely unrestricted model for $P_{x \mid \bc}(k \mid c_{0},c_{1},\dots,c_{p})$ would involve a total number of $(k_{x}-1) \prod_{h=0}^{p} d_{h}$ parameters
which increases exponentially 
and becomes too large to be estimated efficiently with datasets of the sizes typically encountered in practice. 
The issue is further significantly complicated not only by the fact that the mixture component labels associated with the $x_{\ell,i}$'s are latent 
but that the $x_{\ell,i}$'s are also measured with error. 
For our motivating nutritional epidemiology application, for instance, 
for a $d_{0}=3$ dimensional problem with sex ($d_{1}=2)$, ethnicity $(d_{2}=5$) and age ($d_{3}=6$) as the associated predictors and  $k_{x}=30$ components in the mixture, 
the total number of parameters becomes 
$\approx 30 \times 3 \times 2 \times 5 \times 6 = 5400$.  
Without imposing additional model structure, it is practically impossible to estimate this many parameters based on the available error contaminated data.

To this end, we look toward higher order singular value decomposition (HOSVD) inspired conditional tensor factorization techniques that have been greatly successful 
in flexibly yet efficiently modeling high-dimensional conditional probabilities of the type $P_{y \mid \bc}(k \mid c_{1},\dots,c_{p})$ in measurement error free settings, 
where $y$ is a categorical response taking values in the set $\{1,\dots,k_{y}\}$
and $c_{1},\dots,c_{p}$ are associated categorical predictors \citep{yang_dunson:2015}. 
Structuring the transition probabilities $P_{y \mid \bc}(y \mid c_{1},\ldots,c_{p})$ as a $k_{y} \times d_{1} \times \dots \times d_{p}$ dimensional $(p+1)$-way tensor, 
they considered the following HOSVD-type factorization  
\vspace{-5ex}\\
\bse
&&\hspace{-1cm}\textstyle P_{y \mid \bc}(y \mid c_{1},\dots,c_{p})  = \sum_{k_{1}=1}^{k_{y,1}}\cdots\sum_{k_{p}=1}^{k_{y,p}} \lambda_{k_{1},\dots,k_{p}}(y)\prod_{h=1}^{p}\pi_{h,c_{h}}(k_{h}). \label{eq: TFM1}
\ese
\vspace{-5ex}\\
See Figure \ref{fig: HOSVD1}. 
Here $1 \leq k_{y,h} \leq d_{h}$ for all $j$ and the parameters $\lambda_{k_{1},\dots,k_{p}}(y)$ and $\pi_{h,c_{h}}(k_{h})$ are all non-negative and satisfy 
(a) $\sum_{y=1}^{k_{y}}  \lambda_{k_{1},\dots,k_{p}}(y) =1,~~ \text{for each combination}~(k_{1},\dots,k_{p}),  \label{eq: TFM2}$ and 
(b) $\sum_{k_{h}=1}^{k_{y,h}} \pi_{h,c_{h}}(k_{h}) = 1, ~~ \text{for each pair }~(h,c_{h}). \label{eq: TFM3}$
\cite{yang_dunson:2015} further showed that 
any conditional probability tensor can be represented as (\ref{eq: TFM1}) with the parameters satisfying the constraints (a) and (b).

Importantly, when $k_{h}=1$, $\pi_{h,c_{h}}(1)=1$ and $P_{y \mid \bc}(y \mid c_{1},\ldots,c_{p})$ does not vary with $c_{h}$. 
The variable $k_{h}$ thus determines the inclusion of the $h\th$ predictor $c_{h}$ in the model. 
When $k_{y,h}\geq 2$, $c_{h}$ is an important predictor of $y$, and when $k_{y,h}=1$, it does not have any influence on $y$. 
The variable $k_{h}$ also determines the number of latent classes for the $h\th$ predictor $c_{h}$. 
The number of parameters in such a factorization is given by $(k_{y}-1)\prod_{h=1}^{p}k_{y,h} + k_{y} \sum_{h=1}^{p}(k_{y,h}-1)$, 
which will be much smaller than the number of parameters $(k_{y}-1)\prod_{h=1}^{p}d_{h}$ required to specify a full model, if $\prod_{h=1}^{p} k_{y,h} \ll \prod_{h=1}^{p}d_{h}$.

Building on these ideas to model the conditional probabilities $P_{x \mid \bc}(k \mid c_{0},c_{1},\dots,c_{p})$ in our setting, we let 
\vspace{-5ex}\\
\be
\textstyle P_{x \mid \bc}(k \mid c_{0},c_{1},\dots,c_{p}) = \sum_{k_{0}=1}^{k_{x,0}} \sum_{k_{1}=1}^{k_{x,1}} \cdots \sum_{k_{p}=1}^{k_{x,p}}  \lambda_{x,k_{0},k_{1},\dots,k_{p}}(k) \prod_{h=0}^{p}\pi_{x,h,c_{h}} (k_{h}).  \label{eq: model1b}
\ee
\vspace{-5ex}\\
Additionally, we restrict the probabilities $\pi_{x,h,c_{h}}(k_{h})$'s to satisfy $\pi_{x,h,c_{h}}(k_{h})=1$ for one $k_{h}$ and $0$ otherwise, 
allowing each $c_{h}$ to be associated with exactly one latent cluster $k_{h}$, 
simplifying the model structure and thereby facilitating posterior computation and model interpretability while also maintaining full model flexibility.

Introducing latent variables $z_{x,\ell,i}$, $z_{x,0,\ell,i} = z_{x,0}(c_{0,\ell,i})$, and $z_{x,h,\ell,i} = z_{x,h}(c_{h,i})$ for $h=1,\dots,p$, 
we can rewrite the model as 
\vspace{-5ex}\\
\bse
& (x_{\ell,i} \mid z_{x,\ell,i}=k) \sim \TN(x \mid \mu_{k}, \sigma_{k}^{2}, [A,B]),    \\
& (z_{x,\ell,i} \mid z_{x,h,\ell,i}=k_{h}, h=0,\dots,p) \sim \Mult\{\lambda_{x,k_{0},k_{1},\dots,k_{p}}(1),\dots,\lambda_{x,k_{0},k_{1},\dots,k_{p}}(k_{x})\}.
\ese
\vspace{-5ex}

Model specification for the marginal densities of $\bx$ is completed by assigning priors to the model parameters. 
For the mixture kernels $\blambda_{x,k_{0},k_{1},\dots,k_{p}}$, we let 
\vspace{-5ex}\\
\bse
\blambda_{x,k_{0},k_{1},\dots,k_{p}} \sim \Dir\{\alpha_{x}\lambda_{x,0}(1),\dots,\alpha_{x}\lambda_{x,0}(k_{x})\},~~~~~\blambda_{x,0} \sim \Dir(\alpha_{x,0},\dots,\alpha_{x,0}).
\ese
\vspace{-5ex}\\
For the variable selection parameters $k_{x,h}$'s, we assign exponentially decaying priors as
\vspace{-5ex}\\
\bse
p_{0}(k_{x,h}) \propto \exp(-\varphi_{x} k_{x,h}), ~~~h=0,\dots,p.
\ese
\vspace{-5ex}\\
Large values of $k_{x,h}$'s are thus penalized, favoring sparsity. 
For the mixture atoms, we let 
\vspace{-5ex}\\
\bse
(\mu_{x,k},\sigma_{x,k}^{2}) \sim \TN(\mu_{x,0},\sigma_{x,0}^{2},[A,B]) \times \IG(a_{x,\sigma^{2}},b_{x,\sigma^{2}}, [A_{x,\sigma^{2}},B_{x,\sigma^{2}}]),
\ese
\vspace{-5ex}\\
where $\IG(a,b,[A,B])$ denotes an Inverse-Gamma distribution with shape $a$, rate $b$, restricted to the interval $[A,B]$.

We have not imposed any strict identifiability constraints on the mixture components 
as we are only interested (a) in estimating the overall shapes of the marginal densities, which are robust to overfitting and invariant to label switching, 
and (b) in selecting the influential predictors, which are determined by the predictors' influences on the overall distribution of the latent $z_{x,\ell,i}$'s. 
Overfitting could be an issue for the latter problem -- two mixture components can be close enough to be considered practically the same 
but two different levels of a covariate $c_{h}$ may differently prefer one component to the other, spuriously inferring $c_{h}$ to have an important effect on the marginal densities. 
Extensive numerical experiments, however, suggest that such a situation almost never really happens in practice. 
Due to the sparsity inducing properties of the priors for the mixture models, in steady states of our MCMC based implementation, 
the mixture components generally get well separated and the redundant components become near-empty 
in the sense that practically zero probabilities will get assigned to such components. 
We are not invoking any notion of a true number of latent components here, 
but are rather interested in a relatively sparse data adaptive mixture model representation that well approximates
the overall shapes of the marginal densities and allows inference about the influences of the predictors on them.

Next, we consider the problem of modeling $\bR_{\bx}$. 
The correlation structure between the $\bx_{i}$'s may certainly vary with the associated predictors $\bc_{i}$'s. 
The problem of modeling such dependence is, however, an extremely difficult one even in the absence of measurement errors and only gets an order of magnitude more difficult when the $\bx_{i}$'s are all measured with error. 
Empirical explorations also do not seem to suggest any real effect here. 
Practical benefits of accommodating such effects in our model would thus be limited at best, outweighed by the additional computational burden introduced. 
In this article, we thus assume the correlation structures to remain fixed across all predictor combinations. 

We use a model based on spherical coordinate representation of Cholesky factorizations used before in \cite{sarkar2021bayesian,Zhang2011b} 
that allows the involved parameters to be treated separately of each other, simplifying posterior computation 
while guaranteeing the resulting matrix to always be a valid correction matrix. 
To keep this article self-contained, we describe the model below. 
We drop the subscript $\bx$ to keep the notation clean. 
With $\bR=\bV\bV\trans$, 
where $\bV^{d \times d}$ is a lower triangular matrix, 
we have
\vspace{-4ex}\\
\bse
\bV &=& 
\left(\begin{array}{c c c c}
v_{1,1} & 0 & \dots & 0\\
v_{2,1} & v_{2,2} & \dots & 0\\
\vdots & \vdots& \vdots& \vdots \\
v_{d,1} & v_{d,2} & \dots & v_{d,d}
\end{array} \right).
\ese
\vspace{-3ex}\\
We have $r_{\ell,\ell'} = \sum_{k=1}^{\ell}v_{\ell,k}v_{\ell',k}$ for all $\ell \leq \ell'$. 
The restriction that $\bR$ is a correlation matrix then implies $\sum_{k=1}^{\ell}v_{\ell,k}^{2} = 1$ for all $\ell=1,\dots,d$. 
The restrictions are satisfied by the following parameterization
\vspace{-5ex}\\
\bse
&& v_{1,1}=1, \\
&& v_{2,1}=b_{1}, ~ v_{2,2}=\sqrt{1-	b_{1}^{2}},\\
&& v_{3,1} =b_{2}\sin\theta_{1}, ~v_{3,2}=b_{2}\cos\theta_{1},~v_{3,3}=\sqrt{1-b_{2}^{2}},\\
&& v_{\ell,1}=b_{\ell-1}\sin\theta_{i_{1}(\ell)},\\
&& v_{\ell,k}=b_{\ell-1}\cos\theta_{i_{1}(\ell)}\cos\theta_{i_{1}(\ell)+1}\dots \cos\theta_{i_{1}(\ell)+k-2}\sin\theta_{i_{1}(\ell)+k-1}, \\
&&\hspace{9cm} \hbox{for}~k=2,\dots,(\ell-2),\\
&& v_{\ell,\ell-1}=b_{\ell-1}\cos\theta_{i_{1}(\ell)}\cos\theta_{i_{1}(\ell)+1}\dots \cos\theta_{i_{2}(\ell)-1}\cos\theta_{i_{2}(\ell)},~~~~v_{\ell,\ell}=\sqrt{1-b_{\ell-1}^{2}},
\ese
\vspace{-5ex}\\
where $\ell=4,\dots,d$,  
$i_{1}(\ell) = 1+\{1+\dots+(\ell-3)\} = (\ell^{2}-5\ell+8)/2$ and $i_{2}(\ell)=i_{1}(\ell)+(\ell-3) = (\ell^{2}-3\ell+2)/2$,  
$\abs{b_{t}}\leq 1$, $t=1,\dots,(d-1)$, $\abs{\theta_{s}}\leq \pi$, $s=1,\dots,i_{2}(d)$. 
The total number of parameters is $\{1+2+\dots+(d-1)\}=d(d-1)/2$.
We have $\abs{\bR}=\abs{\bV}^{2} = \prod_{\ell=2}^{d}v_{\ell,\ell}^{2} = \prod_{\ell=1}^{d-1}(1-b_{\ell}^{2})$. 
The model for $\bR$ is completed by assigning uniform priors on $b_{t}$'s and $\theta_{s}$'s
\vspace{-5ex}\\
\bse
b_{t} \sim \Unif(-1,1), ~~~~~ \theta_{s} \sim \Unif(-\pi,\pi). 
\ese
\vspace{-5ex}\\
Here $\Unif(a,b)$ denotes a uniform distribution with support $(a,b)$.

\subsection{Modeling the Density $f_{\bepsilon \mid \bc}$}  \label{sec: predictor dependent density of scaled errors}

As in Section \ref{sec: predictor dependent density of interest}, we use a Gaussian copula model to specify the density $f_{\bepsilon \mid \bc}$ 
but the model now has to satisfy mean zero constraints.
Specifically, we let   
\vspace{-5ex}\\
\bse
\textstyle f_{\bepsilon \mid \bc}(\bepsilon \mid \bc) = |\bR_{\bepsilon}|^{-\frac{1}{2}} \exp\left\{-\frac{1}{2}\by_{\bepsilon}\trans(\bR_{\bepsilon}^{-1}- \bI_{d})\by_{\bepsilon}\right\}  \prod_{\ell=1}^{d} f_{\epsilon,\ell \mid \bc}(\epsilon_{\ell} \mid \bc),  \\
\text{subject to}~ \textstyle\eE_{f_{\epsilon,\ell \mid \bc}}(\epsilon_{\ell} \mid \bc) = 0, ~~~\hbox{for}~\ell=1,\dots,d. 
\ese
\vspace{-5ex}\\
Here $F_{\epsilon,\ell \mid \bc}(\epsilon_{\ell} \mid \bc) = \Phi(y_{\epsilon,\ell})$ for all $\ell$ and all $\bc$ 
with $F_{\epsilon,\ell \mid \bc}$ being the cdf corresponding to $f_{\epsilon,\ell \mid \bc}$;
$\by_{\bepsilon} = (y_{\epsilon,1},\dots,y_{\epsilon,d})\trans$;
and $\bR_{\bepsilon}$ is the correlation matrix  between the error components.

Following Section \ref{sec: predictor dependent density of interest} again, we use predictor dependent mixture models with shared atoms to model the marginal densities 
$f_{\epsilon,\ell \mid \bc} (\epsilon \mid c_{1},\dots,c_{p}) =  f_{\epsilon \mid \bc} (\epsilon \mid c_{0}=\ell,c_{1},\dots,c_{p})$ as
\vspace{-4ex}\\
\be
\begin{gathered}
\textstyle f_{\epsilon \mid \bc} (\epsilon \mid c_{0},c_{1},\dots,c_{p}) = \sum_{k=1}^{k_{\epsilon}} P_{\epsilon \mid \bc}(k \mid c_{0},c_{1},\dots,c_{p}) ~ f_{c\epsilon}(\epsilon \mid p_{\epsilon,k},\mu_{\epsilon,k},\sigma_{\epsilon,k,1}^{2},\sigma_{\epsilon,k,2}^{2}),  \\
\textstyle P_{\epsilon \mid \bc}(k \mid c_{0},c_{1},\dots,c_{p}) = \sum_{k_{0}=1}^{k_{\epsilon,0}} \sum_{k_{1}=1}^{k_{\epsilon,1}} \cdots \sum_{k_{p}=1}^{k_{\epsilon,p}}  \lambda_{\epsilon,k_{0},k_{1},\dots,k_{p}}(k) \prod_{h=0}^{p}\pi_{\epsilon,h,c_{h}} (k_{h}).
\end{gathered}
\label{eq: model scaled errors}
\ee
\vspace{-3ex}\\
Here $f_{c\epsilon}(\epsilon\mid p,\mu,\sigma_{1}^{2},\sigma_{2}^{2}) = \{p~\Normal(\epsilon \mid \mu_{1},\sigma_{1}^{2})+(1-p)~\Normal(\epsilon \mid \mu_{2},\sigma_{2}^{2})\}$, with
$\mu_{1}=c_{1}\mu,\mu_{2}=c_{2}\mu$,
 $c_{1}=(1-p)/\{p^2+(1-p)^{2}\}^{1/2}$, $c_{2} = -p/\{p^2+(1-p)^{2}\}^{1/2}$. 
The zero mean constraint on the errors is satisfied, since $p\mu_{1}+(1-p)\mu_{2} = \{pc_{1}+(1-p)c_{2}\}\mu = 0$.
Normal densities are included as special cases with $(p,\mu) = (0.5,0)$ or $(0,0)$ or $(1,0)$.
The mixture atoms $\{(p_{\epsilon,k},\mu_{\epsilon,k},\sigma_{\epsilon,k,1}^{2},\sigma_{\epsilon,k,2}^{2})\}_{k=1}^{k_{\epsilon}}$ are again shared between 
different predictor combinations 
to facilitate dimension reduction and borrowing of of information. 

As in the case of $\bx$, we use a parsimonious conditional tensor factorization based model for the predictor dependent mixture probabilities $P_{\epsilon \mid \bc}(k \mid c_{0},c_{1},\dots,c_{p})$ as  
\vspace{-5ex}\\
\bse
\textstyle P_{\epsilon \mid \bc}(k \mid c_{0},c_{1},\dots,c_{p}) = \sum_{k_{0}=1}^{k_{\epsilon,0}} \sum_{k_{1}=1}^{k_{\epsilon,1}} \cdots \sum_{k_{p}=1}^{k_{\epsilon,p}}  \lambda_{\epsilon,k_{0},k_{1},\dots,k_{p}}(k) \prod_{h=0}^{p}\pi_{\epsilon,h,c_{h}} (k_{h}),  \label{eq: model scaled errors 2}
\ese
\vspace{-5ex}\\ 
where the parameters satisfy $\sum_{k=1}^{k_{\epsilon}} \lambda_{\epsilon,k_{0},k_{1},\dots,k_{p}}(k) =1$ for all $(k_{0},k_{1},\dots,k_{p})$ and  $\pi_{\epsilon,h,c_{h}}(k_{h})=1$ for one $k_{h}$ and $0$ otherwise.

Introducing latent variables $z_{\epsilon,\ell,i,j}$,  $z_{\epsilon,0,\ell,i,j} =z_{\epsilon,0}(c_{0,\ell,i})$, and $z_{\epsilon,h,\ell,i,j} = z_{\epsilon,h}(c_{h,i})$ for $h=1,\dots,p$, we can rewrite the model as 
\vspace{-5ex}\\
\bse
& (\epsilon_{\ell,i,j} \mid z_{\epsilon,\ell,i,j}=k) \sim f_{c\epsilon}(\epsilon_{\ell,i,j} \mid p_{\epsilon,k},\mu_{\epsilon,k},\sigma_{\epsilon,k,1}^{2},\sigma_{\epsilon,k,2}^{2}),    \\
& (z_{\epsilon,\ell,i,j} \mid z_{\epsilon,h,\ell,i,j}=k_{h}, h=0,\dots,p) \sim \Mult\{\lambda_{\epsilon,k_{0},k_{1},\dots,k_{p}}(1),\dots,\lambda_{\epsilon,k_{0},k_{1},\dots,k_{p}}(k_{\epsilon})\}.
\ese
\vspace{-5ex}

We assume hierarchical Dirichlet priors for $\blambda_{\epsilon,k_{0},k_{1},\dots,k_{p}}$ as 
\vspace{-5ex}\\
\bse
\blambda_{\epsilon,k_{0},k_{1},\dots,k_{p}} \sim \Dir\{\alpha_{\epsilon}\lambda_{\epsilon,0}(1),\dots,\alpha_{\epsilon}\lambda_{\epsilon,0}(k_{\epsilon})\},~~~~~\blambda_{\epsilon,0} \sim \Dir(\alpha_{\epsilon,0},\dots,\alpha_{\epsilon,0}).
\ese
\vspace{-5ex}\\
For the variable selection parameters $k_{\epsilon,h}$'s, we assign exponentially decaying priors as
\vspace{-5ex}\\
\bse
p_{0}(k_{\epsilon,h}) \propto \exp(-\varphi_{\epsilon} k_{\epsilon,h}), ~~~h=0,\dots,p.
\ese
\vspace{-5ex}\\
We assume non-informative priors for $(p_{\epsilon,k},\mu_{\epsilon,k},\sigma_{\epsilon,k,1}^{2},\sigma_{\epsilon,k,2}^{2})$ as
\vspace{-5ex}\\
\bse
&& (p_{\epsilon,k},\mu_{\epsilon,k},\sigma_{\epsilon,k,1}^{2},\sigma_{\epsilon,k,2}^{2}) \sim p_{0}(p_{\epsilon,k}) ~ p_{0}(\mu_{\epsilon,k})~p_{0}(\sigma_{\epsilon,k,1}^{2})~ p_{0}(\sigma_{\epsilon,k,2}^{2}) \\
&& ~~~ = \hbox{Unif}(0,1)~ \Normal(0,\sigma_{\epsilon,\mu}^{2})~ \IG(a_{\epsilon},b_{\epsilon})~ \IG(a_{\epsilon},b_{\epsilon}), 
\ese
\vspace{-5ex}\\
where $\hbox{Unif}(\ell,u)$ denotes a uniform distribution on the interval $[\ell,u]$.

As in the case of $\bR_{\bx}$, we assume $\bR_{\bepsilon}$ is  independent of $\bc$ and let $\bR_{\bepsilon}^{d \times d}=((r_{\bepsilon,\ell,\ell'}))=\bV_{\bepsilon}\bV_{\bepsilon}\trans$ and parameterize the elements of $\bV_{\bepsilon}$ using spherical coordinates. 
We assign uniform priors on $b_{\bepsilon,t}, t=1,\dots,d-1$ and $\theta_{\bepsilon,s}, s=1,\dots,i_{2}(d)$ 
\vspace{-5ex}\\
\bse
b_{\bepsilon,t} \sim \Unif(-1,1), ~~~~~ \theta_{\bepsilon,s} \sim \Unif(-\pi,\pi). 
\ese
\vspace{-5ex}

\subsection{Modeling the Variance Functions $v_{\ell}(x_{\ell})$}  \label{sec: predictor dependent var functions}

We model the variance functions $v_{\ell}(x)=s_{\ell}^{2}(x)$ as flexible mixtures of B-splines
\vspace{-5ex}\\
\bse
&& v_{\ell}(x) = s_{\ell}^{2}(x) = \textstyle\sum_{j=1}^{J} b_{j}(x) \exp(\vartheta_{\ell,j}) = \bB_{J}(x) \exp(\bvartheta_{\ell}), \label{eq: models for variance functions} \\
&& \bvartheta_{\ell} \sim \MVN_{J}\{\bzero, (\bSigma_{\vartheta,0}^{-1}+\bP_{\vartheta,0}/\sigma_{\vartheta,0}^{2})^{-1}\},
\ese
\vspace{-5ex}\\
where $\bvartheta_{\ell} = (\vartheta_{\ell,1},\dots,\vartheta_{\ell,J})\trans$ are spline coefficients, 
$\MVN_{J}(\bmu,\bSigma)$ denotes a $J$ dimensional multivariate normal distribution with mean $\bmu$ and covariance $\bSigma$.
We choose $\bP_{\vartheta,0} = \bD_{\vartheta,0}\trans \bD_{\vartheta,0}$, where the $(J-2) \times J$ matrix $\bD_{\vartheta,0}$ is such that $\bD_{\vartheta,0} \bvartheta_{\ell}$ computes the second order differences in $\bvartheta_{\ell}$.
The model thus penalizes $\sum_{j} (\nabla^{2} \vartheta_{\ell,j})^{2} = \bvartheta_{\ell}\trans \bP_{\vartheta,0} \bvartheta_{\ell}$, the sum of squares of second order differences in $\bvartheta_{\ell}$ \citep{Eilers_Marx:1996}. 
The variance parameter $\sigma_{\vartheta,0}^{2}$ models the smoothness of the variance functions, smaller $\sigma_{\vartheta,0}^{2}$ inducing smoother functions.

The methodology proposed here builds on a few diverse topics. 
A high-level overview of the different model components is presented as a box-summary in the supplementary materials for easy reference. 
Brief reviews of a few these topics are also presented in the supplementary materials for easy reference -- 
conditional copula models in Section S.1, 
conditional tensor factorization in Section S.2, and
mixtures with shared atoms in Section S.3.

\begin{figure}[!ht]
\begin{center}
\includegraphics[width=16cm, trim=1cm 1.5cm 0cm 0cm, clip=true]{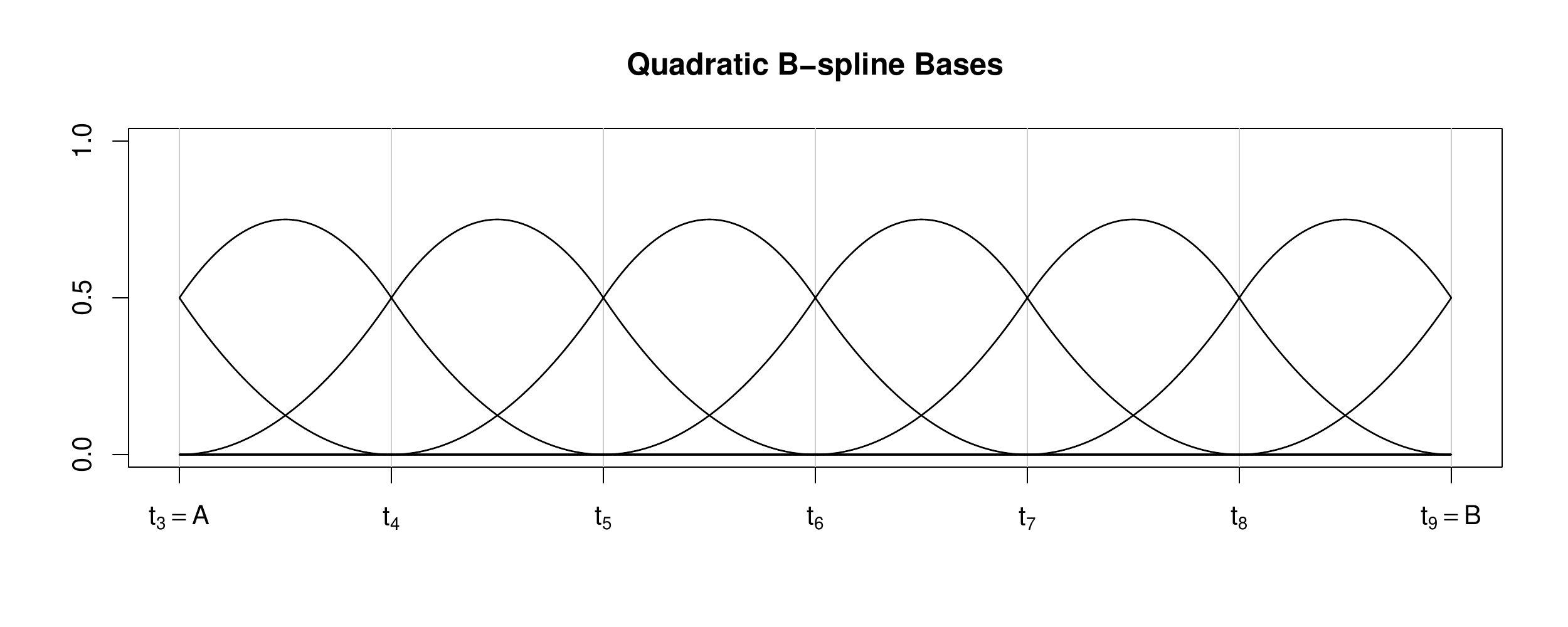}
\end{center}
\caption{\baselineskip=10pt Plot of 9 quadratic $(q=2)$ B-splines on $[A,B]$ defined using $11$ knot points that divide $[A,B]$ into $K=6$ equal subintervals.}
\label{fig: Quadratic B-splines}
\end{figure}

\begin{figure}[!ht]
\centering
\includegraphics[width=11.5cm, trim=2cm 1cm 1cm 1cm, clip=true]{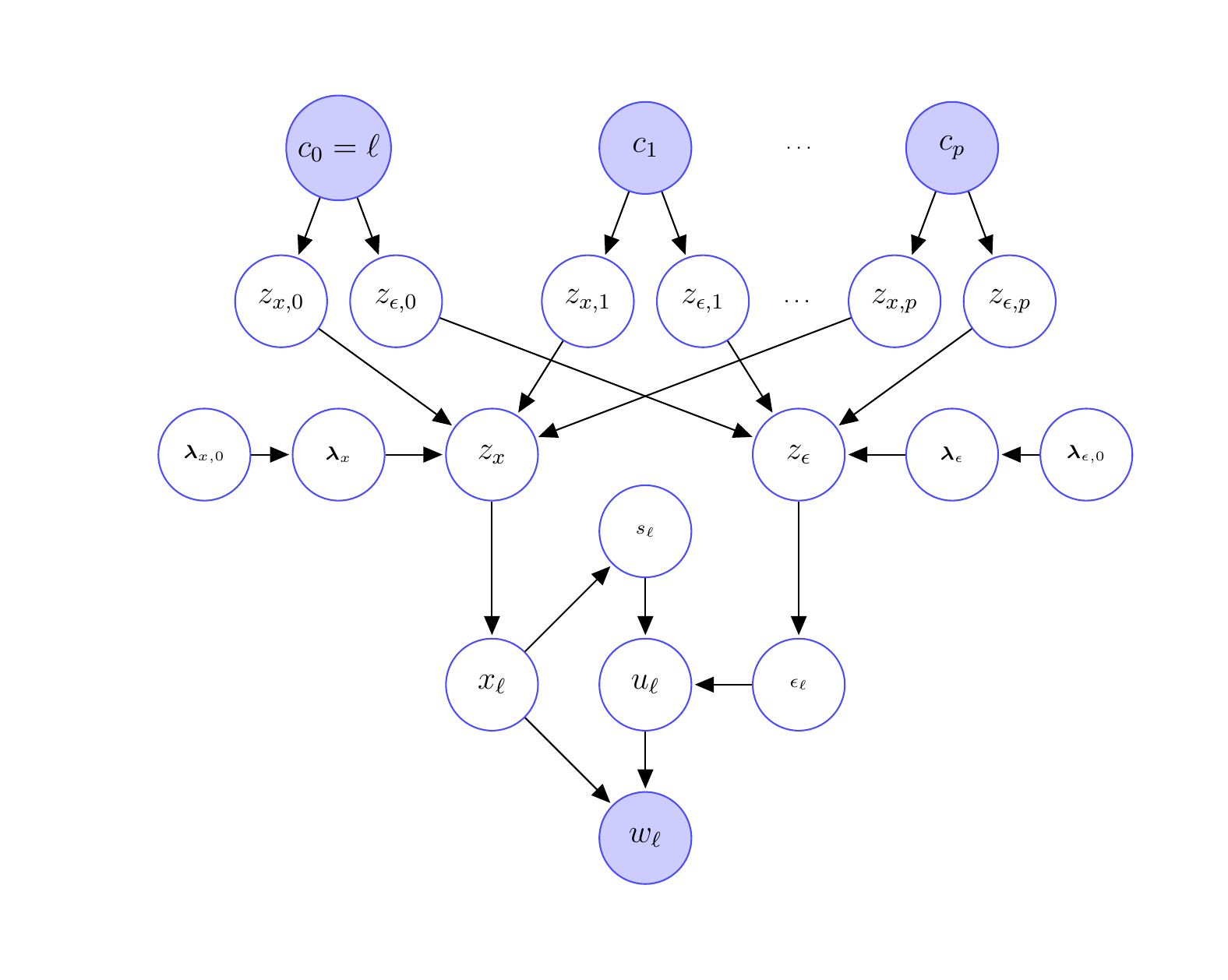}
\caption{\baselineskip=10pt Graphical model depicting the dependency structure 
in the Bayesian hierarchical covariate informed density deconvolution model 
described in Section \ref{sec: models}. 
The unfilled and the shaded nodes signify latent and observable variables, respectively. 
Subject and replicate subscripts ($i$ and $j$, respectively) are suppressed to keep the notation clean. 
}
\label{fig: graphical model 2}
\end{figure}

\section{Applications in Nutritional Epidemiology} \label{sec: applications}

In this section, we discuss the results of our method applied to the EATS data set. 
Specifically, we consider the problem of estimating the distributions of long-term average daily intakes of iron, magnesium and sodium.

\begin{figure}[!ht]
\centering
\includegraphics[width=7cm, trim=0cm 0cm 0cm 0cm, clip=true]{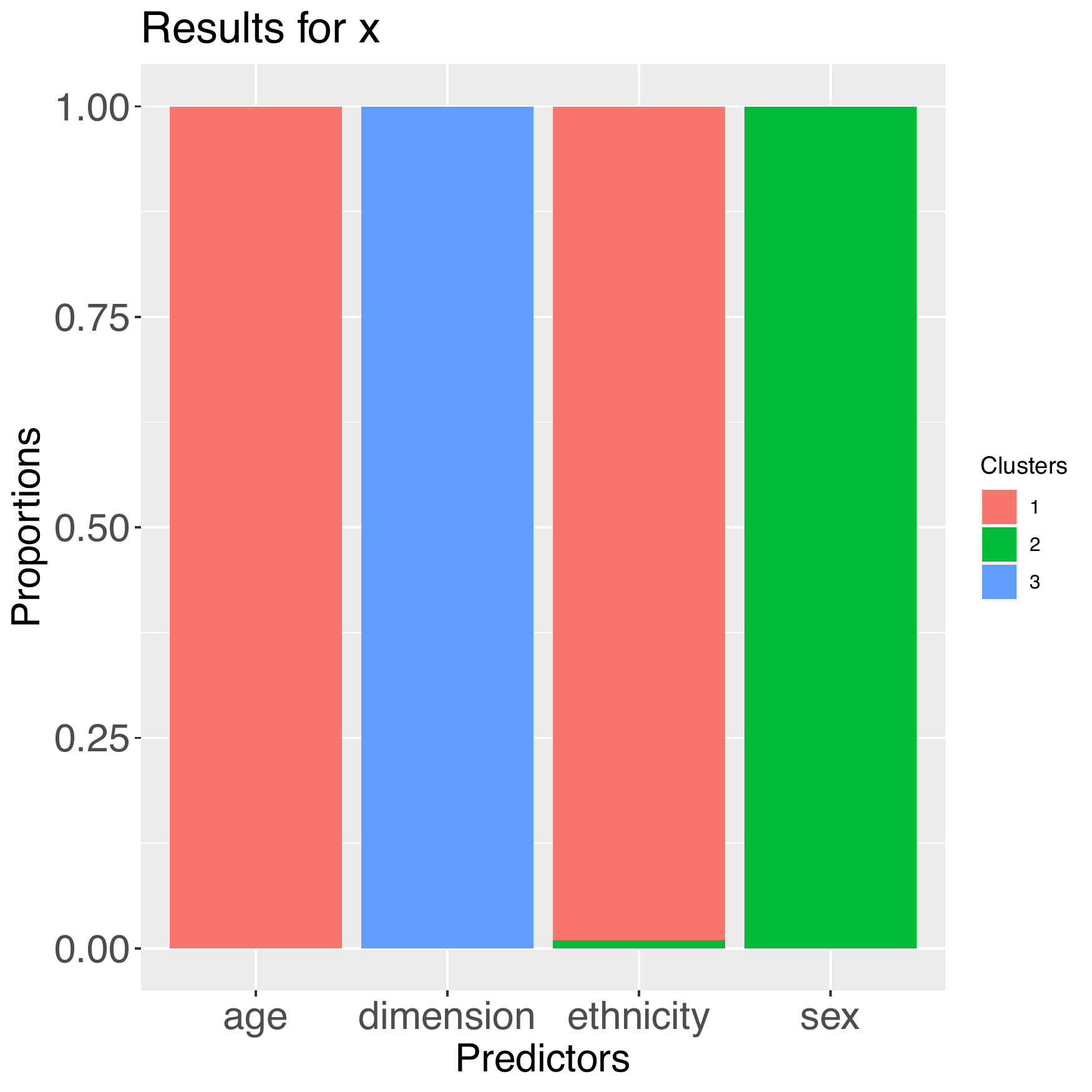}
\quad\quad\quad
\includegraphics[width=7cm, trim=0cm 0cm 0cm 0cm, clip=true]{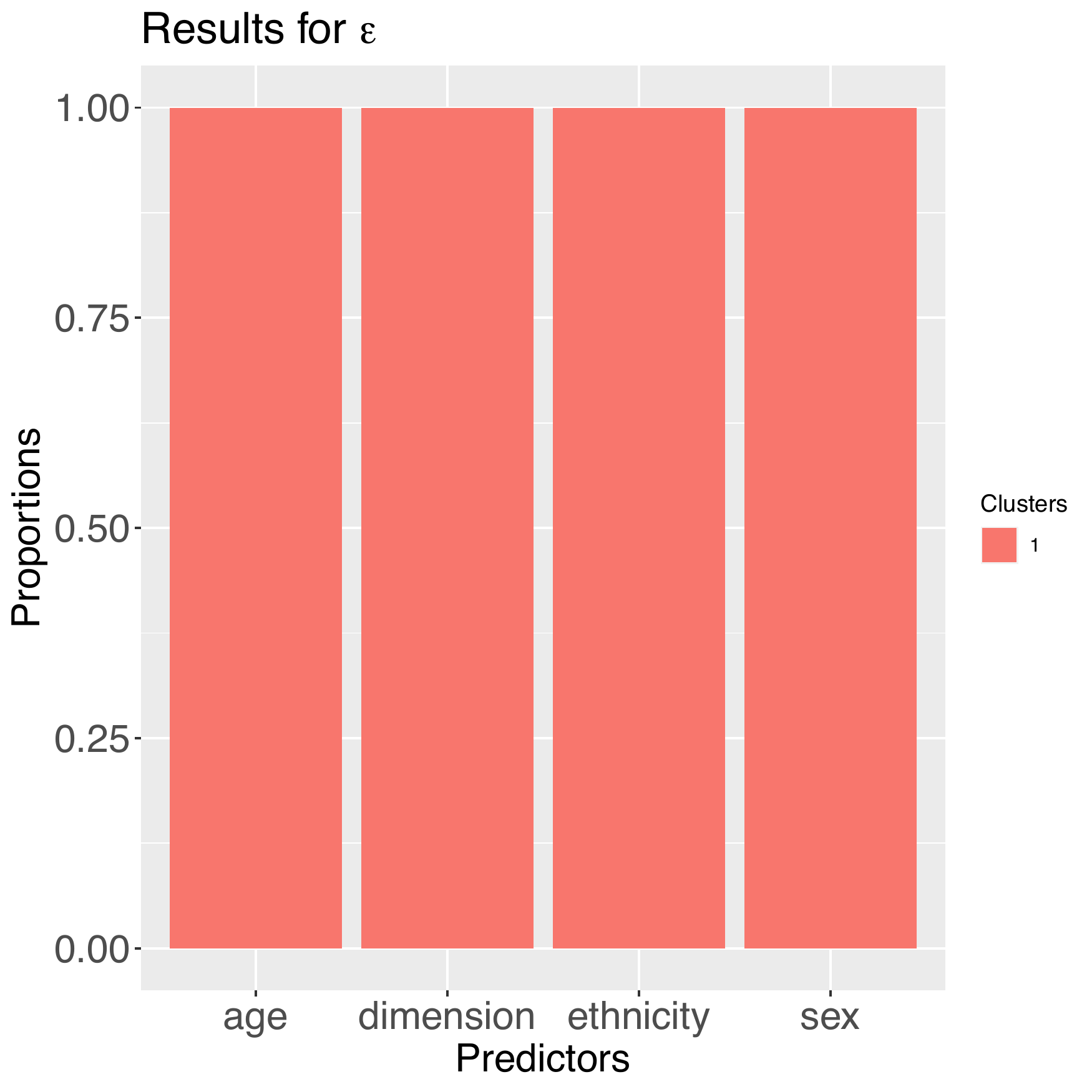}
\caption{\baselineskip=10pt 
Results for the EATS data set showing the estimated probabilities of different numbers of clusters of the associated predictors' levels being included in the model. 
The left panel shows the results for modeling the densities $f_{\bx \mid \bc}$.  
The right panel shows the results for modeling the densities $f_{\bepsilon \mid \bc}$. 
At the median $0.5$ probability level, in the left panel, the dietary component labels and the sex of the subjects are important predictors for modeling the densities $f_{\bx \mid \bc}$, whereas in the right panel, none of the predictors are found to be important for modeling the densities $f_{\bepsilon \mid \bc}$. 
}
\label{fig: EATS Inclusion Probabilities}
\end{figure}

Figure \ref{fig: EATS Inclusion Probabilities} shows the estimated inclusion probabilities of different predictors in the models for $f_{\bx \mid \bc}$ and $f_{\bepsilon \mid \bc}$. 
We recall that a predictor $c_{h}$ is considered important if its levels form at least two clusters, that is, $k_{h} \geq 2$. 
Our MCMC based implementation produces estimates of posterior distribution of the $k_{h}$'s, 
accommodating uncertainly in variable election. 
Using a median probability rule \citep{barbieri2004optimal}, that is, selecting predictors with at least $50\%$ posterior probability of being included in the model, 
the set of significant predictors for the density of main interest $f_{\bx \mid \bc}$ is found to comprise the dimension labels ($c_{0}$) and sex ($c_{1}$). 
For $f_{\bepsilon \mid \bc}$, however, none of the potential predictors were found to be significant. 
The significance of gender is consistent with 
common knowledge that the men on average consume more than women, 
as was also clearly seen the exploratory analysis of Figure \ref{fig: EATS exploratory sex}. 
We must not immediately extend the non-significance of age and ethnicity to the entire population and claim that 
long term dietary intakes do not vary with these covariates at all. 
Based on the finite size EATS data set, however, there is insufficient evidence to claim otherwise. 
The error distributions $f_{\epsilon,\ell \mid \bc}$ all collapsing together, 
while not immediately apparent from the histograms of the `residuals' in Figure \ref{fig: EATS exploratory sex}, 
is consistent with them having very similar right skewed shapes observed in separate univariate analyses (not shown here).

\begin{figure}[!ht]
\centering
\includegraphics[width=11cm, trim=0cm 0.25cm 0cm 0cm, clip=true]{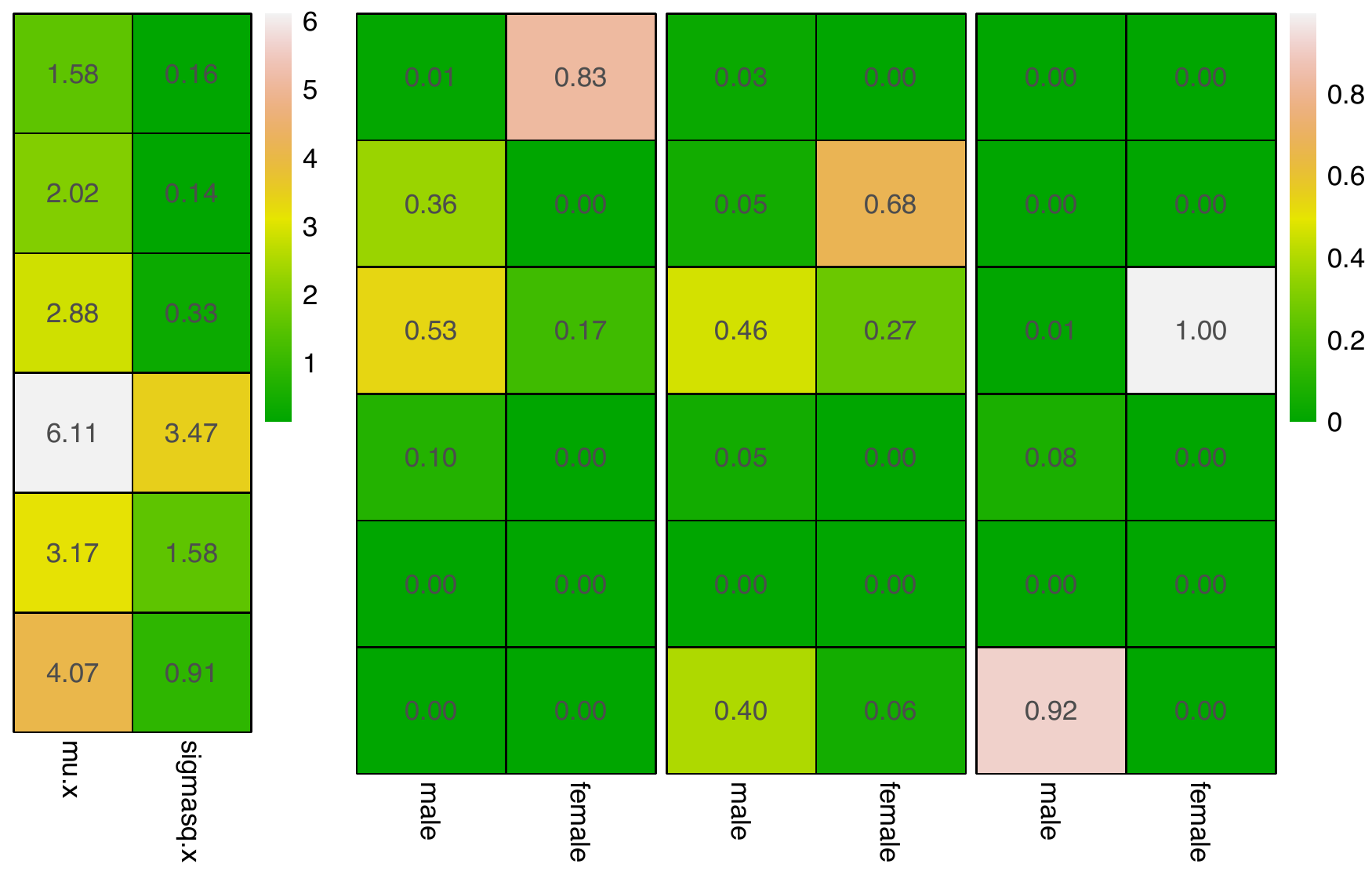}
\caption{\baselineskip=10pt 
Results for the EATS data set. 
These results correspond to the final MCMC iteration but are representative of other iterations in steady state. 
The left panel shows the component specific parameters $(\mu_{x,k}, \sigma_{x,k}^{2})$ 
for the six mixture components that were actually used to model the densities $f_{x,\ell \mid \bc}(x_{\ell} \mid \bc)$.  
The right panel shows the associated `empirical' mixture probabilities $\wh{p}_{x}(k \mid c_{0},c_{1},\dots,c_{p}) = \sum_{i=1}^{n}1\{z_{x,\ell,i}=k, c_{0,\ell,i}=c_{0},c_{1,\ell,i}=c_{1},\dots,c_{p,\ell,i}=c_{p}\}/n$ 
for men and women 
and for the three dietary components, namely iron, magnesium and sodium, from left to right. 
Results for different combinations of dietary component and gender are shown here 
as they are the only predictors found important for $\bx$. 
The mixture probabilities vary significantly between different predictor combinations. 
}
\label{fig: EATS ZX Tables}
\end{figure}

\begin{figure}[!ht]
\centering
\includegraphics[width=12.5cm, trim=0cm 0.25cm 0cm 0cm, clip=true]{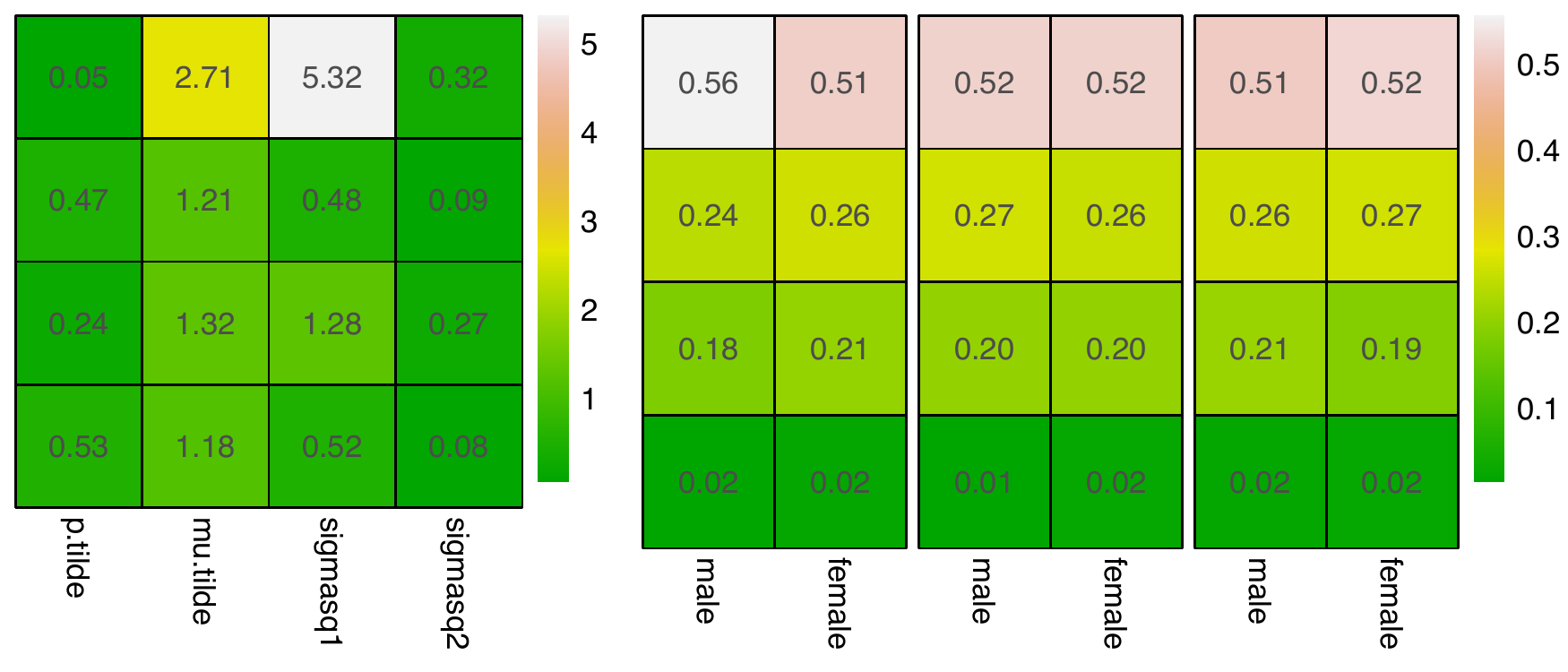}
\caption{\baselineskip=10pt 
Results for the EATS data set. 
These results correspond to the final MCMC iteration but are representative of other iterations in steady state. 
The left panel shows the component specific parameters $(p_{\epsilon,k}, \mu_{\epsilon,k}, \sigma_{\epsilon,k,1}^{2}, \sigma_{\epsilon,k,2}^{2})$ 
for the three mixture components used to model the densities $f_{\epsilon,\ell \mid \bc}(\epsilon_{\ell} \mid \bc)$.  
The right panel shows the associated `empirical' mixture probabilities $\wh{p}_{\epsilon}(k \mid c_{0},c_{1},\dots,c_{p}) = \sum_{i=1}^{n}\sum_{j=1}^{m_{i}}1\{z_{\epsilon,\ell,i,j}=k, c_{0,\ell,i}=c_{0},c_{1,\ell,i}=c_{1},\dots,c_{p,\ell,i}=c_{p}\}/\sum_{i=1}^{n}m_{i}$ 
for men and women 
and for the three dietary components, namely iron, magnesium and sodium, in that order, from left to right. 
Results for different combinations of dietary component and gender are shown here 
for illustrative purposes even though none of the predictors were found important for $\bepsilon$. 
The mixture probabilities do not vary significantly between different predictor combinations. 
}
\label{fig: EATS ZE Tables}
\end{figure}

\begin{figure}[!ht]
\centering
\includegraphics[height=16.5cm, width=15cm, trim=0cm 0cm 0cm 0cm, clip=true]{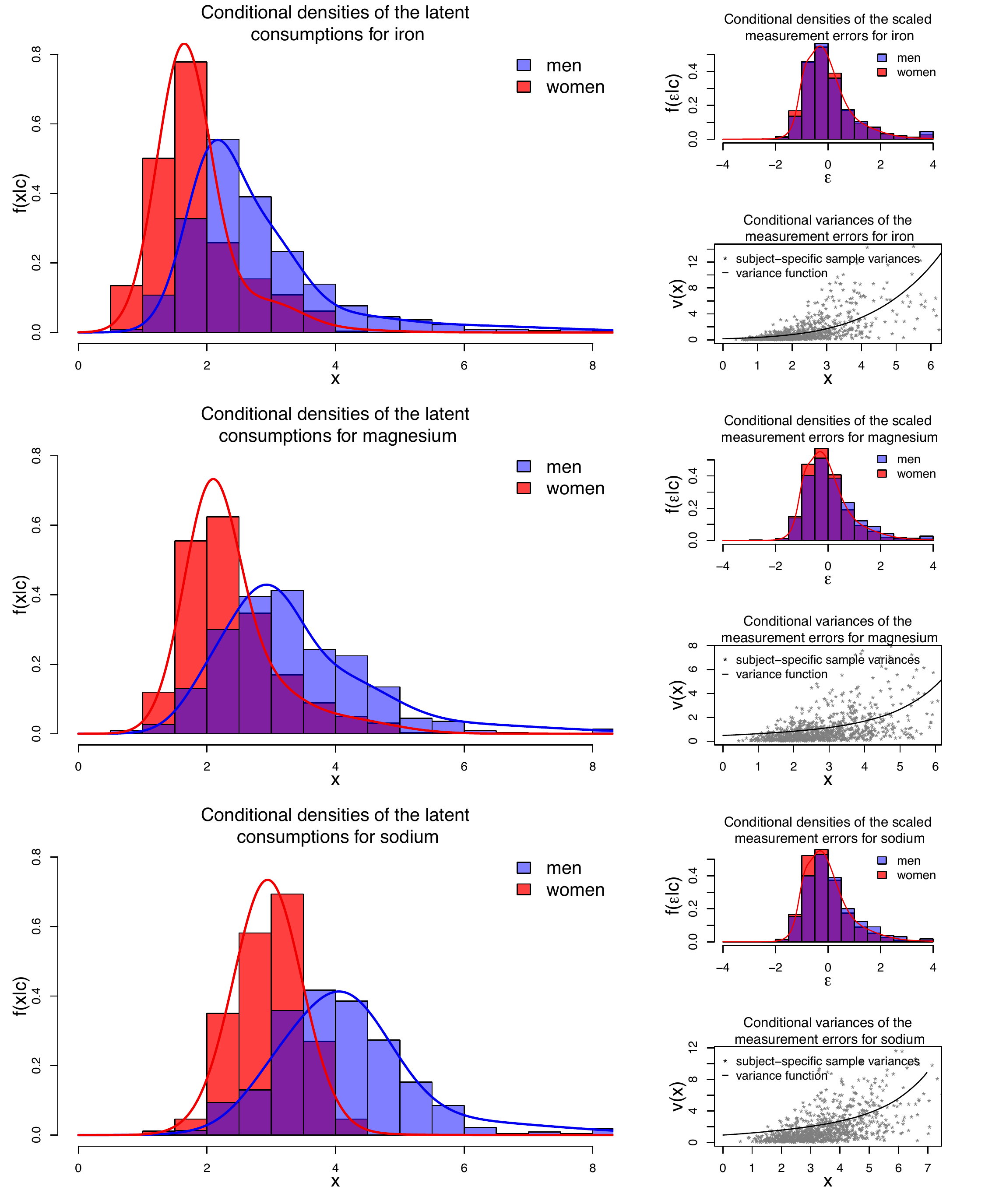}
\caption{\baselineskip=10pt 
Results for the EATS data set obtained by our method. 
From top to bottom, the left panels show the estimated conditional densities $f_{x,\ell \mid \bc}(x_{\ell} \mid \bc)$ 
of iron, magnesium and sodium, respectively. 
The right panels show the associated conditional distributions of the scaled errors $f_{\epsilon,\ell \mid \bc}(\epsilon_{\ell} \mid \bc)$. 
Results for different dietary component and gender combinations are shown here as 
they are the only predictors found important for modeling the densities $f_{x,\ell \mid \bc}(x_{\ell} \mid \bc)$. 
Also shown are the associated variance functions $v_{\ell}(x_{\ell}) = s_{\ell}^{2}(x_{\ell})$. 
}
\label{fig: EATS MARGINALS}
\end{figure}

Figures \ref{fig: EATS ZX Tables} and \ref{fig: EATS ZE Tables} illustrate how the redundant mixture components become near-empty after reaching steady states in our MCMC based implementation. 
While we started with twenty mixture components, only six are finally being used for modeling the densities $f_{x,\ell \mid \bc}$. 
Likewise, while we again started with twenty mixture components, only three are finally being used for modeling the densities $f_{\epsilon,\ell \mid \bc}$. 

Figures \ref{fig: EATS ZX Tables} and \ref{fig: EATS ZE Tables} additionally show how the mixture component specific parameters get shared across different dietary components and predictor combinations in our model and how the associated mixture probabilities vary across these combinations.
For the densities $f_{x,\ell \mid \bc}$, the mixture probabilities clearly vary significantly between men and women as well as between different dietary components, 
hence these variables were selected as important by our method. 
For the densities $f_{\epsilon,\ell \mid \bc}$, on the other hand, 
the mixture probabilities are very similar between men and women as well as between different dietary components, 
hence these covariates were adjudged non-significant by our method.

Figure \ref{fig: EATS MARGINALS} shows the estimated densities $f_{x,\ell \mid \bc}(x_{\ell} \mid \bc)$ 
superimposed over histograms of the corresponding estimated $x_{\ell}$'s obtained by our method. 
Figure \ref{fig: EATS JOINTS} in the supplementary materials 
shows the estimated joint densities $f_{\bx,\ell_{1},\ell_{2} \mid \bc}(x_{\ell_{1}}, x_{\ell_{2}} \mid \bc)$ obtained by our method in the off-diagonal panels. 
The results suggest the model to provide a good fit for the EATS data, 
including especially being able to capture the heavily skewed consumption distributions for men with heavy right tails. 
In comparison, the distributions for women look more symmetric and have much lighter tails.

We compare our results with estimates produced by the method by \cite{Zhang2011b}. 
\cite{Zhang2011b}, strictly speaking, is not a principled deconvolution approach but rather a multi-stage pseudo-Bayesian mixed model approach. 
They use Box-Cox transformations \citep{box1964analysis} of the recalls $w_{\ell,i,j}$ separately for each component. 
The rest of the analysis is done conditional on the estimated Box-Cox parameters, 
assuming the transformed variables to follow a linear mixed model, 
a subject specific random effect component 
and a covariate dependent linear fixed effects component with no interaction terms and an independent error component.  
All covariates are included in the model as there is also no mechanism to select the important ones. 
The random effects components and the errors are modeled using multivariate normal distributions. 
Multivariate normal priors are also assumed for the fixed effects regression coefficients. 
Estimates of the long-term intakes are then obtained via individual transformations back to the original scale. 
As shown in \cite{sarkar2014bayesian, sarkar2021bayesian}, Box-Cox transformations for surrogate observations have severe limitations, 
including almost never being able to produce transformed surrogates that conform to the assumed parametric assumptions, including normality, homoscedasticity, and independence of the errors. 
Single component multivariate normal models are thus highly inadequate for the densities even after transformations.
Estimates of the marginal densities in the original scale are thus obtained not by applying what the model actually implies 
but by applying a univariate kernel density estimation method on the estimated intakes in the observed scale 
thereby mitigating the highly restrictive effects of the inherent parametric assumptions.

For clarity, we summarize the estimates obtained by the method of \cite{Zhang2011b} separately in Figure \ref{fig: EATS MARGINALS Zhang_2011b}, 
moved to the supplementary materials for space limitations. 
The shapes of the estimated densities are in general agreement with those produced by our method. 
The fixed effects regression coefficient estimates are presented in 
Table \ref{tab: EATS Zhang_2011b Results} in the supplementary material. 
With no mechanism to select the important predictors, all covariates are included in the model. 
Taking the exclusion of zero from a $90\%$ central credible interval to be a (somewhat ad-hoc) post-processing rule to 
determine the significance of the associated predictor, 
we can eliminate some, but a good number of coefficients associated with race and age still remain included in the model. 
Exploratory analysis (Figure \ref{fig: EATS exploratory race}) suggests these effects may still be spurious 
- a result of the presence of the  `missing' group 
which is of very small size with only five subjects but includes replicates that look very different from the remaining groups 
(Figure \ref{fig: EATS exploratory race 2} in the supplementary material).

The posterior of our proposed Bayesian hierarchical method did not include age as an important predictor and only included race as important in a small percentage of the MCMC iterations. 
Borrowing information across different predictor groups,  
it is very robust to the presence of small outlying groups such as the  `missing' race.

\section{Discussion} \label{sec: discussion}

In this article, we considered the problem of multivariate density deconvolution in the presence of categorical predictors. 
The problem is important in nutritional epidemiology for estimating long-term intakes of regularly consumed dietary components in the presence of associated demographic variables age, sex and ethnicity. 
We developed a copula based deconvolution approach that focuses on the marginals first and then models the dependence among the components to build the joint densities. 
Our proposed method not only allows the densities to vary flexibly with the associated predictors 
but also allows automatic selection of the most influential predictors. 
Importantly, our proposed method also allows the sets of predictors influencing the density of interest and the density of the measurement errors to potentially be different. 
Applied to our motivating nutritional epidemiology data set, 
we found gender to be an important predictor for the density of long term average intakes of different dietary components.

The applicability of the methodology developed here for covariate informed multivariate densities is not restricted to deconvolution problems but the different model components
can be adapted to other important problems in statistics as well. 
For instance, the methodology developed in Section 2.1 
for modeling $f_{\bx \mid \bc}(\bx \mid \bc)$ can be straightforwardly applied to the problem of
ordinary multivariate density estimation without measurement errors 
in the presence of associated potentially high-dimensional precisely measured covariates. 
Likewise, the methodology developed in Section 2.2 
for modeling $f_{\bepsilon \mid \bc}(\bepsilon \mid \bc)$ can be straightforwardly applied to modeling covariate dependent regression errors 
in the presence of associated potentially high-dimensional precisely measured covariates. 
Section S.8 in the supplementary materials provides additional brief discussions 
and some simulations evaluating the performance of our method for ordinary density estimation problems. 
More rigorous expositions of these problems will be pursued elsewhere.

The methodology developed here is semiparametric in nature, where some model components are highly flexible while some others are highly parametric.
At a first glance, the parametric choices may be perceived as restrictive. 
Deconvolution problems are, however, well known to be extremely difficult ones 
and methods that work for measurement error free settings may not always work for measurement error problems. 
For example, it was shown in \cite{sarkar2014bayesian} that methods that could be adapted to allow all aspects of the error distributions to vary flexibly with covariates, 
e.g., \cite{chung2009nonparametric},  
are not numerically feasible for measurement errors even for moderately large data sets like the EATS, 
and a multiplicative structural assumption $u=s(x)\epsilon$, as considered in our article, 
although in theory more restrictive, is in fact a highly efficient practical choice. 
Likewise, the assumption of covariate independent Gaussian copula can, in principle, be relaxed to include covariates as well as other copula classes. 
In practice, however, these problems are extremely challenging even in measurement error free scenarios \citep{dos2008copula}. 
The Gaussian copula is easy to understand, interpret, and implement and hence is an effective practical choice for deconvolution problems. 

The method of course has other important limits. 
The trick used here to include the component labels as the levels of a categorical covariate 
allowed us to greatly simplify the tensor decomposition computations 
but also restricted each component to be influenced by the same set of important covariates. 
An important direction for future research is to relax this restriction 
to allow different sets of important predictors for different dietary components 
using more flexible partition models. 
Our previous work in \cite{sarkar2021bayesian} also showed that mixtures of truncated normals do not work well for zero-inflated recall data for episodically consumed dietary components 
but requires other modeling strategies to accommodate the sharp boundaries of the densities encountered in such problems. 
Adaptations for episodic components, however, forms a crucial step forward toward a more sophisticated framework for estimating 
the Healthy Eating Index (HEI, https://www.fns.usda.gov/resource/healthy-eating-index-hei), 
a performance measure developed by the US Department of Agriculture (USDA) to assess and promote healthy diets 
\citep{Guenther2008a,krebs2018update}, forming another important direction for future research.

\baselineskip=14pt
\vspace*{-10pt}
\section*{Supplementary Materials}
The supplementary materials details the choice of hyper-parameters and the MCMC algorithm used to sample from the posterior.
R programs implementing the deconvolution methods developed in this article are included as separate files in the supplementary material. 
The EATS data analyzed in Section \ref{sec: applications} can be accessed from National Cancer Institute by arranging a Material Transfer Agreement.
A simulated data set, simulated according to one of the designs described in Section \ref{sec: simulation studies}, 
and a `readme' file providing additional details are also included in the supplementary material.  

\section*{Acknowledgments}
We thank the University of Texas Advanced Computing Center (TACC) for providing computing resources that contributed to the research reported here.

\baselineskip=14pt

\bibliographystyle{natbib}
\bibliography{BNP,ME,Copula,HOMC}

\clearpage\pagebreak\newpage
\pagestyle{fancy}
\fancyhf{}
\rhead{\bfseries\thepage}
\lhead{\bfseries SUPPLEMENTARY MATERIALS}

\begin{center}
\baselineskip=27pt
{\LARGE Supplementary Materials for\\ 
{\bf Bayesian Semiparametric \\
Covariate Informed 
Multivariate Density Deconvolution
}}
\end{center}
%\vskip0.25cm

\vskip 2mm
\begin{center}
Abhra Sarkar\\
abhra.sarkar@utexas.edu \\
Department of Statistics and Data Sciences,
The University of Texas at Austin\\
2317 Speedway D9800, Austin, TX 78712-1823, USA\\
\end{center}

\setcounter{equation}{0}
\setcounter{page}{1}
\setcounter{table}{1}
\setcounter{figure}{0}
\setcounter{section}{0}
\numberwithin{table}{section}
\renewcommand{\theequation}{S.\arabic{equation}}
\renewcommand{\thesubsection}{S.\arabic{section}.\arabic{subsection}}
\renewcommand{\thesection}{S.\arabic{section}}
\renewcommand{\thepage}{S.\arabic{page}}
\renewcommand{\thetable}{S.\arabic{table}}
\renewcommand{\thefigure}{S.\arabic{figure}}
\baselineskip=14pt

\vskip 10mm
Supplementary materials discuss 
copula models where the marginal densities vary with associated precisely measured covariates, 
a brief review of tensor factorization models for easy reference, 
the choice of hyper-parameters and details of the MCMC algorithm we designed to sample from the posterior. 
Supplementary materials also present 
some additional figures summarizing the analysis of the EATS data set. 
Separate files additionally include a synthetic data set simulated according to one of the designs described in Section \ref{sec: simulation studies},  a `readme' file providing additional details of this data set, 
and \texttt{R} programs implementing the deconvolution method developed in this article.

\mybox{Method Overview}{red!40}{red!10}
{
The methodology proposed here builds on many diverse topics. 
A high-level overview of the different model components is presented here for easy reference. 

\begin{enumerate}
\item {\bf Model:} 
\vskip -40pt
\bse
& \bw_{i,j} = \bx_{i}+\bu_{i,j}, ~~~\bu_{i,j}=\bS(\bx_{i})\bepsilon_{i,j},\\
& (\bx_{i} \mid \bc_{i}) \sim f_{\bx \mid \bc}(\bx_{i} \mid \bc_{i}),~~~(\bepsilon_{i,j} \mid \bc_{i}) \sim f_{\bepsilon \mid \bc}(\bepsilon_{i,j} \mid \bc_{i})~\text{with}~\eE(\bepsilon_{i,j} \mid \bc_{i}) = \bzero,\\
& \bS(\bx_{i}) = \diag\{s_{1}(x_{1,i}),\dots,s_{d}(x_{d,i})\}.
\ese

	\begin{enumerate}[labelindent=0pt,labelwidth=0.75em,leftmargin=!]
	\item {\bf Model $f_{\bx \mid \bc}(\bx \mid \bc)$ using copula (Section 2.1 in the main paper).} 
	\bse
	\textstyle f_{\bx \mid \bc}(\bx_{i} \mid \bc_{i}) = c(\bx_{i})\prod_{\ell}f_{x,\ell \mid \bc} (x_{\ell,i} \mid \bc_{i}).
	\ese

		\begin{enumerate}
		\item Model the marginals $f_{x, \ell \mid \bc}$ using mixtures of truncated normals with covariate informed mixture probabilities with atoms shared between different components $\ell$.
		\item Model the covariate informed mixture probabilities using conditional tensor factorization that also allows identification of the important covariates. 
		\item Model the dependence function $c(\bx)$ using a Gaussian copula and its correlation matrix using a polar coordinate parametrization. 
		\end{enumerate}
	
	\item {\bf Model $f_{\boldmath{\epsilon} \mid  \bc}(\bepsilon \mid  \bc)$ using copula (Section 2.2 in the main paper).}
	\bse
	\textstyle f_{\bepsilon \mid \bc}(\bepsilon_{i,j} \mid \bc_{i}) = c(\bepsilon_{i,j})\prod_{\ell}f_{\epsilon,\ell \mid \bc} (\epsilon_{\ell,i,j} \mid \bc_{i}).
	\ese

		\begin{enumerate}
		\item Model the marginals $f_{\epsilon, \ell \mid \bc}$ using mixtures of two-component normals centered to have mean zero with covariate informed mixture probabilities with atoms shared between different components $\ell$.
		\item Model the covariate informed mixture probabilities using conditional tensor factorization that also allows identification of the important covariates. 
		\item Model the dependence function $c(\bepsilon)$ using a Gaussian copula and its correlation matrix using a polar coordinate parametrization.
		\end{enumerate}

	\item {\bf Model $\bS(\bx)$ using B-splines (Section 2.3 in the main paper).}
	Model each component $s_{\ell}^{2}(x_{\ell,i})$ using mixtures of B-splines with large second order differences between the adjacent coefficients penalized to induce smoothness. 
	\end{enumerate}

\item{\bf Prior (in relevant sections in the main paper).} 
Priors, assigned to different model parameters, are described alongside the model components. 
The choice of prior hyper-parameters is discussed in Section S.5 in the supplementary materials.

\item{\bf Posterior (Section S.5 in the supplementary materials).} 
The joint posterior is complex and cannot be computed in closed form nor can be easily sampled from. 
Inference is based on samples from the posterior using an MCMC algorithm. 
\end{enumerate}
}

\clearpage\newpage
\section{Conditional Copula Models} \label{sec: copula model}

The literature on copula models is enormous. 
See, for example, \citelatex{nelsen2007introduction, joe2015dependence, shemyakin2017introduction} and the references therein.

A function $\cC(\bu) = \cC (u_{1},\dots,u_{d}) : [0,1]^{d} \rightarrow [0,1]$ is called a copula 
if $\cC(\bu)$ is a continuous cumulative distribution function (cdf) on $[0,1]^{d}$ 
such that each marginal is a uniform cdf on $[0,1]$. 
That is, for any $\bu \in [0,1]^{d}$,
$\cC(\bu) = \cC (u_{1},\dots,u_{d}) = \Pr(u_{1}\leq u_{1}, \dots,u_{d}\leq u_{d})$ with 
$\cC(1,\dots,1,u_{i},1,\dots,1) = \Pr(u_{i} \leq u_{i}) = u_{i},  i=1,\dots,d$.
If $\{x_{i}\}_{i=1}^{d}$ are absolutely continuous random variables having marginal cdf $\{z_{i}(x_{i})\}_{i=1}^{d}$ 
and marginal probability density functions (pdf) $\{z_{i}(x_{i})\}_{i=1}^{d}$,  joint cdf $H(x_{1},\dots,x_{d})$ and joint pdf $h(x_{1},\dots,x_{d})$, 
then a copula $\cC$ can be defined in terms of $H$ as 
$\cC(u_{1},\dots,u_{d}) = H \left(x_{1}, \dots, x_{d}\right)$ where $u_{i} = z_{i}(x_{i}), i=1,\dots,d$.
It follows that
$h(x_{1},\dots,x_{d}) = c(u_{1},\dots,u_{d})\prod_{i=1}^{d}z_{i}(x_{i})$, 
where $c(u_{1},\dots,u_{d}) = {\partial^{d} \cC(u_{1},\dots,u_{d})}  /  {(\partial u_{1} \dots \partial u_{d})}$.
This defines a copula density $c(\bu)$ in terms of the joint and marginal pdfs of $\{x_{i}\}_{i=1}^{d}$ as
\vspace{-4ex}\\
\be
\textstyle c(u_{1},\dots,u_{d}) = h(x_{1},\dots,x_{d}) / \prod_{i=1}^{d}z_{i}(x_{i}). \label{eq:Copula A2}
\ee
\vspace{-4ex}\\
Conversely, if $\{v_{i}\}_{i=1}^{d}$ are continuous random variables having fixed marginal cdfs $\{F_{i}(v_{i} \mid c_{1},\dots,c_{p})\}_{i=1}^{d}$, 
then their joint cdf $F(v_{1},\dots,v_{d} \mid c_{1},\dots,c_{p})$, with a dependence structure introduced through a copula $\cC$, can be defined as
\vspace{-4ex}\\
\be
F(v_{1},\dots,v_{d} \mid c_{1},\dots,c_{p}) = \cC\{F_{1}(v_{1} \mid c_{1},\dots,c_{p}),\dots,F_{d}(v_{d} \mid c_{1},\dots,c_{p})\} =\cC(u_{1},\dots,u_{d}),         \label{eq:Copula A3} 
\ee
\vspace{-4ex}\\
where $u_{i} = F_{i}(v_{i} \mid c_{1},\dots,c_{p}), i=1,\dots,d$.
If $\{v_{i}\}_{i=1}^{d}$ have marginal densities $\{f_{i}(v_{i} \mid c_{1},\dots,c_{p})\}_{i=1}^{d}$, 
then from (\ref{eq:Copula A3}) it follows that the joint density $f(v_{1},v_{2},\dots,v_{d} \mid c_{1},\dots,c_{p}) $ is given by
\vspace{-4ex}\\
\be
f(v_{1},\dots,v_{d} \mid c_{1},\dots,c_{p}) &= c(u_{1},\dots,u_{d})\prod_{i=1}^{d}f_{i}(v_{i} \mid c_{1},\dots,c_{p}).  \label{eq:Copula A4}
\ee
\vspace{-4ex}\\
With $F_{i}(v_{i}  \mid c_{1},\dots,c_{p}) = u_{i} = z_{i}(x_{i}), i=1,\dots,d$, substitution of the copula density (\ref{eq:Copula A2}) into (\ref{eq:Copula A4}) gives 
\vspace{-4ex}\\
\be
& f(v_{1},\dots,v_{d}  \mid c_{1},\dots,c_{p}) = c(u_{1},\dots,u_{d})\prod_{i=1}^{d}f_{i}(v_{i}  \mid c_{1},\dots,c_{p}) \nonumber\\
& = \bigg\{\frac{h(x_{1},\dots,x_{d})}{\prod_{i=1}^{d}z_{i}(x_{i})}\bigg\}  \prod_{i=1}^{d}f_{i}(v_{i} \mid c_{1},\dots,c_{p}). \label{eq:Copula A5}
\ee
\vspace{-6ex}\\

Equation (\ref{eq:Copula A3}) can be used to define flexible multivariate dependence structure using standard known multivariate densities \citeplatex{sklar1959}. 
Let $\MVN_{d}(\bmu,\bSigma)$ denote a $p$-variate normal distribution with mean vector $\bmu$ and positive semidefinite covariance matrix $\bSigma$. 
An important case is $\bx=(x_{1},\dots,x_{d})\trans \sim \MVN_{d}(\bzero,\bR)$, where $\bR$ is a correlation matrix. 
In this case, $\cC (u_{1},\dots,u_{d}|\bR) = \Phi_{d} \{\Phi^{-1}(u_{1}),\dots,\Phi^{-1}(u_{d})\mid \bR\}$, 
where $\Phi(x) = \Pr\{X \leq x \mid X\sim \Normal(0,1)\}$ and 
$\Phi_{d}(x_{1},\dots,x_{d}|\bR) = \Pr\{x_{1} \leq x_{1}, \dots,x_{d} \leq x_{d} \mid \bx \sim \MVN_{d}(\bzero,\bR)\}$. 
If $\bx \sim N_{d}(\bzero,\bSigma)$, where $\bSigma = ((\sigma_{i,j}))$ is a covariance matrix with $\sigma_{ii}=\sigma_{i}^2$, then defining $\bLambda=\diag(\sigma_{1}^2,\dots,\sigma_{d}^2)$ and $\by = \bLambda^{-\frac{1}{2}}\bx$ and noting that $\bSigma=\bLambda^{1/2} \bR \bLambda^{1/2}$, we have  
\vspace{-4ex}\\
\bse
c(u_{1},\dots,u_{d}) = {\MVN_{d}(\bx \mid \bzero,\bSigma)}   /   {\MVN_{d}(\bx \mid \bzero,\bLambda)}
= |\bLambda|^{1/2} |\bSigma|^{-1/2} \exp\left\{-\bx\trans(\bSigma^{-1}-\bLambda^{-1})\bx/2\right\} \\
= |\bR|^{-1/2} \exp\{-\by\trans(\bR^{-1}-\bI_{d})\by/2\} 
=  {\MVN_{d}(\by\mid \bzero,\bR)}  /  {\MVN_{d}(\by\mid \bzero, \bI_{d})}.
\ese 
\vspace{-4ex}\\
Sticking to the standard normal case, 
a flexible dependence structure between random variables $\{v_{i}\}_{i=1}^{d}$ with given marginals $\{F_{i}(v_{i} \mid c_{1},\dots,c_{p})\}_{i=1}^{d}$ may thus be obtained  
assuming a Gaussian distribution on the latent random variables $\{y_{i}\}_{i=1}^{d}$ 
obtained through the transformations $F_{i}(v_{i} \mid c_{1},\dots,c_{p}) = u_{i} = \Phi(y_{i}), i=1,\dots,d$. 
The joint density of $\bv=(v_{1},\dots,v_{d} \mid c_{1},\dots,c_{p})\trans$ is then given by
\vspace{-4ex}\\
\bse
&& \hspace{-1cm} f(v_{1},\dots,v_{d} \mid c_{1},\dots,c_{p}) = c(u_{1},\dots,u_{d})\prod_{i=1}^{d}f_{i}(v_{i} \mid c_{1},\dots,c_{p}) 
= \frac{\MVN_{d}(\by\mid \bzero,\bR)}   {\MVN_{d}(\by\mid \bzero, \bI_{d})}  \prod_{i=1}^{d}f_{i}(v_{i} \mid c_{1},\dots,c_{p}). 
\ese
\vspace{-4ex}\\
We have 
\vspace{-4ex}\\
\bse
& \Pr(v_{1} \leq v_{1},\dots,v_{d} \leq v_{d} \mid c_{1},\dots,c_{p}) \\
& = \Pr[y_{1} \leq \Phi^{-1}\{F_{1}(v_{1} \mid c_{1},\dots,c_{p})\},\dots,y_{d} \leq \Phi^{-1}\{F_{d}(v_{d} \mid c_{1},\dots,c_{p})\}\mid \bY \sim \MVN_{d}(\bzero,\bR)].
\ese
\vspace{-4ex}\\
For $q\leq d$, with $(y_{1},\dots,y_{q})\trans \sim \MVN_{q}(\bzero,\bR_{q})$, we then have 
\vspace{-4ex}\\
\bse
& \Pr(v_{1} \leq v_{1},\dots,v_{q} \leq v_{q} \mid c_{1},\dots,c_{p}) \\
& = \Pr[y_{1} \leq \Phi^{-1}\{F_{1}(v_{1} \mid c_{1},\dots,c_{p})\},\dots,y_{q} \leq \Phi^{-1}\{F_{q}(v_{q} \mid c_{1},\dots,c_{p})\}\mid \by_{1:q} \sim \MVN_{q}(\bzero,\bR_{q})],
\ese
\vspace{-4ex}\\
implying that the density of $(v_{1},\dots,v_{q})$ will be 
\vspace{-4ex}\\
\bse
& \hspace{-1cm} f(v_{1},\dots,v_{q} \mid c_{1},\dots,c_{p}) = c(u_{1},\dots,u_{q})\prod_{i=1}^{q}f_{i}(v_{i} \mid c_{1},\dots,c_{p}) \\
&= \frac{\MVN_{q}(\by_{1:q} \mid \bzero,\bR_{q})}   {\MVN_{q}(\by_{1:q}\mid \bzero, \bI_{q})}  \prod_{i=1}^{q}f_{i}(v_{i} \mid c_{1},\dots,c_{p}). 
\ese
\vspace{-4ex}

\section{Conditional Tensor Factorization Models} \label{sec: tensor fac review}

\begin{figure}[!ht]
\centering

	\begin{subfigure}{\textwidth}
	\centering
	\includegraphics[width=16cm, trim=1cm 1cm 1cm 1cm, clip=true]{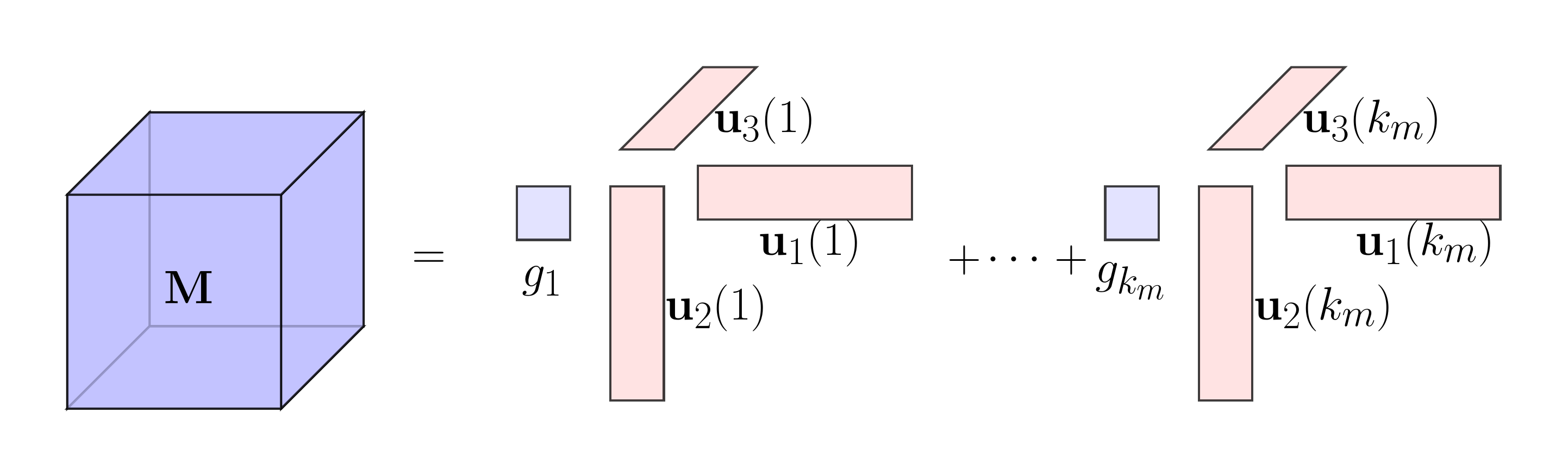}
	\caption{PARAFAC decomposition} 
	\end{subfigure}

	\vskip 5mm
	\begin{subfigure}{\textwidth}
	\centering
	\includegraphics[width=11.5cm, trim=1cm 1cm 1cm 1cm, clip=true]{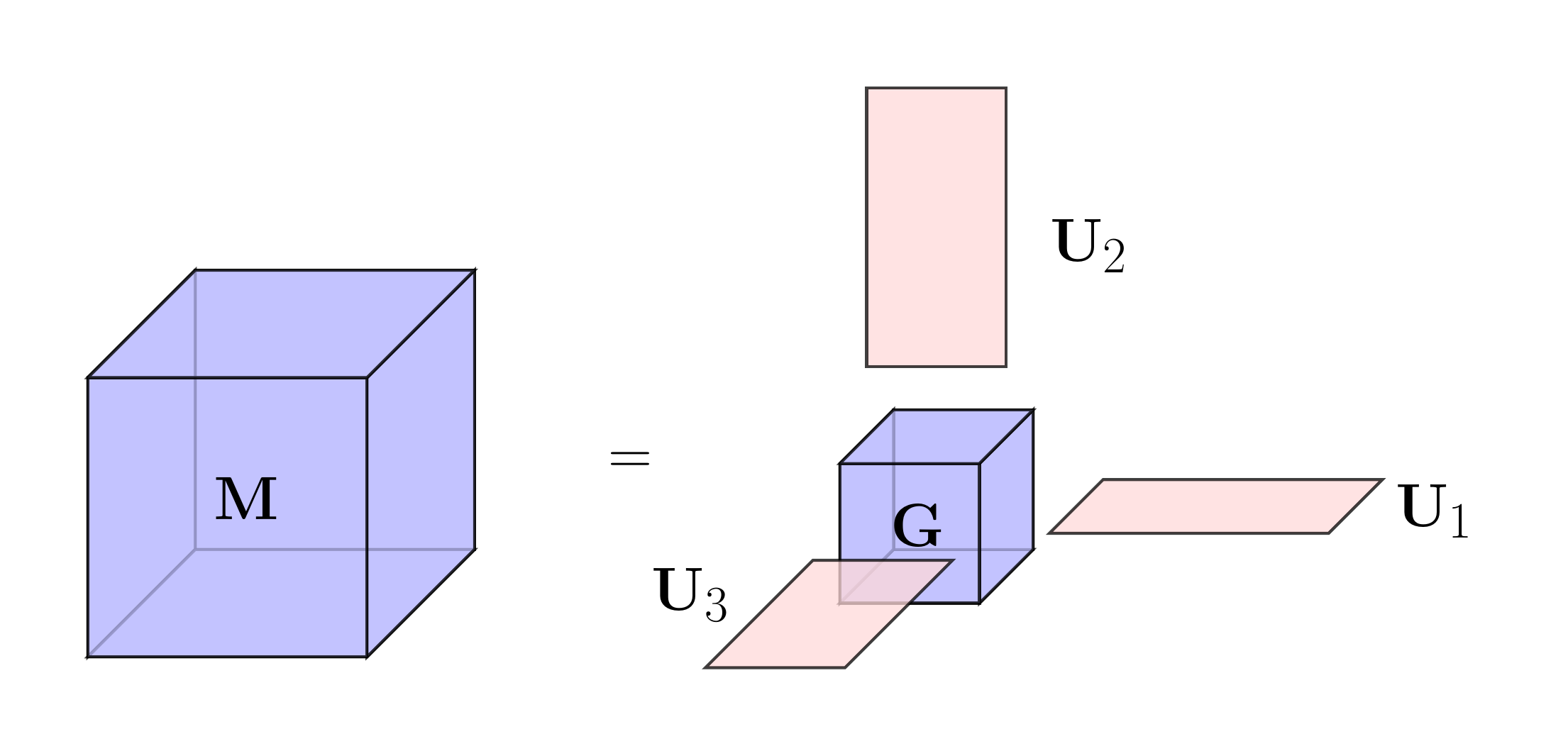}
	\caption{HOSVD/Tucker decomposition} 
	\end{subfigure}

\caption{Pictorial representation of tensor factorizations of a three dimensional tensor $\bM$ 
(a) parallel factor decomposition with vectors 
(b) higher order singular value decomposition with core tensor $\bG$ and mode matrices $\bU^{(h)}, h=1,2,3$.}
\label{fig: HOSVD}
\end{figure}

There is a vast literature on tensor factorization techniques, 
the two most popular approaches being the parallel factor analysis (PARAFAC) and the higher order singular value decomposition (HOSVD). 
The PARAFAC approach \citelatex{harshman:1970} decomposes a $d_{1} \times \dots \times d_{p}$ dimensional tensor $\bM=\{m_{c_{1},\dots,c_{p}}\}$ as the sum of rank one tensors as
\vspace{-4ex}\\
\be
\textstyle m_{c_{1},\dots,c_{p}} = \sum_{k=1}^{k_{m}}g_{k} \prod_{h=1}^{p} u_{h,c_{h}}(k). 
\ee 
\vspace{-4ex}\\
In contrast, the HOSVD approach, proposed by \citelatex{tucker:1966} for three way tensors and extended to multi-way tensors of arbitrary orders by \citelatex{de_lathauwer_etal:2000}, 
would factorize $\bM$ as 
\vspace{-4ex}\\
\be
\textstyle m_{c_{1},\dots,c_{p}} = \sum_{k_{1}=1}^{k_{m,1}}\cdots\sum_{k_{p}=1}^{k_{m,p}}g_{k_{1},\dots,k_{p}} \prod_{h=1}^{p} u_{h,c_{h}}(k_{h}), 
\ee
\vspace{-4ex}\\
where $\bG=\{g_{h_{1},\dots,h_{p}}\}$, a ${k_{m,1} \times \dots \times k_{m,p}}$-dimensional core tensor, captures interactions between the different components and the $k_{m,h} \times d_{h}$ dimensional mode matrices $\bU_{h}=\{u_{h, c_{h}} (k_{h})\}$ consist of the corresponding component specific weights.
See Figure \ref{fig: HOSVD}. 

HOSVD achieves better data compression and requires fewer components compared to the PARAFAC model 
which can be obtained as a special case of HOSVD with the core tensor $\bG$ restricted to being diagonal.

\citelatex{yang_dunson:2015} leveraged these ideas while regressing a categorical response variable $y \in \{1,\dots,k_{y}\}$ on a set of categorical predictors $c_{h} \in \{1,\dots,d_{h}\}$, $h=1,\dots,p$. 
Structuring the conditional probabilities $P_{y \mid \bc}(y\mid c_{h}, h=1,\dots,p)$ as the elements of a $k_{y} \times d_{1} \times \dots \times d_{p}$ dimensional tensor, 
they proposed the following HOSVD-type factorization 
\vspace{-4ex}\\
\be
\textstyle P_{y \mid \bc}(y\mid c_{h},h=1,\dots,p)   =   \sum_{k_{1}=1}^{k_{y,1}}\cdots\sum_{k_{p}=1}^{k_{y,p}} \lambda_{k_{1}\dots k_{p}}(y)\prod_{j=1}^{p}\pi_{h,c_{h}}(k_{h}),  \label{eq: YD CTF}
\ee
\vspace{-4ex}\\
where $1 \leq k_{y,h} \leq d_{h}$ for $h=1,\dots,p$ and the parameters $\lambda_{k_{1},\dots,k_{p}}(y)$ and $\pi_{h,c_{h}}(k_{h})$ are all non-negative and satisfy the constraints 
(a) $\sum_{y=1}^{d_{0}}  \lambda_{k_{1}\dots k_{p}}(y) =1$ for each combination $(k_{1},\dots,k_{p})$, and 
(b) $\sum_{k_{h}=1}^{k_{y,h}} \pi_{h,c_{h}}(k_{h}) = 1$ for each pair $(h,c_{h})$.
See Figure 5 in the main paper. 

\citelatex{yang_dunson:2015} established that any conditional probability tensor can be represented as (\ref{eq: YD CTF}), with the parameters satisfying the constraints (a) and (b). 
When $k_{y,h}=1$, $\pi_{1}^{(h)}(c_{h})=1$ and $P_{y \mid \bc}(y \mid c_{h}, h=1,\dots,p)$ does not vary with $c_{h}$. 
The number of parameters involved in the factorization is given by $(d_{0}-1)\prod_{h=1}^{p}k_{y,h} + d_{0}\sum_{h=1}^{q}(k_{y,h}-1)$, 
which is much less than the number of parameters $(d_{0}-1)\prod_{h=1}^{p}d_{h}$ required to specify a fully parametrized model, if $\prod_{h=1}^{p}k_{y,h} \ll \prod_{h=1}^{p}d_{h}$.

\newpage
\section{Copula Mixture Models with Shared Atoms} \label{sec: shared atoms review}

A mixture model with mixture probabilities $p_{k}$ and mixture kernels $\K(\cdot \mid \btheta_{k})$ parametrized by atoms $\btheta_{k}$ is specified as 
\vspace{-4ex}\\
\bse
\textstyle f_{x}(x) = \sum_{k=1}^{k_{x}} p_{k} \K(x \mid \btheta_{k}). 
\ese
\vspace{-4ex}\\
A Bayesian framework then assigns priors on the mixture probabilities $(p_{1},\dots,p_{k_{x}})$ and the atoms $\{\btheta_{k}\}_{k=1}^{k_{x}}$. 
The number of mixture components $k_{x}$ can be finite \citeplatex[etc.]{fruhwirth2006finite} 
or be a-priori set at $\infty$ in which case the number of `expressed' components can be inferred from the data \citeplatex[etc.]{escobar1995bayesian}. 
Aside from providing a flexible representation of the density $f_{x}$, 
such models also implicitly induce a clustering of the observations $x_{i}, i=1,\dots,n$, 
the $x_{i}$'s associated with the $k\th$ mixture component assumed to belong to the $k\th$ cluster. 

For grouped data $x_{\ell,i}, \ell=1,\dots,d, i=1,\dots,n_{\ell}$, 
with a focus on shared clustering between groups, 
common atoms mixture models 
that keep the atoms fixed between the groups but allow the associated mixture probabilities to vary
can be specified as  
\vspace{-4ex}\\
\bse
\textstyle f_{x,\ell}(x) = \sum_{k=1}^{k_{x}} p_{\ell,k} \K(x \mid \btheta_{k}). 
\ese
\vspace{-4ex}\\
A number of highly sophisticated Bayesian nonparametric hierarchical priors have been developed in the literature 
that induce such models at the observation level 
while may or may not allow additional clustering of the entire group distributions $f_{x,\ell}, \ell=1,\dots,d$. 
See, for example, \citelatex{beraha2021semi,denti2021common} and the references cited therein.  

In deconvolution context, 
where the main focus is not so much on clustering but instead 
on obtaining a flexible representation of the joint density of a multivariate continuous random variable $\bx=(x_{1},\dots,x_{d})\trans$, 
a similar idea with finite mixture models for the marginals and a Gaussian copula for the dependence structure 
was independently developed in \citetlatex{sarkar2021bayesian}, 
where the different component dimensions $\ell=1,\dots,d$ played the roles of different groups 
and the dependence between the components was separately modeled using a Gaussian copula. 
Specifically, they let 
\vspace{-4ex}\\
\bse
\textstyle f_{\bx}(\bx) = |\bR_{\bx}|^{-\frac{1}{2}} \exp\left\{-\frac{1}{2}\by_{\bx}\trans(\bR_{\bx}^{-1}- \bI_{d})\by_{\bx}\right\}  \prod_{\ell=1}^{d} f_{x,\ell}(x_{\ell}), 
\ese
\vspace{-4ex}\\
where 
$F_{x,\ell}(x_{\ell}) = \Phi(y_{x,\ell})$ for all $\ell$, 
$F_{x,\ell}$ being the cdf corresponding to $f_{x,\ell}$;
$\by_{\bx} = (y_{x,1},\dots,y_{x,d})\trans$;
$\Phi(\cdot)$ denotes the cdf of a standard normal distribution; 
and $\bR_{\bx}$ is the correlation matrix between the $d$ components of $\bx$ 
and the marginals were modeled as mixtures of truncated normals with shared atoms $\{(\mu_{k},\sigma_{k}^{2})\}_{k=1}^{k_{x}}$ as 
\vspace{-4ex}\\
\bse
\textstyle f_{x,\ell}(x) = \sum_{k=1}^{k_{x}} p_{\ell,k} \TN(x \mid \mu_{k},\sigma_{k}^{2},[A,B]). 
\ese
\vspace{-4ex}\\
A related model where the marginals were modeled using infinite mixtures of normals, each having its own set of atoms, 
and were linked using a Bernstein polynomial copula had previously appeared in \citelatex{burda2014copula}.

\newpage
\section{Additional Exploratory Figures} \label{sec: exploratory figures}
\begin{figure}[!ht]
\begin{center}
\includegraphics[height=16.5cm, trim=0cm 0cm 0cm 0cm, clip=true]{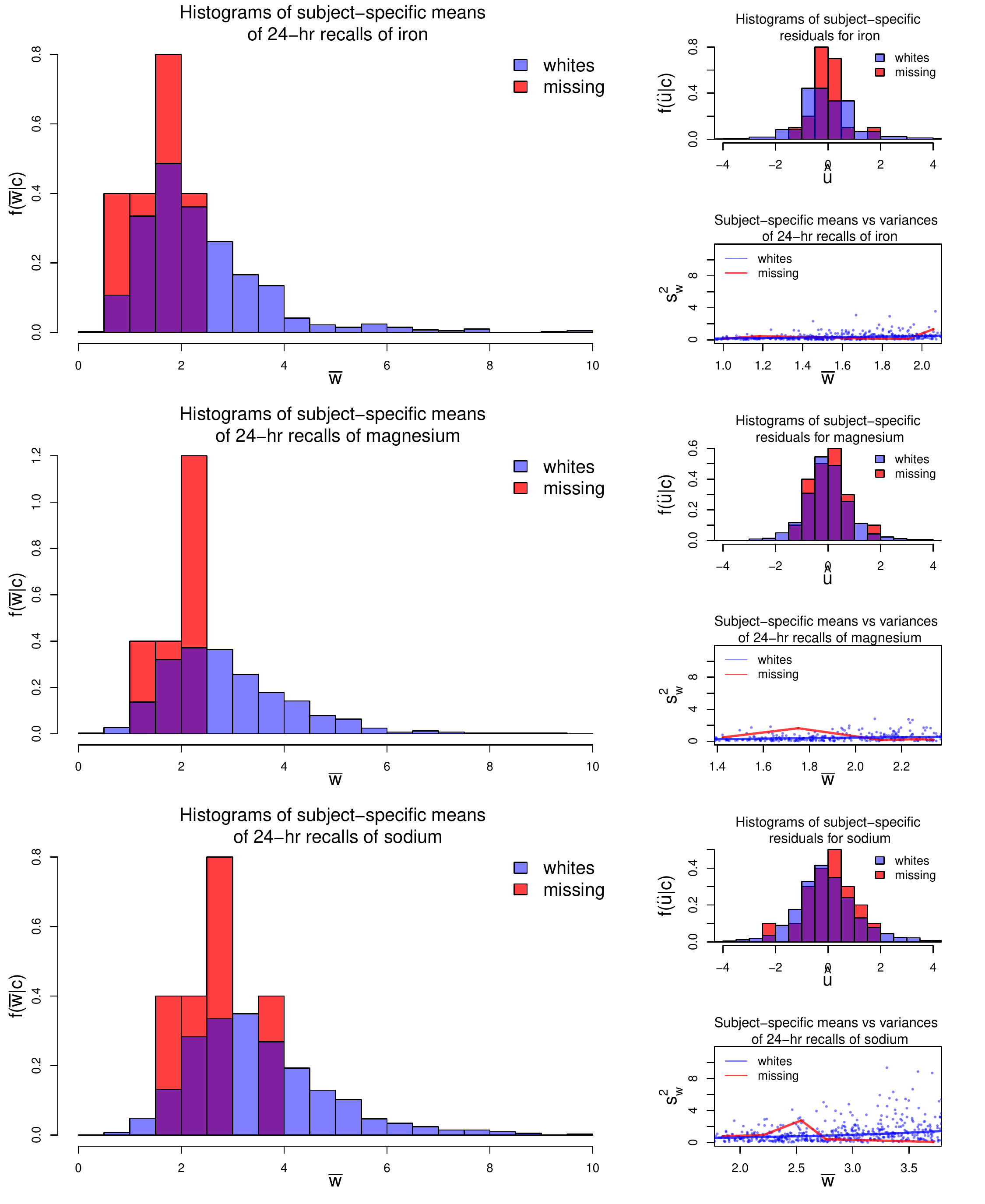}
\end{center}
\caption{\baselineskip=10pt Exploratory plots for iron, magnesium and sodium (from top to bottom). 
Left panels: histograms of subject-specific means $\overline{w}_{\ell,i}$, crude estimates of $x_{\ell,i}$; 
right upper panels: histograms of `residuals' $\wh{u}_{\ell,i,j} = (w_{\ell,i,j} - \overline{w}_{\ell,i})$, crude estimates of $u_{\ell,i,j}$; 
right lower panels: subject-specific means $\overline{w}_{\ell,i}$ vs variances $s_{w,\ell,i}^{2}$, crude estimates of $\var(u_{\ell,i,j} \mid x_{\ell,i})$, superimposed with lowess fits. 
}
\label{fig: EATS exploratory race 2}
\end{figure}

\newpage
\section{\large{Hyper-parameter Choices and Posterior Computation}}  \label{sec: posterior computation}

Samples from the posterior can be drawn using the MCMC algorithm described below. 
In what follows, $\bzeta$ denotes a generic variable that collects the data as well as all parameters of the model, 
including the sampled values of $\bx_{1:n}$ and $\bepsilon_{1:N}$, that are not explicitly mentioned.
Also, the generic notation $p_{0}$ is sometimes used for specifying priors and hyper-priors. 

We now discuss our choices for the prior hyper-parameters and the initial values of the MCMC sampler. 
Carefully chosen starting values facilitate convergence of our sampler.
The starting values of some of the parameters for the multivariate problem are determined by first running samplers for the covariate independent univariate model of \cite{sarkar2014bayesian}.
We describe the hyper-parameter choices and the initial values for the sampler for these univariate models first. 
Unless otherwise mentioned, the prior hyper-parameter choices for similar model components for the multivariate model remain the same as that used for the univariate marginal models. 
We only detail the sampling steps for the multivariate method. 
The steps for the univariate method are detailed in \cite{sarkar2014bayesian}. 

To make the recalls for all the components to be unit free and have a shared support, we transformed the recalls as 
$w_{\ell,i,j} = 20 \times \frac{w_{\ell,i,j}}{\max\{w_{\ell,i,j}\}}$.
The latent $x_{\ell,i}$'s can then be assumed to lie in $[A,B] = [0,10]$, greatly simplifying model specification and hyper-parameter selection. 
As opposed to the non-linear Box-Cox transformations used in the previous literature, including  \cite{Zhang2011b}, which often result in loss of information and introduce bias, 
we only make linear scale transformations here that preserve all features of the original data points. 

For the univariate samplers for the marginal components, 
we used the subject-specific sample means $\overline{w}_{\ell,1:n}$ as the starting values for $x_{\ell,1:n}$.
The appropriate number of mixture components in a mixture model depends on the flexibility of the component mixture kernels as well as on specific demands of the particular application at hand. 
With appropriately chosen mixture kernels, univariate mixture models with 5-10 components have often been found to be sufficiently flexible. 
Detailed guidelines on selecting the number of mixture components for the specific context of deconvolution problems can be found in Section S.1 and S.6 in the supplementary materials of \cite{sarkar2018bayesian}. 
Based on such guidelines, 
we used $J_{\ell}=12$ equidistant knot points for the B-splines supported on $[A,B] = [0,10]$ for modeling the variance functions, 
and $k_{x,\ell}=10$ truncated normals for modeling the densities. 
We also allowed $k_{\epsilon,\ell}=10$ mixture components for the mixtures modeling the densities of the scaled errors. 
For the Dirichlet prior hyper-parameters, we set $\alpha_{x} = 1/k_{x}$, $\alpha_{\epsilon} = 1/k_{\epsilon}$. 
The hyper-parameters for the smoothness inducing parameters are set to be mildly informative as 
$a_{\vartheta}=10, b_{\vartheta}=1$. 
Introducing latent mixture component allocation variables $\bz_{x,1:d,1:n}$, $\bz_{\epsilon,1:d,1:N}$ and $\bz2_{\epsilon,1:d,1:N}$, we can write  
\vspace{-4ex}\\
\bse
& (x_{\ell,i} \mid z_{x,\ell,i}=k,\mu_{x,\ell,k},\sigma_{x,\ell,k}^{2}) \sim \TN(x_{\ell,i} \mid \mu_{x,\ell,k},\sigma_{x,\ell,k}^{2},[A,B]), ~~~\hbox{and}\\
& (\epsilon_{\ell,i,j} \mid z_{\epsilon,\ell,i,j}=k,z2_{\epsilon,\ell,i,j}=t,\mu_{\epsilon,\ell,k,t},\sigma_{\epsilon,\ell,k,t}^{2}) \sim \Normal(\epsilon_{\ell,i,j} \mid \mu_{\epsilon,\ell,k,t},\sigma_{\epsilon,\ell,k,t}^{2}),~~~\ell=1,\dots,d.
\ese
\vspace{-4ex}\\
The mixture labels $z_{x,\ell,i}$'s, and the component specific parameters $\mu_{x,\ell,k}$'s and $\sigma_{x,\ell,k}$'s are initialized by fitting a $k$-means algorithm with $k=k_{x}$.
The parameters of the distribution of scaled errors are initialized at values that correspond to the special standard normal case.
The initial values of the smoothness inducing parameters are set at $\sigma_{\vartheta,\ell}^{2} = \sigma_{\xi,\ell}^{2} = \sigma_{\xi,\ell}^{2} = 0.1$. 
The associated mixture labels $z_{\epsilon,\ell,i,j}$'s are thus all initialized at $z_{\epsilon,\ell,i,j}=1$.
The initial values of $\bvartheta_{\ell}$'s are obtained by maximizing 
\vspace{-4ex}\\
\bse
\ell(\bvartheta_{\ell}\mid \sigma_{\vartheta,\ell}^{2},\overline{\bw}_{\ell,1:n}) = -\frac{\bvartheta_{\ell}\trans \bP_{\ell} \bvartheta_{\ell}}{2\sigma_{\vartheta,\ell}^{2}} - \sum_{i=1}^{n} \frac{1}{2 s_{\ell}^{2}(\overline{w}_{\ell,i},\bvartheta_{\ell})} \sum_{j=1}^{m_{i}}(w_{\ell,i,j}-\overline{w}_{\ell,i})^{2}
\ese
\vspace{-4ex}\\
with respect to $\bvartheta_{\ell}$.

We now discuss how we set the initial values of the sampler for the multivariate method. 
The starting values of the $w_{\ell,i,j}$'s, $x_{\ell,i}$'s, $u_{\ell,i,j}$'s, $\bvartheta_{\ell}$'s were all set at the corresponding estimates returned by the univariate samplers. 
We set the number of shared atoms of the mixture models for the densities $f_{x,\ell}$ and $f_{\epsilon,\ell}$ at $k_{x} = k_{\epsilon} = \max\{5d,20\}$.  
We set $\alpha_{x,\ell}=\alpha_{\epsilon,\ell}=1$. 
The atoms of the mixtures of truncated normals for the marginal densities $f_{x,\ell}$ of the regular components are shared, 
so are the atoms of the mixture models for the univariate marginals $f_{\epsilon,\ell}$ of the scaled errors, 
and hence these parameters could not be initialized directly using the univariate model output. 
We initialized these parameters by iteratively sampling them from their posterior full conditionals $100$ times, keeping the estimated $x_{\ell,i}$'s fixed.   
Adopting a similar strategy, we initialized the parameters specifying the densities $f_{\epsilon,\ell}$ of the scaled errors 
by iteratively sampling them from their posterior full conditionals $100$ times, keeping the estimated errors $u_{\ell,i,j}$'s fixed.   
Finally, the parameters specifying $\bR_{\bx}$ and $\bR_{\bepsilon}$ were set at values that correspond to the special case $\bR_{\bx}=\bR_{\bepsilon}=\bI_{d}$.

We set the variable selection parameters $k_{x,0}=d$ and $k_{x,h}=1$ for all $h=1,\dots,p$.
Likewise, $k_{\epsilon,0}=d$ and $k_{\epsilon,h}=1$ for all $h=1,\dots,p$. 
So only the component label is considered important initially. 
The values of the associated latent variables are set accordingly. 
Specifically, we let $z_{x,0,\ell,i}=\ell$ for all $i=1,\dots,n$ and $z_{x,h,\ell,i}=1$ for all $h=1,\dots,p; \ell=1,\dots,d; i=1,\dots,n$. 
Likewise, $z_{\epsilon,0,\ell,i,j}=\ell$ for all $i=1,\dots,n;j=1,\dots,m_{i}$ and $z_{\epsilon,h,\ell,i,j}=1$ for all $h=1,\dots,p; \ell=1,\dots,d; i=1,\dots,n;j=1,\dots,m_{i}$.

In our sampler for the multivariate problem, 
we first update the parameters specifying the different marginal densities using a pseudo-likelihood that ignores the contribution of the copula. 
The parameters characterizing the copula and the latent $\bx_{i}$'s are then updated using the exact likelihood function conditionally on the parameters obtained in the first step. 
We then update the parameters of the marginal densities again and so forth. 
A more appealing approach would have been to perform joint estimation of the marginal distributions and the copula functions. 
Joint estimation algorithms, most involving carefully designed Metroplis-Hastings (M-H) moves, have been proposed in much simpler settings in \citelatex{pitt2006efficient, wu2014bayesian, wu2015bayesian} etc. 
Designing such moves for our complex deconvolution problem is a daunting task. 
Importantly, the results of \citelatex{dos2008copula} suggest that two-stage approaches often perform just as good as joint estimation procedures, validating their use for practical reasons.

Our sampler for the multivariate model iterates between the following steps. 

\begin{enumerate}[topsep=0ex,itemsep=2ex,partopsep=2ex,parsep=0ex, leftmargin=0cm, rightmargin=0cm, wide=3ex]
\item {\bf Updating the parameters specifying $f_{x \mid \bc}$:}	
	We have
	\vspace{-4ex}\\
	\bse
	 & \Pr(\blambda_{x,k_{0},k_{1},\dots,k_{p}} \mid \bzeta) = \textstyle \Dir\{\alpha_{x}\lambda_{x,0}(1)+n_{x,k_{0},k_{1},\dots,k_{p}}(1),\dots,\alpha_{x}\lambda_{x,0}(k_{x})+n_{x,k_{0},k_{1},\dots,k_{p}}(k_{x})\}, \\
	& P_{x \mid \bc}(z_{x,\ell,i}=k \mid z_{x,0,\ell,i}=k_{0},z_{x,1,\ell,i}=k_{1},\dots,z_{x,p,\ell,i}=k_{p}, \bzeta) \\
	& \propto \lambda_{x,k_{0},k_{1},\dots,k_{p}}(k) \times \TN(x_{\ell,i}\mid \mu_{x,k},\sigma_{x,k}^{2},[A,B]), 
	\ese
	\vspace{-4ex}\\
	where $n_{x,k_{0},k_{1},\dots,k_{p}}(k) = \sum_{\ell=1}^{d}\sum_{i=1}^{n}1(z_{x,\ell,i}=k, z_{x,0,\ell,i}=k_{0},z_{x,1,\ell,i}=k_{1},\dots,z_{x,p,\ell,i}=k_{p})$.
	To update $\blambda_{x,0}$, 
	mimicking ideas presented in \citelatex{sarkar2020bayesianhohmm}, 
	for $s=1,\dots,n_{x,k_{0},k_{1},\dots,k_{p}}(k)$, 
	we first sample an auxiliary variable $\omega_{s}$ as
	\vspace{-4ex}\\
	\bse
	\omega_s \mid \bzeta \sim \Bern\left\{\frac{\alpha_{x}\lambda_{x,0}(k)}{s-1+\alpha_{x}\lambda_{x,0}(k)}\right\}.
	\ese
	\vspace{-4ex}\\
	We then set $m_{x,k_{0},k_{1},\dots,k_{p}}(k)=\sum_{s}\omega_{s}$, 
	$m_{x,0}(k)=\sum_{(k_{0},k_{1},\dots,k_{p})}m_{x,k_{0},k_{1},\dots,k_{p}}(k)$.
	Finally, we sample $\blambda_{x,0}$ as
	\vspace{-4ex}\\
	\bse
	\blambda_{x,0} \mid \bzeta \sim \Dir\{\alpha_{x,0}/k_{x}+m_{x,0}(1),\dots,\alpha_{x,0}/k_{x}+m_{x,0}(k_{x})\}. 
	\ese
	\vspace{-4ex}\\
	The full conditionals of  $\mu_{x,k}$ and $\sigma_{x,k}^{2}$ are given by
	\vspace{-4ex}\\
	\bse
	& \textstyle \Pr(\mu_{x,k}\mid \bzeta) \propto p_{0}(\mu_{x,k}) \times \prod_{\ell=1}^{d}\prod_{\{i: z_{x,\ell,i}=k\}}\TN(x_{\ell,i}\mid \mu_{x,k},\sigma_{x,k}^{2},[A,B]), \\
	& \textstyle \Pr(\sigma_{x,k}^{2} \mid \bzeta) \propto p_{0}(\sigma_{x,k}^{2}) \times \prod_{\ell=1}^{d} \prod_{\{i: z_{x,\ell,i}=k\}}\TN(x_{\ell,i}\mid \mu_{x,k},\sigma_{x,k}^{2},[A,B]).
\ese
	\vspace{-4ex}\\
	These parameters are updated by Metropolis-Hastings (MH) steps with the proposals  
	$q(\mu_{x,k}  \to  \mu_{x,k,new})=\Normal(\mu_{x,k,new} \mid \mu_{x,k},\sigma_{x,\ell,\mu}^{2})$ 
	and $q(\sigma_{x,k}^{2}  \to  \sigma_{x,k,new}^{2})=\TN(\sigma_{x,k,new}^{2}\mid \sigma_{x,k}^{2},\sigma_{x,\ell,\sigma}^{2},[\max\{0,\sigma_{x,k}^{2}-1\},\sigma_{x,k}^{2}+1])$, respectively.

\item {\bf Updating the covariate selection parameters $\bk_{x}$ and $\bz_{x}$:}	
The current values of $\bk_{x}=(k_{x,0},k_{x,1},\dots,k_{x,p})\trans$ and $\bz_{x,\ell,i}=(z_{x,0,\ell,i},z_{x,1,\ell,i},\dots,z_{x,p,\ell,i})\trans$ induce 
a partition of the $d_{h}$ levels of $c_{h}$ into $k_{x,h}$ clusters $\{\C_{x,h,r}: r=1,\dots,k_{x,h}\}$ with each cluster $\C_{x,h,r}$ corresponding to the latent class $z_{x,h}=r$. 
Integrating out $\blambda_{x,k_{0},k_{1},\dots,k_{p}}$, 
conditional on the cluster configurations $\C_{x}=\{\C_{x,h,r}: h=0,\dots,p, r=1,\dots,k_{x,h}\}$, we have 
\be
& \hspace{-0.75cm} \Pr(\bz_{x} \mid \C_{x}, \bzeta) =  \prod_{(k_{0},k_{1},\dots,k_{p})}    \frac{\beta\{\alpha_{x}\lambda_{x,0}(1)+n_{x,k_{0},k_{1},\dots,k_{p}}(1), \dots, \alpha_{x}\lambda_{x,0}(k_{x})+n_{x,k_{0},k_{1},\dots,k_{p}}(k_{x})\}}{\beta\{\alpha_{x}\lambda_{x,0}(1),\dots,\alpha_{x}\lambda_{x,0}(k_{x})\}}, \label{eq: marginal likelihood}
\ee
where $\beta(a_{1},\dots,a_{r}) = \prod_{j=1}^{r}\Gamma(a_{j})/\Gamma(a_{1}+\dots+a_{r})$ is the Beta function.
Given the current values of $k_{x,h}$ and the current clusters $\C_{x}=\{\C_{x,h,r}: h=0,\dots,p, r=1,\dots,k_{x,h}\}$, we do the following for $h=0,\dots,p$.
%\State
If $k_{x,h}<c_{h}$, we propose to increase $k_{x,h}$ to $(k_{x,h}+1)$. 
If $k_{x,h}>1$, we propose to decrease $k_{x,h}$ to $(k_{x,h}-1)$.
For $1<k_{x,h}<c_{h}$, the moves are proposed with equal probabilities. 
For $k_{x,h}=1$, the increase move is selected with probability $1$.
For $k_{x,h}=c_{h}$, the decrease move is selected with probability $1$. 
%\State
If an increase move is proposed, we randomly split a cluster into two. 
%\State
If a decrease move is proposed, we randomly merge two clusters into a single one. 
%\State
We accept the proposed moves with acceptance rates based on the marginal likelihood (\ref{eq: marginal likelihood}). 
%if $\sum_{\ell}k_{\ell}>q$.
%\State
Finally, we set $\bz_{x}$ to be the cluster allocation variables determined by the updated cluster mappings.

\item {\bf Updating the parameters specifying $f_{\epsilon \mid \bc}$:} 
	We have
	\vspace{-4ex}\\
	\bse
	& \Pr(\blambda_{\epsilon,k_{0},k_{1},\dots,k_{p}} \mid \bzeta) = \textstyle \Dir\{\alpha_{\epsilon}\lambda_{\epsilon,0}(1)+n_{\epsilon,k_{0},k_{1},\dots,k_{p}}(1),\dots,\alpha_{\epsilon}\lambda_{\epsilon,0}(k_{\epsilon})+n_{\epsilon,k_{0},k_{1},\dots,k_{p}}(k_{\epsilon})\},\\
	& P_{\epsilon \mid \bc}(z_{\epsilon,\ell,i,j}=k \mid  z_{\epsilon,0,\ell,i,j}=k_{0},z_{\epsilon,1,\ell,i,j}=k_{1},\dots,z_{\epsilon,p,\ell,i,j}=k_{p}, \bzeta) \\
	& \propto \lambda_{\epsilon,k_{0},k_{1},\dots,k_{p}}(k) \times f_{w_{\ell}\mid x_{\ell}}(w_{\ell,i,j}\mid p_{\epsilon,k},\mu_{\epsilon,k},\sigma_{\epsilon,k,1}^{2},\sigma_{\epsilon,k,2}^{2},\bzeta), 
	\ese
	\vspace{-4ex}\\
	where $n_{\epsilon,k_{0},k_{1},\dots,k_{p}}(k)=\sum_{\ell=1}^{d}\sum_{i=1}^{n}\sum_{j=1}^{m_{i}}1(z_{\epsilon,\ell,i,j}=k, z_{\epsilon,0,\ell,i,j}=k_{0},z_{\epsilon,1,\ell,i,j}=k_{1},\dots,z_{\epsilon,p,\ell,i,j}=k_{p})$. 
	To update $\blambda_{\epsilon,0}$, 
	following the same ideas for sampling $\blambda_{x,0}$ above, 
	for $s=1,\dots,n_{\epsilon,k_{0},k_{1},\dots,k_{p}}(k)$, we first sample an auxiliary variable $\omega_{s}$ as
	\vspace{-4ex}\\
	\bse
	\omega_{s} \mid \bzeta \sim \Bern\left\{\frac{\alpha_{\epsilon}\lambda_{\epsilon,0}(k)}{s-1+\alpha_{\epsilon}\lambda_{\epsilon,0}(k)}\right\}.
	\ese
	\vspace{-4ex}\\
	We set $m_{\epsilon,k_{0},k_{1},\dots,k_{p}}(k)=\sum_{s}\omega_{s}$, 
	$m_{\epsilon,0}(k)=\sum_{(k_{0},k_{1},\dots,k_{p})}m_{\epsilon,k_{0},k_{1},\dots,k_{p}}(k)$, and sample $\blambda_{\epsilon,0}$ as
	\vspace{-4ex}\\
	\bse
	\blambda_{\epsilon,0} \mid \bzeta \sim \Dir\{\alpha_{\epsilon,0}/k_{\epsilon}+m_{\epsilon,0}(1),\dots,\alpha_{\epsilon,0}/k_{\epsilon}+m_{\epsilon,0}(k_{\epsilon})\}. 
	\ese
	\vspace{-4ex}\\
	We propose a new $(p_{\epsilon,k},\mu_{\epsilon,k},\sigma_{\epsilon,k,1}^{2},\sigma_{\epsilon,k,2}^{2})$ with the proposal
	$q\{ \btheta_{\epsilon,k} = (p_{\epsilon,k},\mu_{\epsilon,k},\sigma_{\epsilon,k,1}^{2},\sigma_{\epsilon,k,2}^{2})  \rightarrow (p_{\epsilon,k,new},\mu_{\epsilon,k,new},\sigma_{\epsilon,k,1,new}^{2},\sigma_{\epsilon,k,2,new}^{2}) = \btheta_{\epsilon,k,new}\} =
\hbox{TN}(p_{\epsilon,k,new}\mid p_{\epsilon,k},\sigma_{p,\epsilon}^{2},[0,1])~\times~
\Normal(\mu_{\epsilon,k,new}\mid \mu_{\epsilon,k}, \sigma_{\epsilon,\mu}^{2})~\times~
\hbox{TN}(\sigma_{\epsilon,k,1,new}^{2}\mid \sigma_{\epsilon,k,1}^{2},\sigma_{\epsilon,\sigma}^{2},[0,\infty))~\times~
\hbox{TN}(\sigma_{\epsilon,k,2,new}^{2}\mid \sigma_{\epsilon,k,2}^{2},\sigma_{\epsilon,\sigma}^{2},[0,\infty))$.
	We update $\btheta_{k}$ to the proposed value $\btheta_{k,new}$ with probability 
	\vspace{-4ex}\\
	\bse
	\min\bigg\{1, \frac{q(\btheta_{\epsilon,k,new} \rightarrow \btheta_{\epsilon,k})}
		      {q(\btheta_{\epsilon,k} \rightarrow \btheta_{\epsilon,k,new})}
	    \frac{\prod_{\ell=q+1}^{2q+p}\prod_{\{i,j: c_{\epsilon,\ell,i,j}=k\}} f_{w_{\ell}\mid \wt{x}_{\ell}}(w_{\ell,i,j}\mid \btheta_{\epsilon,k,new},\bzeta) ~ p_{0}(\btheta_{\epsilon,k,new})}
	      {\prod_{\ell=q+1}^{2q+p} \prod_{\{i,j: c_{\epsilon,\ell,i,j}=k\}}f_{w_{\ell}\mid \wt{x}_{\ell}}(w_{\ell,i,j}\mid \btheta_{\epsilon,k},\bzeta) ~ p_{0}(\btheta_{\epsilon,k})}\bigg\}.
	\ese
	\vspace{-4ex}

\item {\bf Updating the covariate selection parameters $\bk_{\epsilon}$ and $\bz_{\epsilon}$:}	
The values of $\bk_{\epsilon}=(k_{\epsilon,0},k_{\epsilon,1},\dots,k_{\epsilon,p})\trans$ and $\bz_{\epsilon,\ell,i,j}=(z_{\epsilon,0,\ell,i,j},z_{\epsilon,1,\ell,i,j},\dots,z_{\epsilon,p,\ell,i,j})\trans$ 
are updated mimicking the same strategy used to update $\bk_{x}$ and $\bz_{x}$.

\item {\bf Updating the parameters specifying $v_{\ell}$: } 
The full conditional of each $\bvartheta_{\ell}$ is 
\vspace{-4ex}\\
\bse
\Pr(\bvartheta_{\ell} \mid \bw_{\ell,1:N},\bzeta) \propto p_{0}(\bvartheta_{\ell}) \times \prod_{i,j} f_{w_{\ell}\mid x_{\ell}}(w_{\ell,i,j}\mid x_{\ell,i},\bvartheta_{\ell},\bzeta).
\ese
\vspace{-4ex}\\
We use M-H sampler with random walk proposal
$q(\bvartheta_{\ell} \rightarrow \bvartheta_{\ell,new}) = \MVN(\bvartheta_{\ell,new}\mid \bvartheta_{\ell},\bSigma_{\vartheta,\ell})$.

\item {\bf Updating the values of $\bx$:}
The full conditionals for $\bx_{i}$ are given by
\vspace{-4ex}\\
\bse
&&\hspace{-1cm}(\bx_{i}\mid \bc, \bzeta) 
\propto f_{\bx \mid \bc}(\bx_{i}\mid \bc,\bzeta) \times \textstyle\prod_{j=1}^{m_{i}} f_{\bw\mid {\bx}}(\bw_{i,j} \mid {\bx}_{i}, \bzeta)  \\
&& \textstyle = |\bR_{\bx}|^{-1/2} \exp\left\{-\frac{1}{2}\by_{\bx,i}\trans(\bR_{\bx}^{-1}-\bI_{d})\by_{\bx,i}\right\}\prod_{\ell=1}^{d}f_{x,\ell \mid \bc}(x_{\ell,i}\mid \bc,\bzeta)  \\
&& ~~~~~~ \textstyle\times ~ \prod_{j=1}^{m_{i}} \left[ |\bR_{\bepsilon}|^{-1/2} \exp\left\{-\frac{1}{2}\by_{\epsilon,i,j}\trans(\bR_{\bepsilon}^{-1}-\bI_{d})\by_{\epsilon,i,j}\right\}\prod_{\ell=1}^{d}f_{w_{\ell}\mid x_{\ell}}(w_{\ell,i,j}\mid x_{\ell,i},\bzeta) \right],
\ese
\vspace{-4ex}\\
where $F_{x,\ell}(x_{\ell,i}\mid\bzeta)=\Phi(y_{x,\ell,i})$ and $F_{\epsilon,\ell}\{(w_{\ell,i,j}-x_{\ell,i})/s_{\ell}(x_{\ell,i})\mid\bzeta\}=\Phi(y_{\epsilon,\ell,i,j})$. 
The full conditionals do not have closed forms. 
MH steps with independent truncated normal proposals for each component are used within the Gibbs sampler.

\item {\bf Updating the parameters specifying the copula: } 
We have $F_{x,\ell}(x_{\ell,i}\mid\bzeta)=\Phi(y_{x,\ell,i})$ for all $i=1,\dots,n$ and $\ell=1,\dots,d$.
Conditionally on the parameters  specifying the marginals, $\by_{\bx,1:d,1:n}$ are thus known quantities. 
We plug-in these values and use that $(\by_{\bx,i}\mid\bR_{\bx}) \sim \MVN_{d}(\bzero,\bR_{\bx})$ to update $\bR_{\bx}$.  
The full conditionals of the parameters specifying $\bR_{\bx}$ do not have closed forms. 
We use M-H steps to update these parameters. 
\begin{enumerate}[topsep=2ex,itemsep=0ex,partopsep=2ex,parsep=0ex, leftmargin=0cm, rightmargin=0cm, wide=3ex]
\item
For $t=1,\dots,(d-1)$, we discretized the values of $b_{x,t}$ to the set $\{-0.99+2\times 0.99(m-1)/(M-1)\}$, where $m=1,\dots,M$ and we chose $M=41$. 
A new value $b_{x,t,new}$ is proposed at random from the set comprising the current value of $b_{x,t}$ and its two neighbors. 
The proposed value is accepted with probability $\min\{1,a(b_{x,t,new})/a(b_{x,t})\}$, where 
\vspace{-4ex}\\
\bse
\textstyle a(b_{x,t}) = (1-b_{x,t}^{2})^{-n/2} \times \exp\left\{- (1/2)\sum_{i=1}^{n}\sum_{j=1}^{m_{i}}\by_{\bx,i,j}\trans\{\bSigma_{\bx}(b_{x,t},\bzeta)\}^{-1}\by_{\bx,i,j}\right\}. 
\ese
\vspace{-8ex}\\
\item
For $s=1,\dots,(d-1)(d-2)/2$, we discretized the values of $\theta_{x,s}$ to the set $\{-3.14+2\times 3.14 (m-1)/(M-1)\}$, where $m=1,\dots,M$ and $M=41$. 
A new value $\theta_{s,new}$ is proposed at random from the set comprising the current value and its two neighbors. 
The proposed value is accepted with probability $\min\{1,a(\theta_{x,s,new})/a(\theta_{x,s})\}$, where 
\vspace{-4ex}\\
\bse
\textstyle a(\theta_{x,s}) = \exp\left\{- (1/2)\sum_{i=1}^{n}\sum_{j=1}^{m_{i}}\by_{\bx,i,j}\trans\{\bSigma_{\bx}(\theta_{x,s},\bzeta)\}^{-1}\by_{\bx,i,j}\right\}. 
\ese
\vspace{-5ex}
\end{enumerate}
The parameters specifying $\bR_{\bepsilon}$ are updated in a similar fashion.

\end{enumerate}

With carefully chosen initial values and proposal densities for the MH steps, we were able to achieve quick convergence for the MCMC samplers. 
For our proposed method, $5,000$ MCMC iterations were run in each case with the initial $3,000$ iterations discarded as burn-in. 
The remaining samples were further thinned by a thinning interval of $5$. 
We programmed in {R}.
With $n=1000$ subjects and $m_{i}=3$ proxies for each subject, on an ordinary desktop, $5,000$ MCMC iterations required approximately $3$ hours to run.

\section{Simulation Studies} \label{sec: simulation studies}

We focus here on comparisons with our main competitor, the method of \cite{Zhang2011b}. 
Simulation scenarios to perform these comparisons were designed as follows. 

We mimic some aspects of the real data set analyzed in Section \ref{sec: applications} and the design in \cite{sarkar2021bayesian} as closely as possible 
while modifying some others to illustrate the flexibility and efficiency of the proposed method. 
We chose $n=965$, $m_{i}=3$ replicates per subject, and $d = 3$ dimensional $\bx$. 
While our method scales quite well to much higher dimensional problems, 
with $3$ total components the results can be conveniently graphically summarized. 

We maintained the same distribution of the covariates as in the EATS data (Figure \ref{fig: EATS exploratory cat dists}).

To generate the true $x_{\ell,i}$'s for $\ell=1,\dots,d$, we (a) first sampled $\bx_{i}^{\triangle} \sim \MVN_{d}(\bzero,\bR_{\bx})$, 
(b) then, set $\bx_{i}^{\triangle\triangle}=\Phi(\bx_{i}^{\triangle})$, (c) finally, set $x_{\ell,i} = F_{TN,mix}^{-1}(x_{\ell,i}^{\triangle\triangle} \mid \bpi_{x,\ell},\bmu_{x,\ell},\bsigma_{x,\ell}^{2}, x_{\ell,L},x_{\ell,U})$, 
where $F_{TN,mix}(x \mid \bpi,\bmu,\bsigma^{2},x_{L},x_{U}) = \sum_{k=1}^{k}\pi_{k}F_{TN}(x \mid \mu_{k},\sigma_{k}^{2},x_{L},x_{U})$. 
The marginal distributions are thus mixtures of truncated normal distributions and hence can take widely varying shapes (Figure \ref{fig: Sim Study MARGINALS}) 
while the correlation between different components is $\bR_{\bx}$. 
We set 
\vspace{-4ex}\\
\bse
%k_{\bx}=3,~
& 
\hskip -0.5cm ~\bR_{\bx} = \left(\begin{array}{c c c}
1 & 0.7 & 0.7^2 \\
 & 1 & 0.7 \\
 &  & 1 
\end{array} \right),
~\bmu_{\bx} =  \left(\begin{array}{c}
\bmu_{x,1}^{\trans} \\
\bmu_{x,2}^{\trans} \\
\bmu_{x,3}^{\trans}
\end{array} \right) 
= 
\left(\begin{array}{c c c c}
1.5 & 1.5 & 3.0 & 5.0 \\
2.0 & 2.0 & 4.0 & 5.0 \\
2.0 & 3.0 & 4.0 & 5.0 \\
\end{array} \right),\\
& x_{\ell,L}=0,~x_{\ell,U}=10 ~\text{for all}~\ell, 
~\text{and}~
~\sigma_{x,\ell,k}^{2}=0.75^{2}~\text{for all}~\ell,k. 
\ese
\vspace{-5ex}\\
We assume, as we have seen in the case of the real data application, that only $c_{1}=$ gender ($c_{1} = 1 \equiv$ man, $c_{1}=2 \equiv$ woman) is an important predictor for $f_{x,\ell \mid \bc}$. 
We set
\vspace{-4ex}\\
\bse
\left(\begin{array}{c}
\Pr(1 \mid c_{0},1,c_{2},\dots,c_{p}) \\
\Pr(2 \mid c_{0},1,c_{2},\dots,c_{p}) \\
\Pr(3 \mid c_{0},1,c_{2},\dots,c_{p}) \\
\Pr(4 \mid c_{0},1,c_{2},\dots,c_{p})
\end{array}\right)
=  \left(\begin{array}{c}
0.10 \\
0.40 \\
0.20 \\
0.30 
\end{array}\right) ~\text{and} 
~\left(\begin{array}{c}
\Pr(1 \mid c_{0},2,c_{2},\dots,c_{p}) \\
\Pr(2 \mid c_{0},2,c_{2},\dots,c_{p}) \\
\Pr(3 \mid c_{0},2,c_{2},\dots,c_{p}) \\
\Pr(4 \mid c_{0},2,c_{2},\dots,c_{p})
\end{array}\right)
=  \left(\begin{array}{c}
0.40 \\
0.40 \\
0.00 \\
0.20 
\end{array}\right)
\ese
%\vspace{-4ex}\\
for all $(c_{0},c_{2},\dots,c_{p})$. 
The densities $f_{x,\ell \mid \bc}$'s also vary between different dimensions $c_{0}=\ell$.  
This becomes clearer from a rearrangement of the distinct values of the components of $\bmu_{\bx}$ and associated mixture probabilities 
for different combinations of $(c_{0},c_{1})$ in Table \ref{tab: sim design}.  
\vspace{0.25cm}
\begin{table}[!ht]
\begin{center}
\footnotesize
\begin{tabular}{|c|c|c c c c c|}
\hline
\multirow{3}{20pt}{Dim}		& \multirow{3}{20pt}{Sex}	& \multicolumn{5}{|c|}{$\mu_{x,k}$} 	\\ \cline{3-7}
						&	& 1.5 & 2.0 & 3.0 & 4.0 & 5.0 \\ \cline{3-7}
						&	& \multicolumn{5}{|c|}{Associated Probabilities} \\ \hline \hline
\multirow{2}{20pt}{~~1}		& M	& 0.50 & 0.00 & 0.20 & 0.00 & 0.30	\\
						& W	& 0.80 & 0.00 & 0.00 & 0.00 & 0.20	\\\cline{1-7}
\multirow{2}{20pt}{~~2}		& M	& 0.00 & 0.50 & 0.00 & 0.20 & 0.30	\\
						& W	& 0.00 & 0.80 & 0.00 & 0.00 & 0.20	\\\cline{1-7}
\multirow{2}{20pt}{~~3}		& M	& 0.00 & 0.10 & 0.40 & 0.20 & 0.30 	\\
						& W	& 0.00 & 0.40 & 0.40 & 0.00 & 0.20	\\\cline{1-7}
\hline
\end{tabular}
\caption{\baselineskip=10pt 
Distinct values of the components of $\bmu_{\bx}$ and associated mixture probabilities 
for different combinations of $(c_{0},c_{1})=\text{(dimension, sex)}$ used in the simulation design. 
}
\label{tab: sim design}
\end{center}
\end{table}
\vspace{-10pt}

\begin{figure}[!ht]
\centering
\includegraphics[width=7cm, trim=0cm 0cm 0cm 0cm, clip=true]{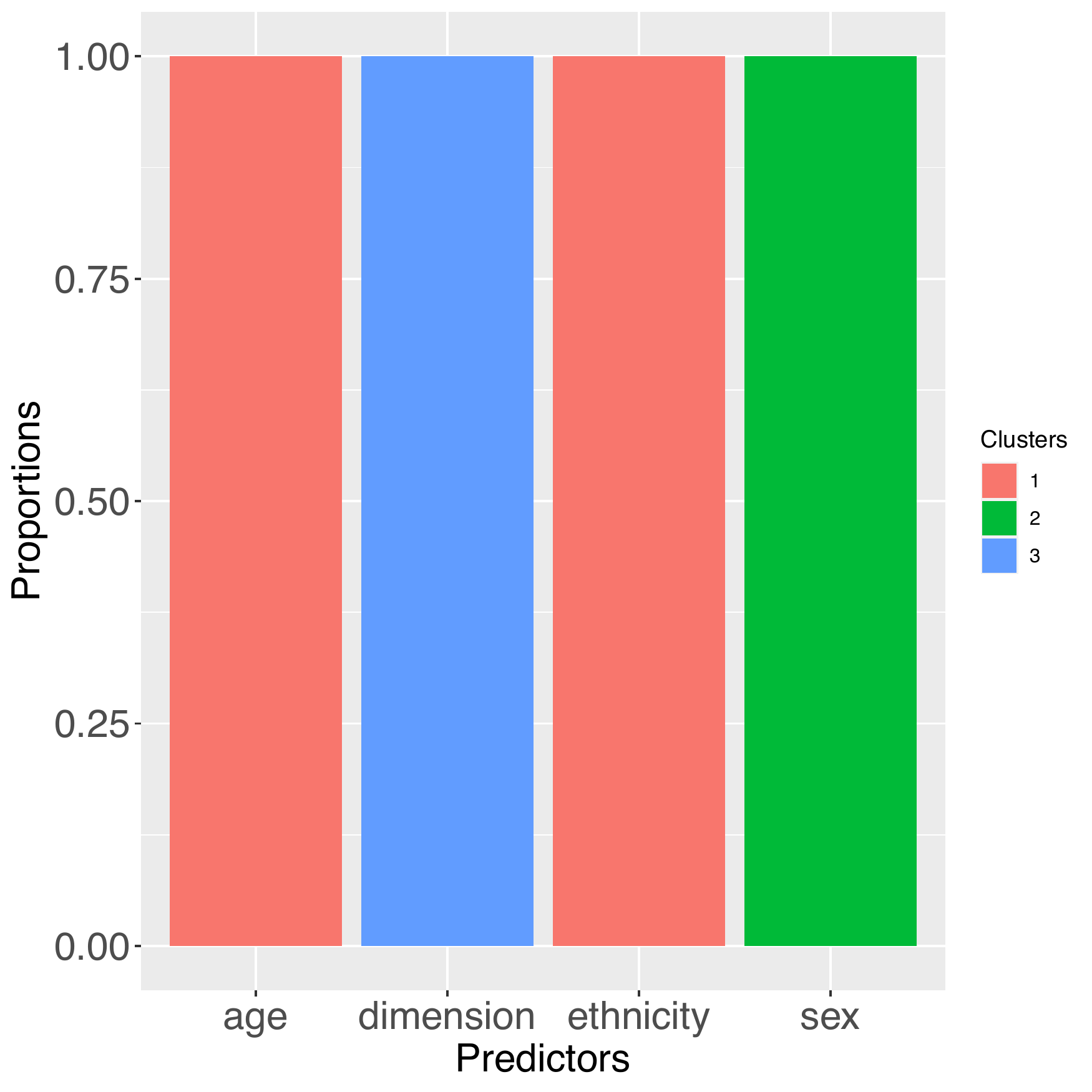}
\quad\quad\quad
\includegraphics[width=7cm, trim=0cm 0cm 0cm 0cm, clip=true]{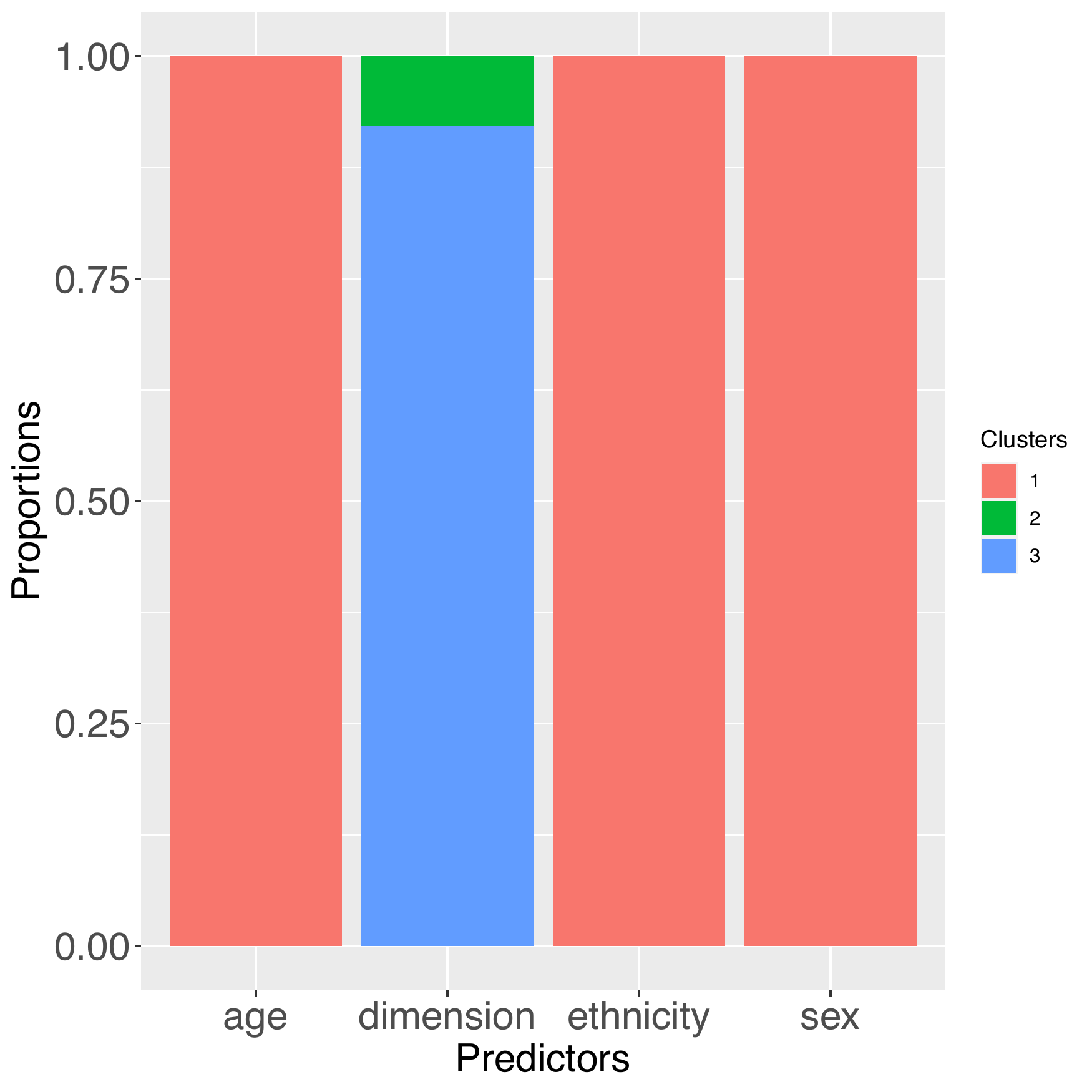}
\caption{\baselineskip=10pt 
Results for the synthetic data set corresponding to the $25\th$ percentile of the average ISEs, 
showing the estimated probabilities of different numbers of clusters of the associated predictors' levels being included in the model. 
The left panel shows the results for modeling the densities $f_{\bx \mid \bc}$.  
The right panel shows the results for modeling the densities $f_{\bepsilon \mid \bc}$. 
At the median $0.5$ probability level, in the left panel, the component labels and a binary covariate proxying for sex of the subjects are important predictors for modeling the densities $f_{\bx \mid \bc}$, whereas in the right panel, only the component labels are important for modeling the densities $f_{\bepsilon \mid \bc}$. 
The results are consistent with the true simulation scenario.
}
\label{fig: Sim Study Inclusion Probabilities}
\end{figure}

We used a similar procedure to simulate the true scaled errors $\epsilon_{\ell,i,j}$'s, $\ell=1,\dots,d$. 
Following \cite{sarkar2021bayesian}, we (a) first sampled $\bepsilon_{i,j}^{\triangle} \sim \MVN_{d}(\bzero,\bR_{\bepsilon})$, 
(b) then, set $\bepsilon_{i,j}^{\triangle\triangle}=\Phi(\bepsilon_{i}^{\triangle})$, (c) finally, set $\epsilon_{\ell,i,j} = F_{\epsilon,\ell,mix,scaled}^{-1}(\epsilon_{\ell,i,j}^{\triangle\triangle} \mid \bpi_{\epsilon,\ell}, \btheta_{\epsilon,\ell})$. 
Here for $\ell=1,2$, $F_{\epsilon,\ell,mix,scaled}$ is a scaled version of 
$F_{\epsilon,\ell,mix}(\epsilon \mid \bpi_{\epsilon,\ell}, \btheta_{\epsilon,\ell}) = \sum_{k=1}^{k_{\epsilon,\ell}}\pi_{\epsilon,\ell,k}F_{c\epsilon}(\epsilon \mid p_{\epsilon,\ell,k},\mu_{\epsilon,\ell,k},\sigma_{\epsilon,\ell,k,1}^{2},\sigma_{\epsilon,\ell,k,2}^{2})$. 
with $\btheta_{\epsilon,\ell}=\{(p_{\epsilon,\ell,k},\mu_{\epsilon,\ell,k},\sigma_{\epsilon,\ell,k,1}^{2},\sigma_{\epsilon,\ell,k,2}^{2})\}_{k=1}^{k_{\epsilon,\ell}}$. 
And, for $\ell=3$, $F_{\epsilon,\ell,mix,scaled}$ is a scaled version of $F_{\epsilon,\ell,mix}(\epsilon \mid \bpi_{\epsilon,\ell}, \btheta_{\epsilon,\ell}) = \sum_{k=1}^{k_{\epsilon,\ell}}\pi_{\epsilon,\ell,k}F_{\tiny{\Laplace}}(\epsilon \mid m_{\epsilon,\ell,k}, b_{\epsilon,\ell,k})$ 
with $\btheta_{\epsilon,\ell}=\{(m_{\epsilon,\ell,k},b_{\epsilon,\ell,k})\}_{k=1}^{k_{\epsilon,\ell}}$, 
adjusted to have mean zero and variance $1$. 
This way, the marginal distributions can take widely varying shapes (Figure \ref{fig: Sim Study MARGINALS}) 
while the marginal correlation between different components is $\bR_{\bepsilon}$. 
In this case, we set
\vspace{-4ex}\\
\bse
& 
~\bR_{\bepsilon} = \left(\begin{array}{c c c}
1 & 0.5 & 0.5^2 \\
 & 1 & 0.5 \\
 &  & 1 
\end{array} \right),
~\bpi_{\epsilon,\ell} =  \left(\begin{array}{c}
0.25 \\
0.50 \\
0.25 
\end{array}\right) ~\text{for all}~\ell,\\ 
&~\btheta_{\bepsilon} =  \left(\begin{array}{c}
\btheta_{\epsilon,1}^{\trans} \\
\btheta_{\epsilon,2}^{\trans} \\
\btheta_{\epsilon,3}^{\trans}
\end{array} \right) 
= 
\left(\begin{array}{c c c}
(0.4,2,2,1) & (0.4,2,2,1) & (0.4,2,2,1) \\
(0.5,0,0.25,0.25) & (0.5,0,0.25,0.25) & (0.5,0,5,5) \\
(0,1) & (0,1) & (0,1) 
\end{array} \right).
\ese
The representations with $k_{\epsilon,\ell}=3$ components above are more than what are really needed to describe the particular assumed truths -- 
we are effectively using a single component mixture of two-component scaled normals for $f_{\epsilon,1}$ producing a bimodal error distribution, 
a two component $(0.75,0.25)$ mixture of two-component scaled normals for $f_{\epsilon,2}$ producing a unimodal but heavier tailed error distribution, and finally 
a single component scaled Laplace for $f_{\epsilon,3}$ producing a unimodal error distribution with a spike at zero (Figure \ref{fig: Sim Study MARGINALS}).
It is clear, however, that such 3-component models are capable of generating a very wide variety of shapes, including multimodality heavy-tails etc., for the error distributions. 
Unlike the real data application, we thus chose the marginal distributions of the measurement errors associated with the three components to all be different.

Finally, we set $w_{\ell,i,j}=x_{\ell,i}+u_{\ell,i,j}$ with $u_{\ell,i,j}=s_{\ell}( x_{\ell,i})\epsilon_{\ell,i,j}$ and $s_{\ell}( x_{\ell}) =  x_{\ell}/3$ for each $\ell$.

\begin{figure}[!ht]
\centering
\includegraphics[width=11cm, trim=0cm 0.25cm 0cm 0cm, clip=true]{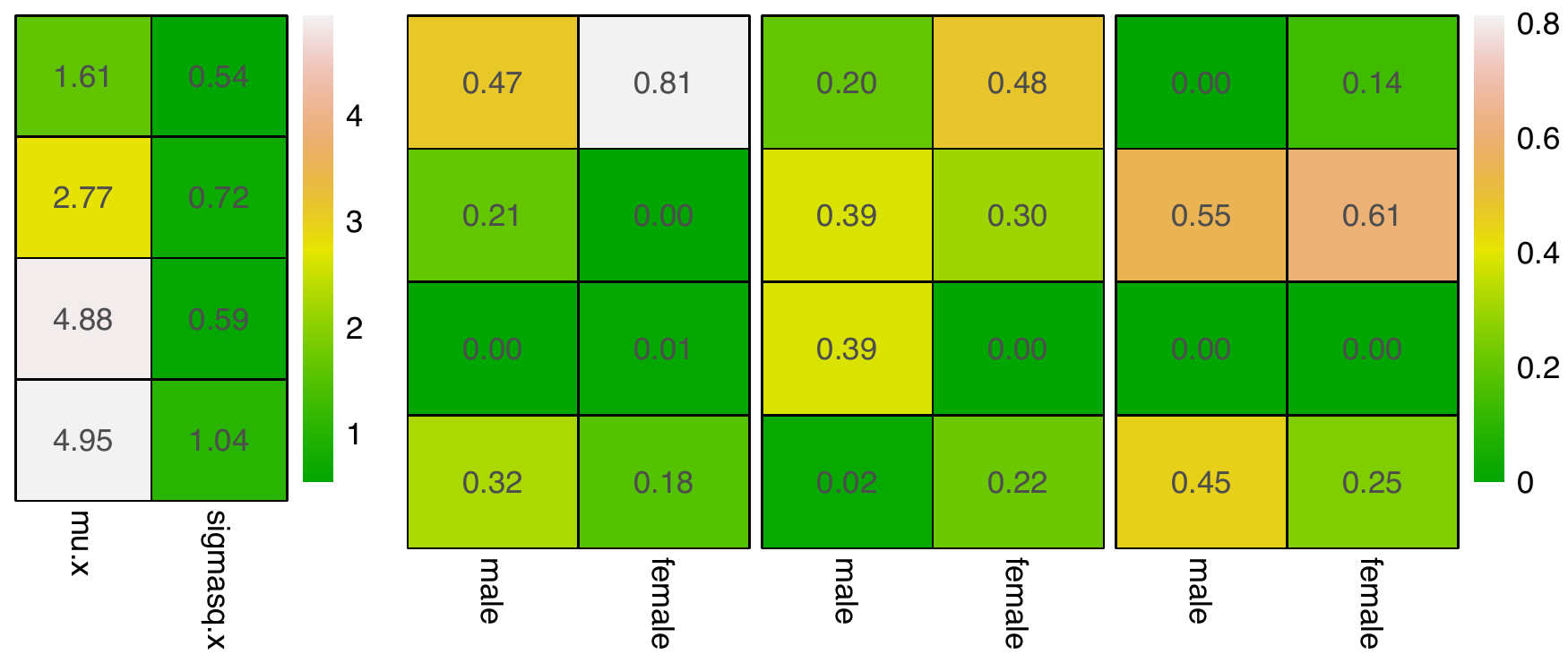}
\caption{\baselineskip=10pt 
Results for the synthetic data set corresponding to the $25\th$ percentile of the average ISEs. 
These results correspond to the final MCMC iteration but are representative of other iterations in steady state. 
The left panel shows the component specific parameters $(\mu_{x,k}, \sigma_{x,k}^{2})$ 
for the six mixture components that were actually used to model the densities $f_{x,\ell \mid \bc}(x_{\ell} \mid \bc)$.  
The right panel shows the associated `empirical' mixture probabilities $\wh{p}_{x}(k \mid c_{0},c_{1},\dots,c_{p}) = \sum_{i=1}^{n}1\{z_{x,\ell,i}=k, c_{0,\ell,i}=c_{0},c_{1,\ell,i}=c_{1},\dots,c_{p,\ell,i}=c_{p}\}/n$ 
for `men' and `women' 
and for the three components, from left to right. 
Results for different combinations of component and `gender' are shown here 
as they are the only predictors important for $\bx$. 
Consistent with the simulation truth, the mixture probabilities vary significantly between these predictor combinations. 
}
\label{fig: Sim Study ZX Tables}
\end{figure}

\begin{figure}[!ht]
\centering
\includegraphics[width=12.5cm, trim=0cm 0.25cm 0cm 0cm, clip=true]{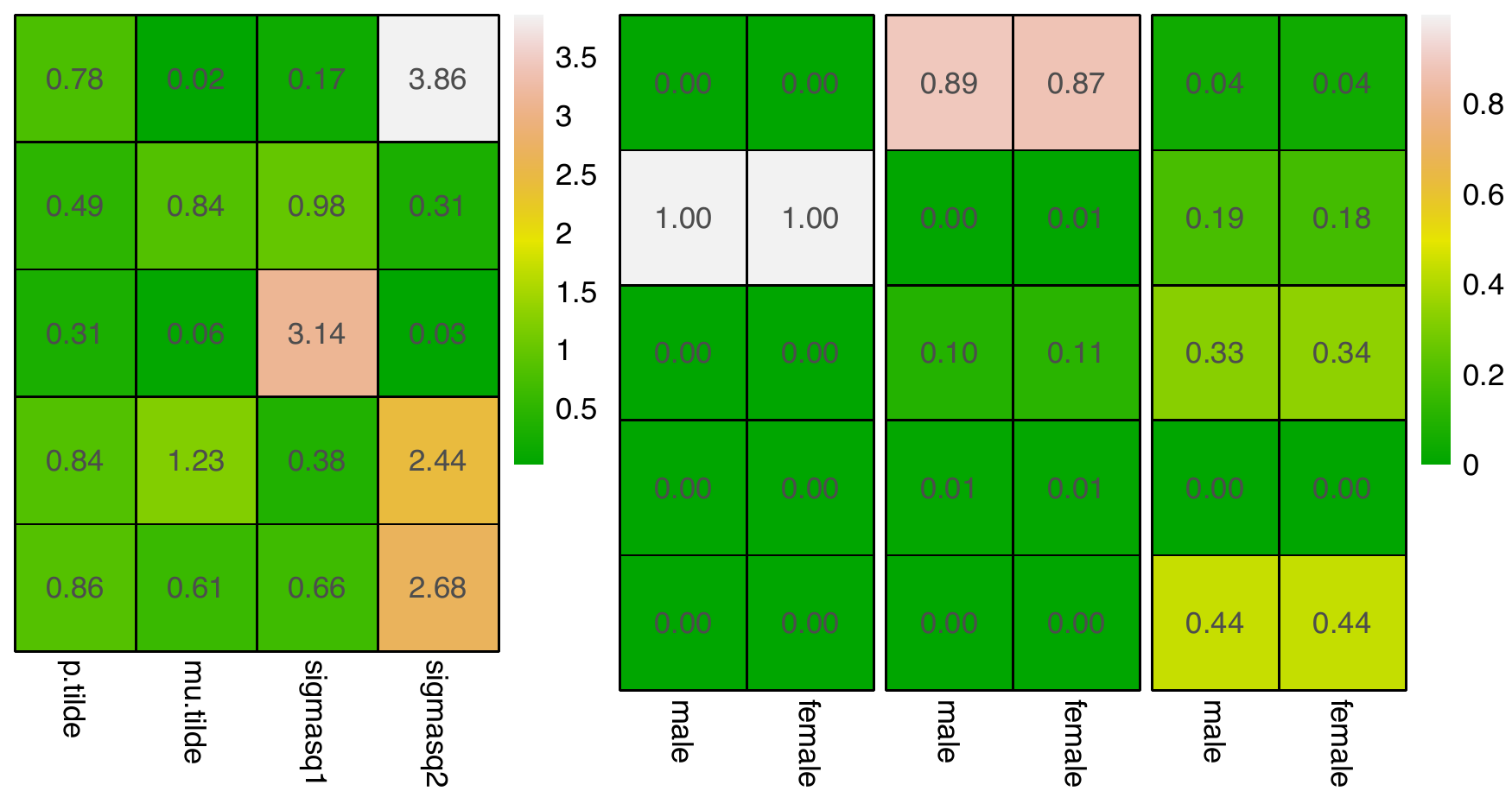}
\caption{\baselineskip=10pt 
Results for the synthetic data set corresponding to the $25\th$ percentile of the average ISEs. 
These results correspond to the final MCMC iteration but are representative of other iterations in steady state. 
The left panel shows the component specific parameters $(p_{\epsilon,k}, \mu_{\epsilon,k}, \sigma_{\epsilon,k,1}^{2}, \sigma_{\epsilon,k,2}^{2})$ 
for the three mixture components used to model the densities $f_{\epsilon,\ell \mid \bc}(\epsilon_{\ell} \mid \bc)$.  
The right panel shows the associated `empirical' mixture probabilities $\wh{p}_{\epsilon}(k \mid c_{0},c_{1},\dots,c_{p}) = \sum_{i=1}^{n}\sum_{j=1}^{m_{i}}1\{z_{\epsilon,\ell,i,j}=k, c_{0,\ell,i}=c_{0},c_{1,\ell,i}=c_{1},\dots,c_{p,\ell,i}=c_{p}\}/\sum_{i=1}^{n}m_{i}$ 
for `men' and `women' 
and for the three components, from left to right. 
Results for different combinations of component and `gender' are shown here. 
Consistent with the simulation truth, the mixture probabilities vary significantly between different components but not between the gender categories. 
}
\label{fig: Sim Study ZE Tables}
\end{figure}

\begin{figure}[h!]
\centering
\includegraphics[height=16.5cm, trim=0cm 0cm 0cm 0cm, clip=true]{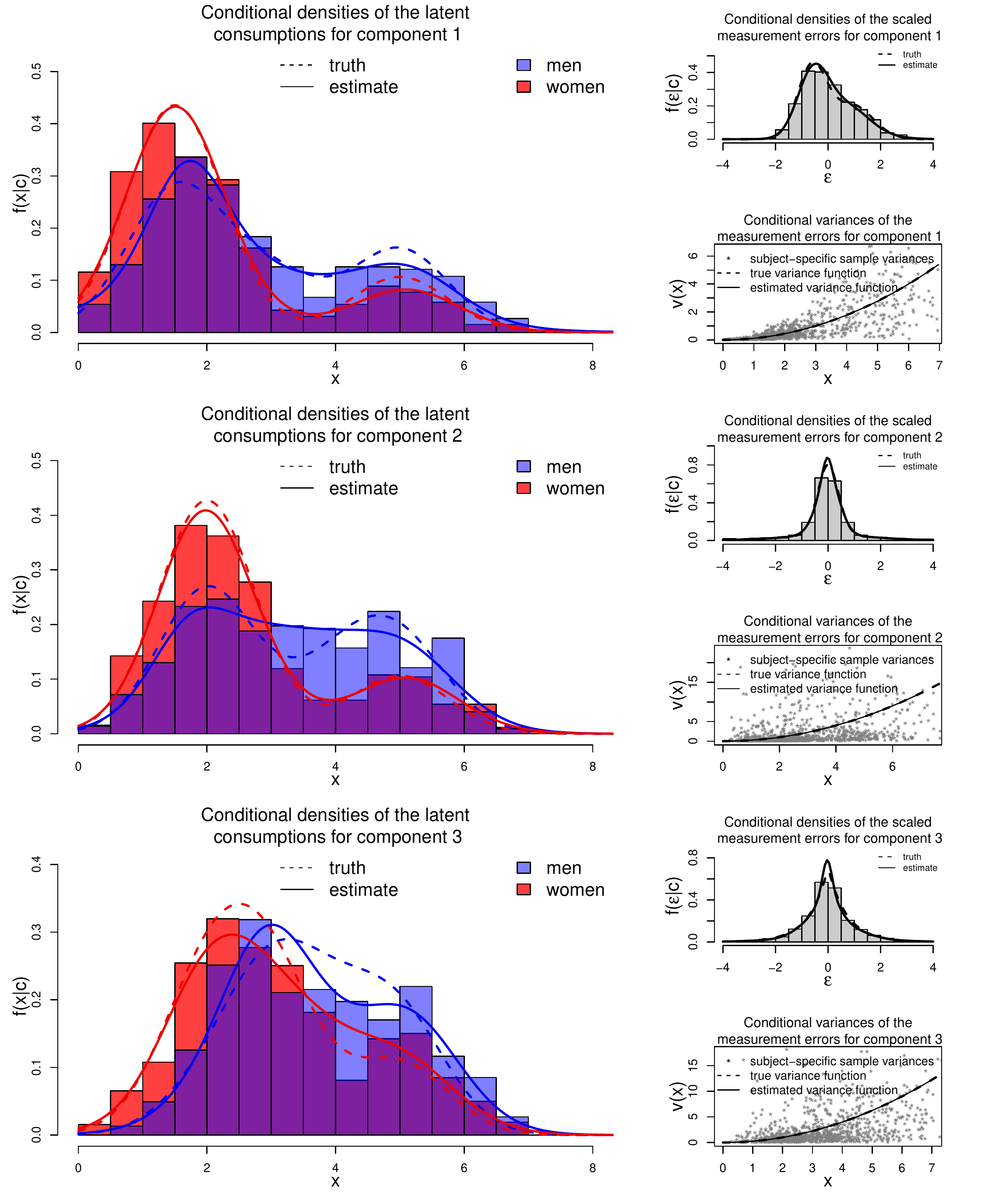}
\caption{\baselineskip=10pt 
Results for the synthetic data set corresponding to the $25\th$ percentile of the average ISEs. 
From top to bottom, the left panels show the estimated conditional densities $f_{x,\ell \mid \bc}(x_{\ell} \mid \bc)$ 
obtained by our method %(solid lines) 
and the corresponding truths. %(dashed lines).  
The right panels show the associated conditional distributions of the scaled errors $f_{\epsilon,\ell \mid \bc}(\epsilon_{\ell} \mid \bc)$ 
obtained by our method %(solid lines) 
and the corresponding truths. %(dashed lines).  
Results for different component and `gender' combinations are shown here as 
they are the only predictors important for modeling the densities $f_{x,\ell \mid \bc}(x_{\ell} \mid \bc)$. 
%Results for men are shown in blue, for women are shown in red. 
Also shown are the associated variance functions $v_{\ell}(x_{\ell}) = s_{\ell}^{2}(x_{\ell})$, estimated by our method %(solid lines) 
and the corresponding truths.% (dashed lines). 
}
\label{fig: Sim Study MARGINALS}
\end{figure}

\vspace{0.25cm}
\begin{table}[!ht]
\begin{center}
\footnotesize
\begin{tabular}{|c|c|c|c|}
\hline
\multirow{2}{50pt}{Component}	& \multirow{2}{20pt}{Sex}		& \multicolumn{2}{|c|}{Median ISE $\times 1000$} 	\\ \cline{3-4}
						& &{Zhang, et al. (2011)} 			& Our Proposed Method	\\ \hline \hline
\multirow{2}{20pt}{1} 	& M	& 6.08 		& \bf{2.99} 	\\
				& W	& 4.69		& \bf{2.75} 	\\\cline{1-4}
\multirow{2}{20pt}{2} 	& M	& 18.66		& \bf{4.92}  	\\
				& W	& 15.61		& \bf{2.52}	\\\cline{1-4}
\multirow{2}{20pt}{3} 	& M	& 10.06 		& \bf{5.88}  	\\
				& W	& 18.84		& \bf{3.70} 	\\\cline{1-4}
\hline
\end{tabular}
\caption{\baselineskip=10pt 
Median integrated squared error (MISE) performance 
of the covariate informed density deconvolution method developed in this article 
compared with the methods of \cite{Zhang2011b}. 
Here M and W are abbreviations for `men' and `women', respectively.
\vspace{-20pt}
}
\label{tab: MISEs 1}
\end{center}
\end{table}

The integrated squared error (ISE) of estimation of $f_{x,\ell}$ by $\wh{f}_{x,\ell}$ is defined as $ISE = \int \{f_{x,\ell}(x_{\ell})-\widehat{f}_{x,\ell}(x_{\ell})\}^{2}dx_{\ell}$.
A Monte Carlo estimate of ISE for the $b\th$ simulated data set is given by
$ISE_{est} = \sum_{m=1}^{M}\{f_{x,\ell}(x_{m})-\widehat{f}_{x,\ell}^{(b)}(x_{m})\}^{2} \Delta_{m}$, 
where $\{x_{m}\}_{m=1}^{M}$ are grid points covering $[A,B]$, the support of each $x_{\ell}$. 
Table \ref{tab: MISEs 1} reports the median ISEs (MISEs) for estimating the trivariate joint densities and the univariate marginals  
obtained by our method, compared with the method of \cite{Zhang2011b}.  
The MISEs reported here are all based on $B=100$ simulated data sets.
{As Table \ref{tab: MISEs 1} shows, our method outperforms \cite{Zhang2011b} in all cases, often significantly.} 
Our method also produces detailed information about the distributions of the intakes as well as the distributions of the associated measurement errors, 
including specifically which predictors are mostly important in influencing these distributions. 
Additional graphical summaries for the synthetic data set corresponding to the $25\th$ percentile of the average ISEs are provided below.

Figure \ref{fig: Sim Study Inclusion Probabilities} shows the estimated inclusion probabilities of different predictors in the models for $f_{\bx \mid \bc}$ and $f_{\bepsilon \mid \bc}$. 
Consistent with the simulation truth, 
the set of significant predictors for the density of main interest $f_{\bx \mid \bc}$ is found to comprise the dimension labels ($c_{0}$) and `gender' ($c_{1}$), 
and, for $f_{\bepsilon \mid \bc}$, the set of significant predictors comprises only the dimension labels ($c_{0}$).

Figures \ref{fig: Sim Study ZX Tables} and \ref{fig: Sim Study ZE Tables} show how the redundant mixture components become empty after reaching steady states of the MCMC algorithm. 
They also show how the mixture component specific parameters get shared across different components and predictor combinations. 
For the densities $f_{x,\ell \mid \bc}$, the mixture probabilities vary significantly between `men' and `women' as well as between different dietary components. 
For the densities $f_{\epsilon,\ell \mid \bc}$, on the other hand, 
the mixture probabilities vary between different dietary components but are very similar between `men' and `women'. 
These results are all consistent with the simulation truth.

Figure \ref{fig: Sim Study MARGINALS} shows the estimated densities $f_{x,\ell \mid \bc}(x_{\ell} \mid \bc)$ 
superimposed over histograms of the corresponding estimated $x_{\ell}$'s obtained by our method. 
Figure \ref{fig: Sim Study JOINTS M} and \ref{fig: Sim Study JOINTS F} repeat the univariate densities $f_{x,\ell \mid \bc}(x_{\ell} \mid \bc)$ in the diagonal panels 
but also show the estimated joint densities $f_{\bx,\ell_{1},\ell_{2} \mid \bc}(x_{\ell_{1}}, x_{\ell_{2}} \mid \bc)$ obtained by our method in the off-diagonal panels separately for men and women, respectively. 
The results suggest the model to provide a good fit for the simulated data, 
including being able to capture skewness, multimodality and heavy tails. 

Results obtained by the method of \cite{Zhang2011b} in numerical experiments are discussed in Section \ref{sec: additional simulation results} of the supplementary material. 
The results summarized in the supplementary materials are for the synthetic data set corresponding to the $25\th$ percentile of the average ISEs across all predictor combinations. 
The fixed effects regression coefficient estimates in the transformed scale for the data set are presented in 
Table \ref{tab: Sim Study Zhang_2011b Results} in the supplementary material. 
With no mechanism to select the important predictors, all covariates are included in the model. 
Interestingly, however, a $90\%$ central credible interval based post-processing rule to 
determine the significance of the associated predictor still produces 
a good number of spurious significant coefficients 
- which we believe is an artifact of the bias introduced in the analysis due to non-linear transformations, strong parametric assumptions, exclusion of interaction effects, etc.
Figure \ref{fig: Sim Study MARGINALS Zhang_2011b} in the supplementary materials shows the estimated univariate densities which, 
as in the case of the real data application, are in general agreement with the estimates produced by our method but are much worse compared to ours in terms of finer details. 
To reiterate, (a) the main reason the highly restrictive parametric assumptions of \cite{Zhang2011b} 
do not get reflected as much in the final density estimates is 
because these estimates are obtained by a separate kernel density estimation approach applied to the estimated intakes in the original scale;
(b) additionally, the approach of \cite{Zhang2011b} does not provide any insight into the distribution of the measurement errors, including their shapes, how they may be influenced by the associated predictors, etc.

\newpage
\section{Additional Results for the EATS Data Set}

\begin{figure}[!ht]
\centering
\includegraphics[height=16.5cm, width=16cm, trim=0cm 0cm 0cm 0.5cm, clip=true]{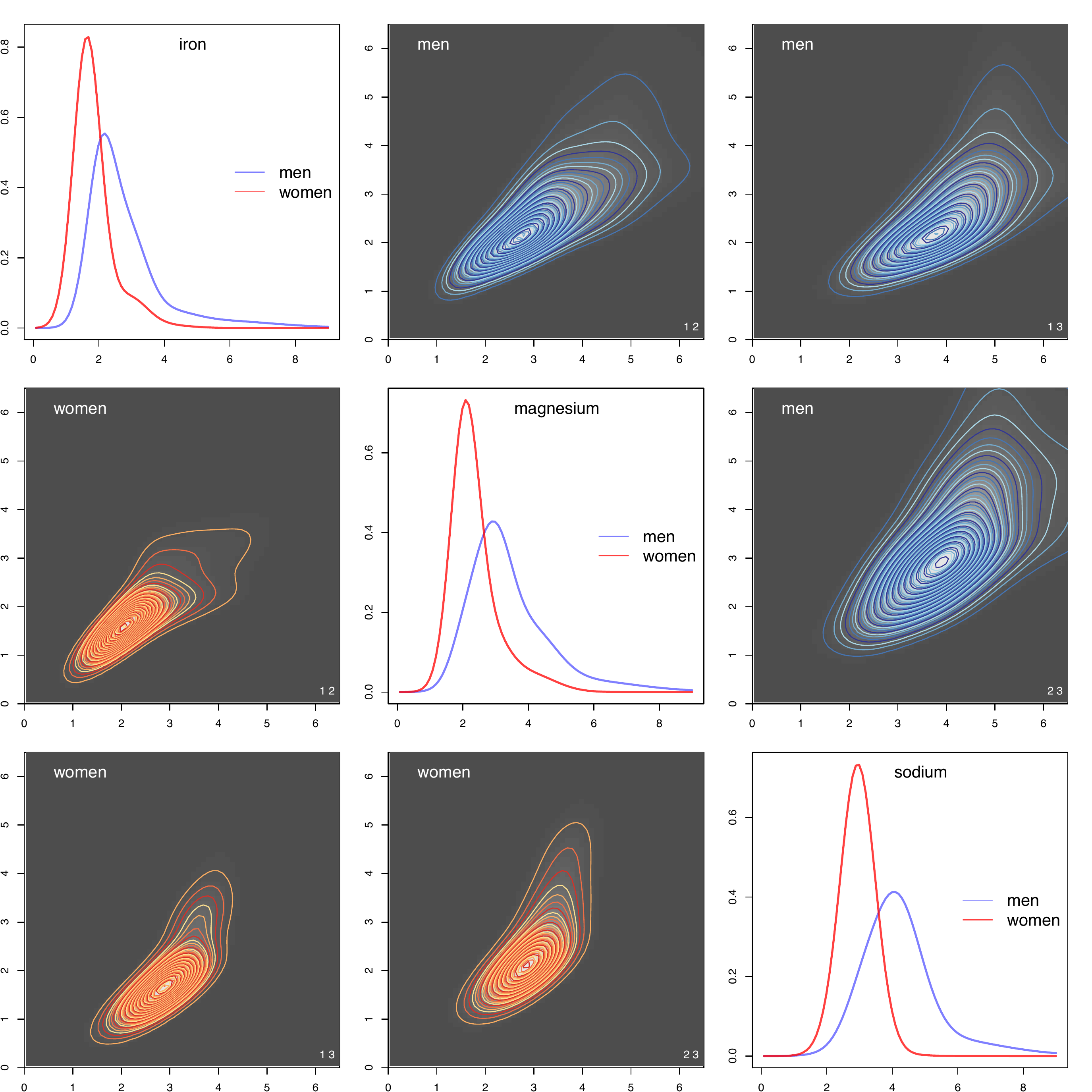}
\caption{\baselineskip=10pt 
Results for the EATS data set obtained by our method. 
Results for different dietary component and gender combinations are shown here as 
they are the only predictors found important for modeling the densities $f_{x,\ell \mid \bc}(x_{\ell} \mid \bc)$. 
The diagonal panels show the estimated conditional densities $f_{x,\ell \mid \bc}(x_{\ell} \mid \bc)$ 
of iron, magnesium and sodium, respectively. 
The off-diagonal panels show the estimated joint densities $f_{\bx,\ell_{1},\ell_{2}\mid \bc}(x_{\ell_{1}}, x_{\ell_{2}} \mid \bc)$.
Results for men are shown in blue, for women are shown in red. 
Axis labels are suppressed to allow more space for the individual panels.
}
\label{fig: EATS JOINTS}
\end{figure}

\begin{figure}[!ht]
\centering
\includegraphics[height=16.5cm, width=10cm, trim=0cm 0cm 0cm 0cm, clip=true]{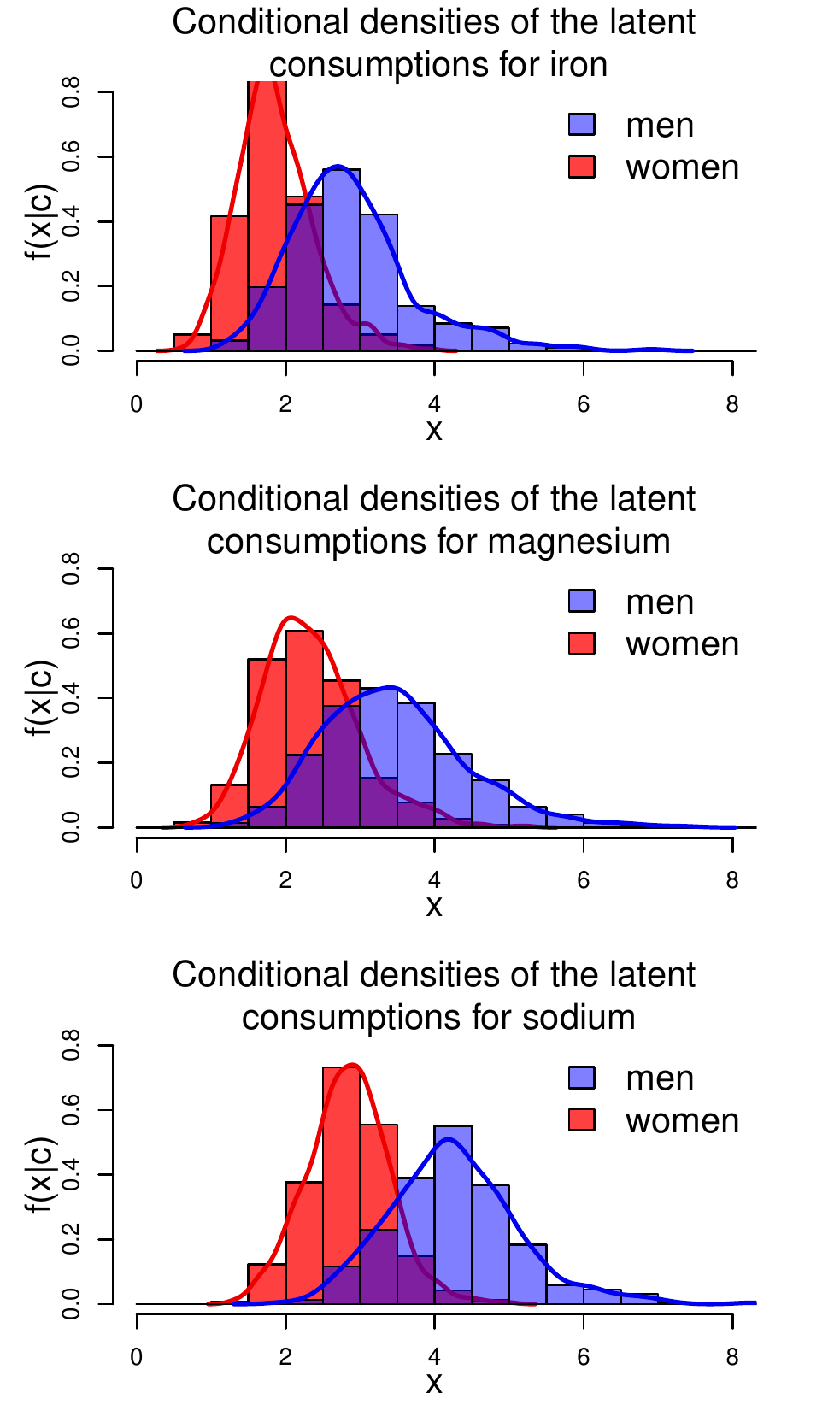}
\caption{\baselineskip=10pt 
Results for the EATS data set produced by the method of \cite{Zhang2011b}. 
From top to bottom, the left panels show the estimated densities of iron, magnesium and sodium, respectively. 
Results for men are shown in blue, for women are shown in red. 
}
\label{fig: EATS MARGINALS Zhang_2011b}
\end{figure}

\vspace{0.25cm}
\begin{table}[!ht]
\begin{center}
\footnotesize
\begin{tabular}{|c|c|c|c|}
\hline
						& Iron 				& Magnesium 			& Sodium 		\\ \hline \hline
\multirow{2}{40pt}{Intercept}	& \bf{-1.6949}			& \bf{-2.0106}			& \bf{-2.1528} \\
						& (-2.4356,   -0.9409)	& (-2.8563,   -1.1225)	& (-2.8852,   -1.4361) \\
\multirow{2}{40pt}{Woman}	& \bf{1.0028}   			& \bf{1.0608}    			& \bf{1.1138} \\
						& (0.9180,    1.0966)		& (0.9516,    1.1653)  	& (1.0292,    1.2034) \\
\multirow{2}{40pt}{Race 1}		& \bf{0.9754}  			& \bf{1.1526}    			& \bf{0.8766}   \\
						& (0.3134,    1.6014)		& (0.3603,    1.9438)  	& (0.2316,    1.5447) \\
\multirow{2}{40pt}{Race 2}		& 0.6977  				& 0.6774    			& \bf{0.8989} \\
						& (-0.0177,    1.3439)	& (-0.1347,    1.5296)  	& (0.2633,    1.5974) \\
\multirow{2}{40pt}{Race 3}		& \bf{0.8953}  			& \bf{1.1218}    			& \bf{0.7712} 	 \\
						& (0.2107,    1.5648)		& (0.3038,   1.9787)   	& (0.0820,    1.4438) \\
\multirow{2}{40pt}{Race 4}		& \bf{0.9703}  			& \bf{1.1668}    			& \bf{0.7059} 	 \\
						& (0.2442,   1.6625)		& (0.3320,    1.9847)  	& (0.0196,   1.3802) \\
\multirow{2}{40pt}{Age 1}		& 0.4554  				& 0.4151   			& \bf{1.0209} \\
						& (-0.0302,    0.9439)	& (-0.0810,    0.9439)  	& (0.5497,    1.4854) \\
\multirow{2}{40pt}{Age 2}		& 0.3299  				& 0.4570    			& \bf{0.8306}	 \\
						& (-0.1165,    0.8084)	& (-0.0109,    0.9666)  	& (0.3454,    1.2670) \\
\multirow{2}{40pt}{Age 3}		& 0.2595  				& 0.3936   			& \bf{0.7061} \\
						& (-0.2279,    0.7423)	& (-0.0954,    0.8919)  	& (0.2275,    1.1361) \\
\multirow{2}{40pt}{Age 4}		& 0.1847  				&  0.3623   			& \bf{0.6949}	 \\
						& (-0.2870,    0.6661)	& (-0.1358,    0.8610)  	& (0.2022,    1.1429) \\
\multirow{2}{40pt}{Age 5}		& 0.1954  				& 0.4332    			& \bf{0.6903}	 \\
						& (-0.3022,    0.7023)	& (-0.0778,    0.9654)  	& (0.1878,   1.1354) \\
\cline{1-4}
\hline
\end{tabular}
\caption{\baselineskip=10pt 
Results for the EATS data set produced by the method of \cite{Zhang2011b}. 
Posterior means of the fixed effects regression effects and associated $90\%$ credible intervals. 
The estimates with credible intervals not including zero are highlighted in bold.  
}
\label{tab: EATS Zhang_2011b Results}
\end{center}
\end{table}
\vspace{-10pt}

\clearpage
\newpage
\section{Additional Results for the Simulation Experiments} \label{sec: additional simulation results}

\begin{figure}[!ht]
\centering
\includegraphics[height=16.5cm, width=16cm, trim=0cm 0cm 0cm 0.5cm, clip=true]{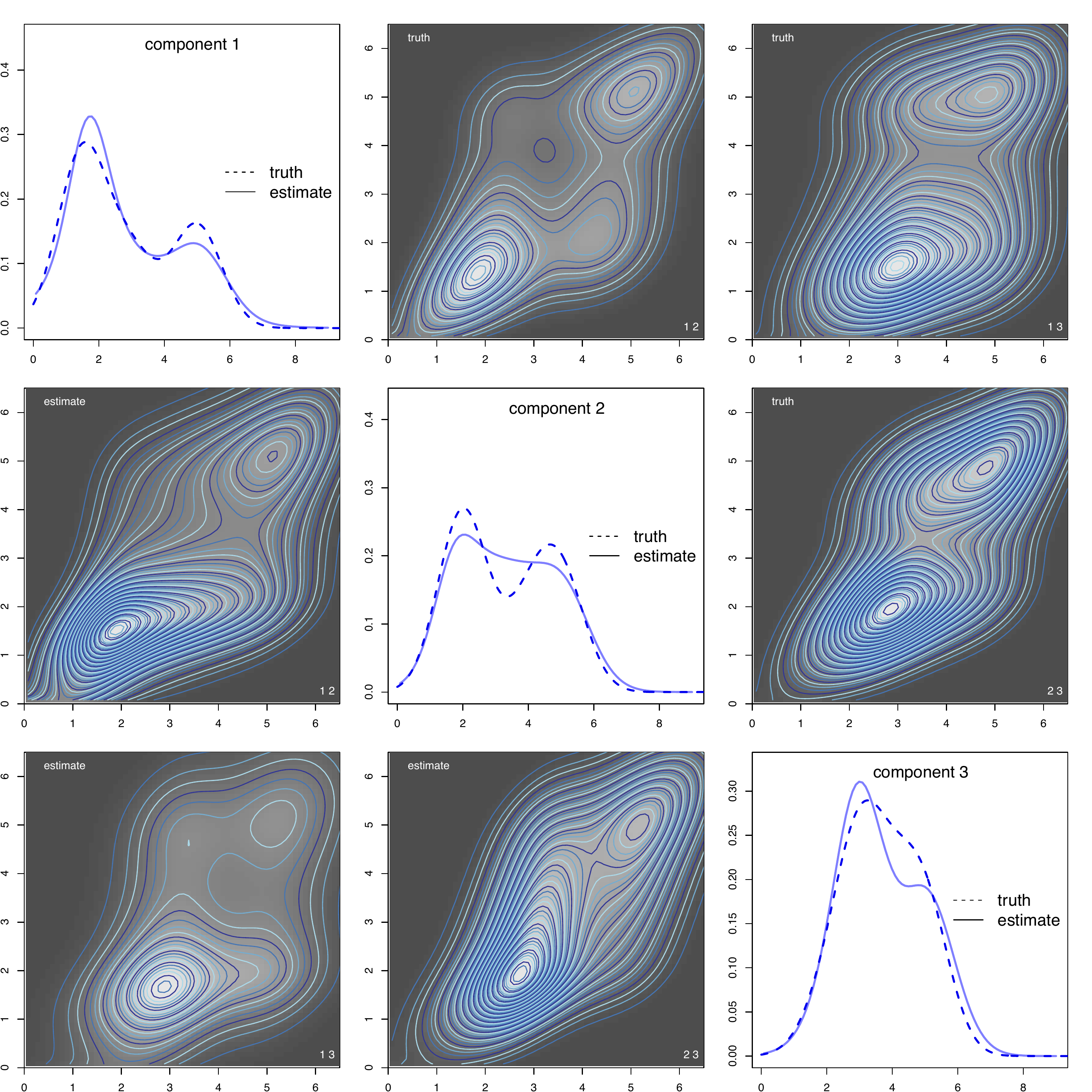}
\caption{\baselineskip=10pt 
Results for the synthetic data set corresponding to the 25th percentile of the average ISEs. 
Results for `men' are shown here. 
The diagonal panels show %the true (dashed lines) 
one dimensional marginal densities and the corresponding estimates %(solid lines) 
produced by our method. 
The off-diagonal panels show the contour plots of the true two-dimensional densities $f_{\bx,\ell_{1},\ell_{2}\mid \bc}(x_{\ell_{1}}, x_{\ell_{2}} \mid \bc)$ %(upper triangular panels) 
and the corresponding estimates obtained by our method. %(lower triangular panels).
Axis labels are suppressed to allow more space for the individual panels.
}
\label{fig: Sim Study JOINTS M}
\end{figure}

\begin{figure}[!ht]
\centering
\includegraphics[height=16.5cm, width=16cm, trim=0cm 0cm 0cm 0.5cm, clip=true]{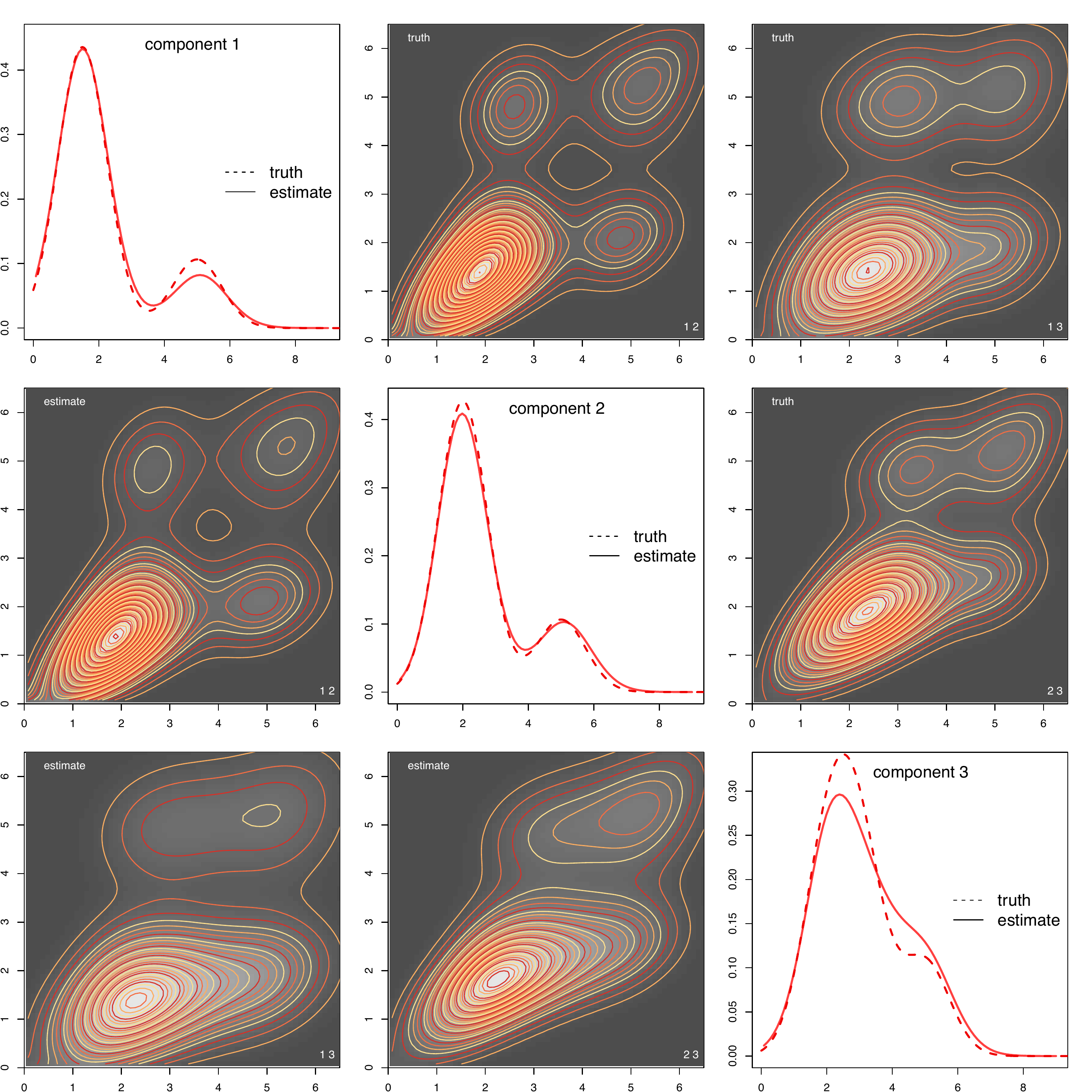}
\caption{\baselineskip=10pt 
Results for the synthetic data set corresponding to the 25th percentile of the average ISEs. 
Results for `women' are shown here. 
The diagonal panels show %the true (dashed lines) 
one dimensional marginal densities and the corresponding estimates %(solid lines) 
produced by our method. 
The off-diagonal panels show the contour plots of the true two-dimensional densities $f_{\bx,\ell_{1},\ell_{2}\mid \bc}(x_{\ell_{1}}, x_{\ell_{2}} \mid \bc)$ %(upper triangular panels) 
and the corresponding estimates obtained by our method. %(lower triangular panels).
Axis labels are suppressed to allow more space for the individual panels.
}
\label{fig: Sim Study JOINTS F}
\end{figure}

\clearpage
\newpage
\begin{figure}[!ht]
\centering
\includegraphics[height=16.5cm, width=10cm, trim=0cm 0cm 0cm 0.5cm, clip=true]{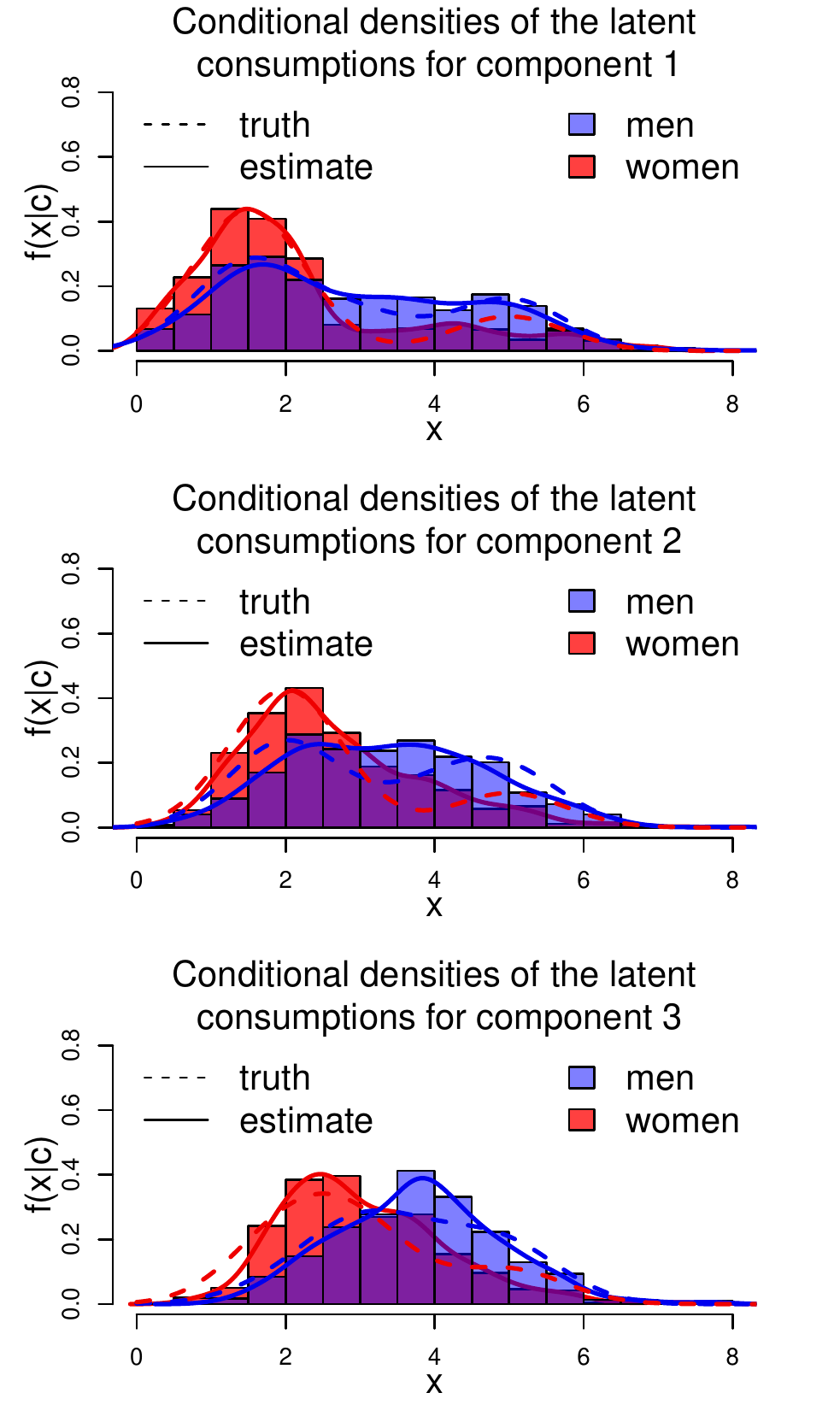}
\caption{\baselineskip=10pt 
Results for the synthetic data set corresponding to the $25\th$ percentile of the average ISEs produced by the method of \cite{Zhang2011b}. 
The panels show the estimated conditional densities $f_{x,\ell \mid \bc}(x_{\ell} \mid \bc)$ 
obtained by the method of \cite{Zhang2011b} %(solid lines) 
and the corresponding truths. %(dashed lines).  
Results for different component and gender combinations are shown here as 
they are the only predictors important for modeling the densities $f_{x,\ell \mid \bc}(x_{\ell} \mid \bc)$. 
}
\label{fig: Sim Study MARGINALS Zhang_2011b}
\end{figure}

\vspace{0.25cm}
\begin{table}[!ht]
\begin{center}
\footnotesize
\begin{tabular}{|c|c|c|c|}
\hline
						& Component 1  		& Component 2 		& Component 3 		\\ \hline \hline
\multirow{2}{40pt}{Intercept}	& -0.1166				& \bf{-0.8776}			& -0.1128 \\
						& (-1.2525,    0.9613)	& (-1.7734,   -0.0620)	& (-0.8528,    0.6691) \\
\multirow{2}{40pt}{Woman}	& \bf{0.6092}   			& \bf{0.5062}    			& \bf{0.5023} \\
						& (0.4852,    0.7252)		& (0.3980,    0.6159)  	& (0.3960,    0.6003) \\
\multirow{2}{40pt}{Race 1}		& {0.2858}  			& {0.6613}    			& {0.1167}   \\
						& (-0.6955,    1.2389)	& (-0.0601,    1.4242)  	& (-0.5912,    0.8163) \\
\multirow{2}{40pt}{Race 2}		& 0.3730  				& \bf{0.7778}    			& {0.3307} \\
						& (-0.6361,    1.4630)	& (0.0377,    1.5489)  	& (-0.4025,    1.0383) \\
\multirow{2}{40pt}{Race 3}		& {0.2222}  			& \bf{0.9318}    			& {0.3871} 	 \\
						& (-0.8744,    1.2838)	& (0.1707,    1.7002)   	& (-0.3557,    1.0959) \\
\multirow{2}{40pt}{Race 4}		& {0.7873}  			& \bf{0.8241}    			& {0.2596} 	 \\
						& (-0.3523,    1.7383)	& (0.0159,    1.6462)  	& (-0.4826,    0.9952) \\
\multirow{2}{40pt}{Age 1}		& {-0.5335}  			& {-0.0646}    			& {-0.2986} 	 \\
						& (-1.2414,    0.1857)	& (-0.6731,    0.5503)  	& (-0.8749,    0.2256) \\
\multirow{2}{40pt}{Age 2}		& -0.4836 				& -0.0865   			& {-0.2780} \\
						& (-1.1630,    0.2302)	& (-0.6755,    0.4994)  	& (-0.8378    0.2431) \\
\multirow{2}{40pt}{Age 3}		& -0.5214  			& -0.0637    			& {-0.2779}	 \\
						& (-1.1801,    0.1904)	& (-0.6242,    0.5048)  	& (-0.8315,    0.2064) \\
\multirow{2}{40pt}{Age 4}		& -0.2687 				& 0.1233   			& {-0.2016} \\
						& (-0.9604,    0.4463)	& (-0.4948,    0.7186)  	& (-0.7821,    0.3007) \\
\multirow{2}{40pt}{Age 5}		& -0.5619 				&  -0.0868   			& {-0.2880}	 \\
						& (-1.3525,    0.1676)	& (-0.6703,    0.4953)  	& (-0.8680,    0.2599) \\
\cline{1-4}
\hline
\end{tabular}
\caption{\baselineskip=10pt 
Results for the synthetic data set corresponding to the 25th percentile of the average ISEs 
produced by the method of \cite{Zhang2011b}. 
Posterior means of the fixed effects regression effects and associated $90\%$ credible intervals. 
The estimates with credible intervals not including zero are highlighted in bold.  
}
\label{tab: Sim Study Zhang_2011b Results}
\end{center}
\end{table}
\vspace{-10pt}

\clearpage\newpage
\section{Multivariate Density Regression} \label{sec: mvt dens est}

The applicability of the methodology developed herein for 
modeling covariate informed multivariate densities is not restricted exclusively to deconvolution problems 
but the different model components
can be adapted to other important statistics problems as well. 

For example, the methodology developed in Section 2.1 in the main paper for modeling $f_{\bx \mid \bc}(\bx \mid \bc)$ can be straightforwardly applied to the (order-of-magnitude simpler) problem of
ordinary multivariate density estimation without measurement errors 
in the presence of associated potentially high-dimensional precisely measured covariates. 
Likewise, the methodology developed in Section 2.2 in the main paper for modeling $f_{\bepsilon \mid \bc}(\bepsilon \mid \bc)$ can be straightforwardly applied to modeling covariate dependent regression errors 
in the presence of associated potentially high-dimensional precisely measured covariates.

As a matter of illustration, we discuss here the problem of covariate dependent density estimation 
using the methodology developed in Section 2.1 in the main paper. 
More rigorous exposition of this problem as well as the problem of modeling covariate dependent multivariate regression errors may be pursued separately elsewhere.

Specifically, we have precisely measured observations $\bx_{i}=(x_{1,i},\dots,x_{d,i})\trans$ and associated precisely measured covariates $\bc_{i} = (c_{1,i},\dots,c_{p,i})\trans$ for $n$ observational units $i=1,\dots,n$, 
and the goal is to estimate $f_{\bx \mid \bc}(\bx \mid \bc)$.

The statistics literature on univariate density estimation is enormous; 
and the problem of covariate density estimation, sometimes referred to as density regression, has also received attention 
\citelatex[see, e.g.,][and the references therein]{payne2020conditional}. 
The literature on multivariate density estimation, in contrast, is small 
and the literature on multivariate density regression almost non-existent. 
There does exist a body of works on conditional copula estimation. 
See, for example, \citelatex{gijbels2011conditional} and the subsequent works citing this paper. 
The main focus here is often on modeling the conditional dependence patterns. 
Additionally, most of these works are restricted to simple bivariate settings and/or single covariates.  
An optimal transport based approach was recently considered in \citelatex{tabak2020conditional}. 
Variational Bayes inference for a mixture model with multivariate normal component kernels 
with their means and the mixture probabilities varying with associated covariates has been considered in \citelatex{dao2021flexible}. 
Copula based models with flexible marginals have previously been shown to outperform mixtures of multivariate kernels in realistic scenarios \citeplatex{sarkar2021bayesian}.
Additionally, it is not clear how scalable these methods are for high-dimensional problems -- high-dimensional $\bx$ as well as high-dimensional $\bc$. 
Our method, on the other hand, can efficiently accommodate high-dimensional $\bx$'s by sparse shared atoms mixture models as well as high-dimensional $\bc$'s by sparse conditional tensor factorization techniques. 

We now report the numerical performance of our method for ordinary density estimation problems in simulation experiments. 
To our knowledge, software for the alternatives cited above are not publicly available, 
we restrict ourselves to evaluating the numerical performance of our method alone. 
We simulated from the same true $f_{\bx \mid \bc}(\bx \mid \bc)$ 
that we had considered for the deconvolution problem in Section \ref{sec: simulation studies} in the main paper. 
As in the case of the deconvolution problem, we set $n=965$, and $d=3$, 
and the same number $p=3$ and distribution of covariates as seen in Figure \ref{fig: EATS exploratory cat dists} in the main paper. 
Other details can be found there in the main paper.

\vspace{0.25cm}
\begin{table}[!ht]
\begin{center}
\footnotesize
\begin{tabular}{|c|c|c|}
\hline
\multirow{2}{50pt}{Component}	& \multirow{2}{20pt}{Sex}		& \multicolumn{1}{|c|}{Median ISE $\times 1000$} 	\\ \cline{3-3}
						& 		& Our Model in Section 2.1	\\ \hline \hline
\multirow{2}{20pt}{1} 	& M	& \bf{0.112} 	\\
				& W	& \bf{0.129} 	\\\cline{1-3}
\multirow{2}{20pt}{2} 	& M	& \bf{0.081}  	\\
				& W	& \bf{0.108}		\\\cline{1-3}
\multirow{2}{20pt}{3} 	& M	& \bf{0.046}  	\\
				& W	& \bf{0.033} 	\\\cline{1-3}
\hline
\end{tabular}
\caption{\baselineskip=10pt 
Median integrated squared error (MISE) performance
of the model developed in Section 2.1 in the main paper applied to covariate informed DENSITY ESTIMATION problem.
Here M and W are abbreviations for `men' and `women', respectively.
\vspace{-20pt}
}
\label{tab: DENSITY ESTIMATION MISEs 1}
\end{center}
\end{table}

Table \ref{tab: MISEs 1} reports the median ISEs (MISEs) for estimating the trivariate joint densities and the univariate marginals  
obtained by our method. 
The MISEs reported here are based on $B=100$ simulated data sets.
{As expected, the MISEs are now an order-of-magnitude smaller than the corresponding MISEs for the deconvolution problem reported in Table \ref{tab: MISEs 1} in the main paper.

Figure \ref{fig: DENSITY ESTIMATION Sim Study Inclusion Probabilities} shows the estimated inclusion probabilities of different predictors in the models for $f_{\bx \mid \bc}$. 
Consistent with the simulation truth, 
the set of significant predictors for the density of main interest $f_{\bx \mid \bc}$ is found to comprise the dimension labels ($c_{0}$) and `gender' ($c_{1}$), 
and, for $f_{\bepsilon \mid \bc}$, the set of significant predictors comprises only the dimension labels ($c_{0}$).

Figures \ref{fig: DENSITY ESTIMATION Sim Study ZX Tables} shows 
how the mixture component specific parameters get shared across different components and predictor combinations. 
Consistent with the simulation truth, for the densities $f_{x,\ell \mid \bc}$, the mixture probabilities vary significantly between `men' and `women' as well as between different  components. 
Figure \ref{fig: DENSITY ESTIMATION Sim Study MARGINALS} shows the estimated densities $f_{x,\ell \mid \bc}(x_{\ell} \mid \bc)$ obtained by our method 
superimposed over histograms of the corresponding $x_{\ell}$'s. 
Figure \ref{fig: DENSITY ESTIMATION Sim Study JOINTS M} and \ref{fig: DENSITY ESTIMATION Sim Study JOINTS F} repeat the univariate densities $f_{x,\ell \mid \bc}(x_{\ell} \mid \bc)$ in the diagonal panels 
and also show the estimated joint densities $f_{\bx,\ell_{1},\ell_{2} \mid \bc}(x_{\ell_{1}}, x_{\ell_{2}} \mid \bc)$ obtained by our method in the off-diagonal panels separately for `men' and `women', respectively. 
These graphical summaries suggest the model to provide an excellent fit for the simulated data. 

As also discussed in the concluding section of the main paper, 
our construction implies each component depends on the same of important covariates 
which can be restrictive in real world applications. 
The problem can be addressed by constructing more flexible partition structures. 
In the absence of measurement errors, the copula function can potentially also be modeled more flexibly, 
including possibly allowing it to vary with the covariates. 
The problem of modeling flexible regression errors can be similarly addressed. 
These research directions are being pursued separately elsewhere.

\begin{figure}[!ht]
\centering
\includegraphics[width=7cm, trim=0cm 0cm 0cm 0cm, clip=true]{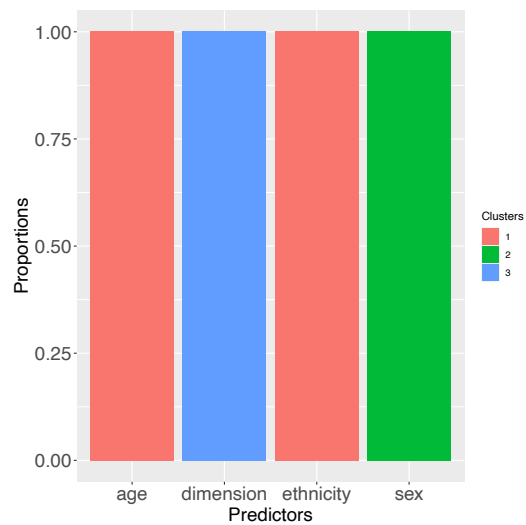}
\caption{\baselineskip=10pt 
Results for DENSITY ESTIMATION for the synthetic data set corresponding to the $25\th$ percentile of the average ISEs, 
showing the estimated probabilities of different numbers of clusters of the associated predictors' levels being included in the model 
for the densities $f_{\bx \mid \bc}$.  
At the median $0.5$ probability level, the component labels and the sex of the subjects are important predictors for modeling the densities $f_{\bx \mid \bc}$. 
The results are consistent with the true simulation scenario.
}
\label{fig: DENSITY ESTIMATION Sim Study Inclusion Probabilities}
\end{figure}

\begin{figure}[!ht]
\centering
\includegraphics[width=11cm, trim=0cm 0.25cm 0cm 0cm, clip=true]{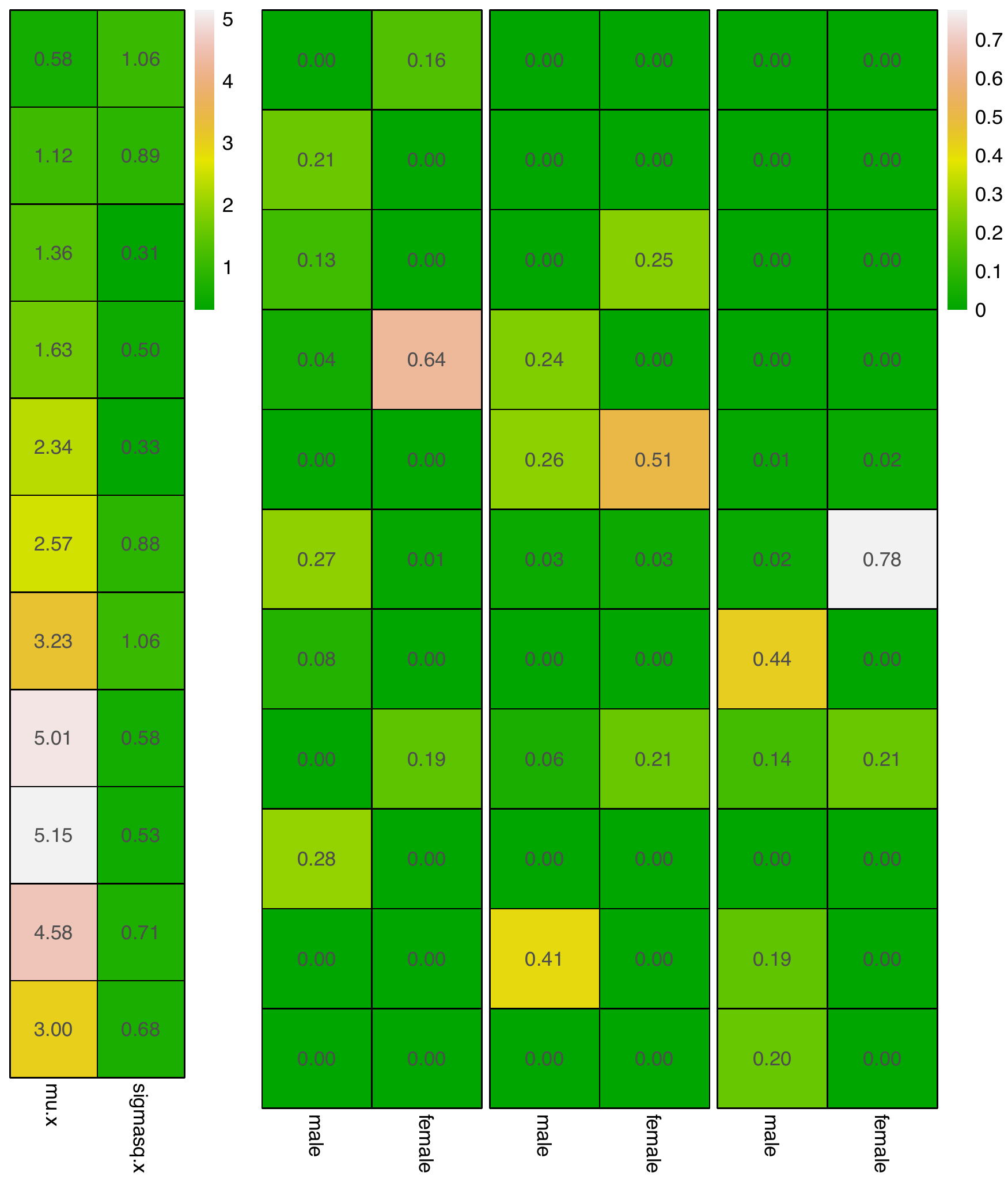}
\caption{\baselineskip=10pt 
Results for DENSITY ESTIMATION for the synthetic data set corresponding to the $25\th$ percentile of the average ISEs. 
These results correspond to the final MCMC iteration but are representative of other iterations in steady state. 
The left panel shows the component specific parameters $(\mu_{x,k}, \sigma_{x,k}^{2})$ 
for the eleven mixture components that were actually used to model the densities $f_{x,\ell \mid \bc}(x_{\ell} \mid \bc)$.  
The right panel shows the associated `empirical' mixture probabilities $\wh{p}_{x}(k \mid c_{0},c_{1},\dots,c_{p}) = \sum_{i=1}^{n}1\{z_{x,\ell,i}=k, c_{0,\ell,i}=c_{0},c_{1,\ell,i}=c_{1},\dots,c_{p,\ell,i}=c_{p}\}/n$ 
for `men' and `women' 
and for the three components, from left to right. 
Results for different combinations of component and `gender' are shown here 
as they are the only predictors important for $\bx$. 
The mixture probabilities vary significantly between these predictor combinations. 
}
\label{fig: DENSITY ESTIMATION Sim Study ZX Tables}
\end{figure}

\begin{figure}[h!]
\centering
\includegraphics[height=16.5cm, trim=0cm 0cm 0cm 0cm, clip=true]{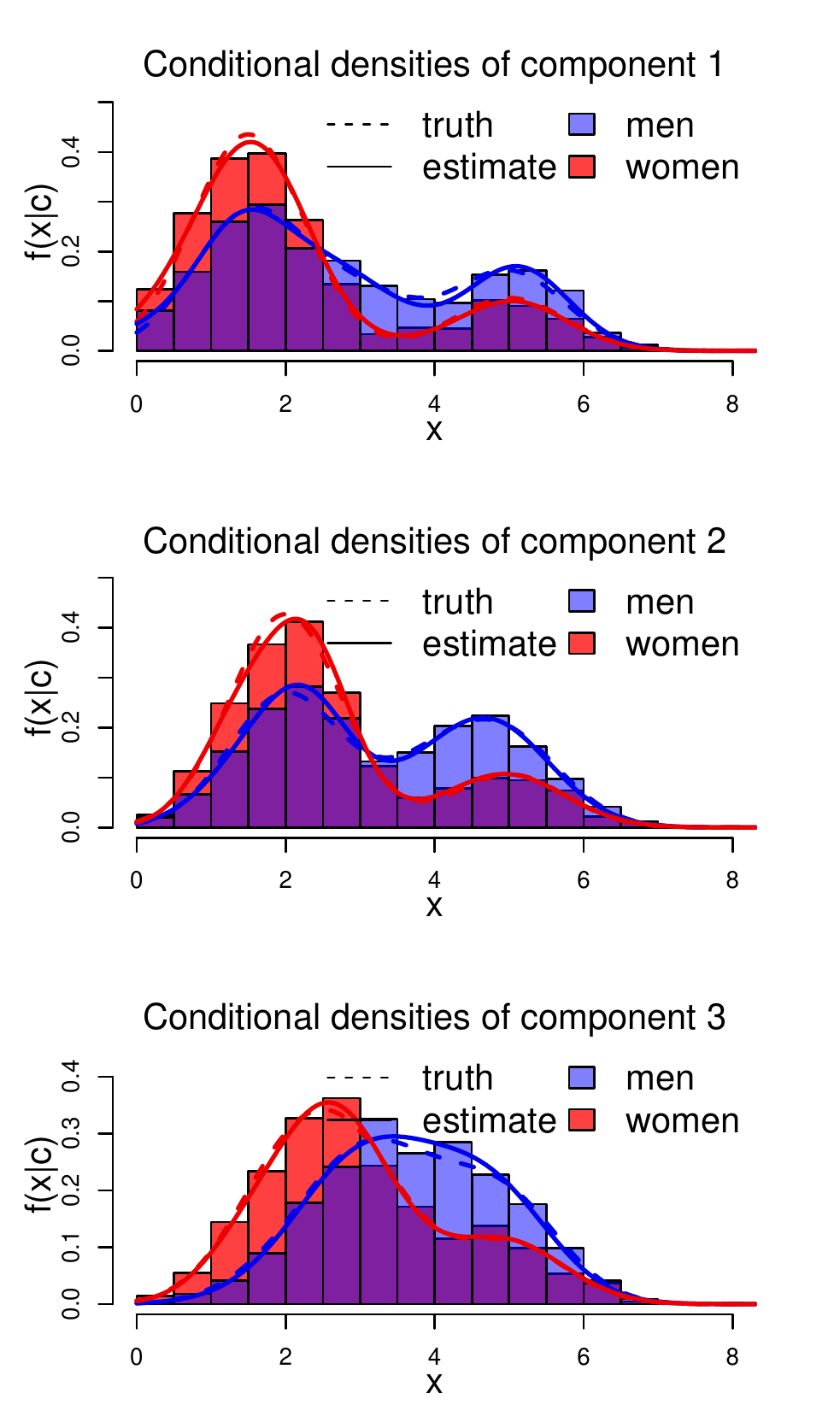}
\caption{\baselineskip=10pt 
Results for DENSITY ESTIMATION for the synthetic data set corresponding to the $25\th$ percentile of the average ISEs. 
From top to bottom, the left panels show the estimated conditional densities $f_{x,\ell \mid \bc}(x_{\ell} \mid \bc)$ 
obtained by our method %(solid lines) 
and the corresponding truths. %(dashed lines).  
Results for different component and `gender' combinations are shown here as 
they are the only predictors important for modeling the densities $f_{x,\ell \mid \bc}(x_{\ell} \mid \bc)$. 
}
\label{fig: DENSITY ESTIMATION Sim Study MARGINALS}
\end{figure}

\begin{figure}[!ht]
\centering
\includegraphics[height=16.5cm, width=16cm, trim=0cm 0cm 0cm 0.5cm, clip=true]{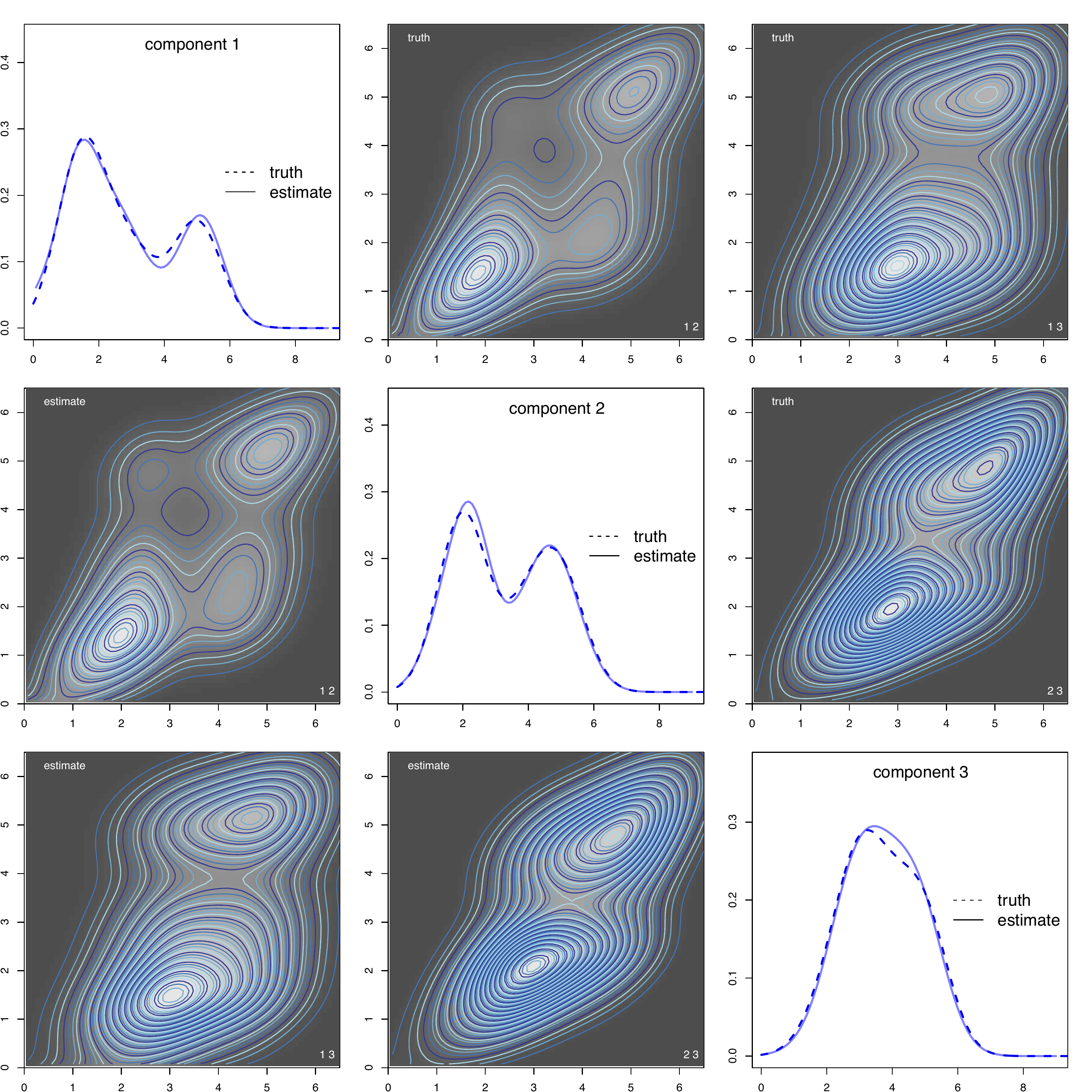}
\caption{\baselineskip=10pt 
Results for DENSITY ESTIMATION for the synthetic data set corresponding to the 25th percentile of the average ISEs. 
Results for `men' are shown here. 
The diagonal panels show %the true (dashed lines) 
one dimensional marginal densities and the corresponding estimates %(solid lines) 
produced by our method. 
The off-diagonal panels show the contour plots of the true two-dimensional densities $f_{\bx,\ell_{1},\ell_{2}\mid \bc}(x_{\ell_{1}}, x_{\ell_{2}} \mid \bc)$ %(upper triangular panels) 
and the corresponding estimates obtained by our method. %(lower triangular panels).
Axis labels are suppressed to allow more space for the individual panels.
}
\label{fig: DENSITY ESTIMATION Sim Study JOINTS M}
\end{figure}

\begin{figure}[!ht]
\centering
\includegraphics[height=16.5cm, width=16cm, trim=0cm 0cm 0cm 0.5cm, clip=true]{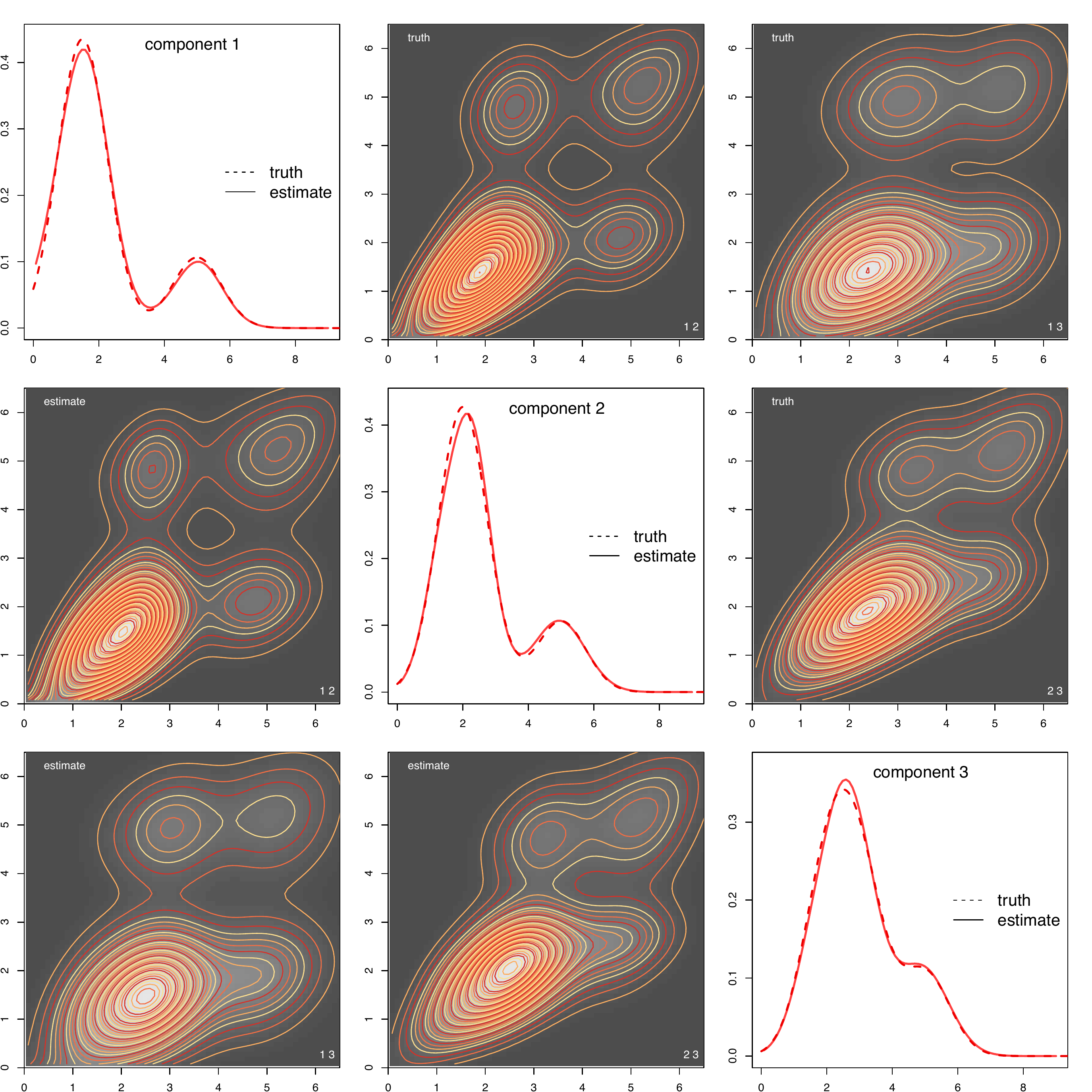}
\caption{\baselineskip=10pt 
Results for DENSITY ESTIMATION for the synthetic data set corresponding to the 25th percentile of the average ISEs. 
Results for `women' are shown here. 
The diagonal panels show %the true (dashed lines) 
one dimensional marginal densities and the corresponding estimates %(solid lines) 
produced by our method. 
The off-diagonal panels show the contour plots of the true two-dimensional densities $f_{\bx,\ell_{1},\ell_{2}\mid \bc}(x_{\ell_{1}}, x_{\ell_{2}} \mid \bc)$ %(upper triangular panels) 
and the corresponding estimates obtained by our method. %(lower triangular panels).
Axis labels are suppressed to allow more space for the individual panels.
}
\label{fig: DENSITY ESTIMATION Sim Study JOINTS F}
\end{figure}

\clearpage
\bibliographystylelatex{natbib}
\bibliographylatex{BNP,ME,Copula,HOMC}

\end{document}